\documentclass[a4paper,11pt]{article}
\usepackage{jcappub}

\usepackage[latin1]{inputenc}
\usepackage{amsmath, amsfonts, amssymb, hyperref, graphicx, multirow, subfigure, bbold}
\hypersetup{colorlinks=True,citecolor=blue}
\usepackage[usenames,dvipsnames,svgnames,table]{xcolor}
\usepackage[normalem]{ulem}
\usepackage{caption}
\usepackage{array}
\usepackage{aas_macros}

\newcommand{\beq}{\begin{equation}}
\newcommand{\eeq}{\end{equation}}
\newcommand{\beqry}{\begin{eqnarray}}
\newcommand{\eeqry}{\end{eqnarray}}
\newcommand{\beqrys}{\begin{subequations}\begin{eqnarray}}
\newcommand{\eeqrys}{\end{eqnarray}\end{subequations}}
\newcommand{\fqu}{\begin{bmatrix} Q \\ U \end{bmatrix}}

\newcommand{\qutox}{\begin{bmatrix} \mathbb{1} & i\mathbb{1}  \\  \mathbb{1} & -i\mathbb{1}   \end{bmatrix}}
\newcommand{\qutoxd}{\begin{bmatrix} \mathbb{1} & \mathbb{1}  \\  -i\mathbb{1} & i\mathbb{1}   \end{bmatrix}}

\newcommand{\ymatc}{\begin{bmatrix} _{+2}Y^* & 0  \\  0 & _{-2}Y^*  \end{bmatrix}}
\newcommand{\yzmat}{\begin{bmatrix} _{0}Y & 0  \\  0 & _{0}Y  \end{bmatrix}}

\newcommand{\bmat}{\begin{bmatrix}}
\newcommand{\emat}{\end{bmatrix}}
\newcommand{\mdi}{\mathcal{D}/\mathcal{I}}
\newcommand{\mm}{\mathcal{M}}
\newcommand{\md}{\mathcal{D}}
\newcommand{\mi}{\mathcal{I}}
\def\vp#1{$\bar{P}_{#1}$ }
\def\vs{$\bar{S}~$}
\def\eq#1{{Eq.~(\ref{#1})}}
\def\sec#1{{Sec.~\ref{#1}}}
\def\fig#1{{Fig.~\ref{#1}}}

\def\ymatc#1{\begin{bmatrix} _{+2}Y^{\dag}_{#1} & 0  \\  0 & _{-2}Y^{\dag}_{#1}  \end{bmatrix}}
\def\yzmat#1{\begin{bmatrix} _{0}Y_{#1} & 0  \\  0 & _{0}Y_{#1}  \end{bmatrix}}

\begin{document}
\title{Real-space computation of $E$/$B$-mode maps I: Formalism, Compact Kernels, and Polarized Filaments}

\author[a,b]{Aditya Rotti}
\author[a]{and Kevin Huffenberger}
\affiliation[a]{Department of Physics, Florida State University, Keen Physics Building, Tallahassee, Florida 32306, USA}
\affiliation[b]{Jodrell Bank Center for Astrophysics, University of Manchester, Oxford Road, Manchester M13 9PL, UK}
\emailAdd{adityarotti@gmail.com}
\emailAdd{khuffenberger@fsu.edu}

\abstract{
  \noindent {We derive full-sky, real-space operators that convert between polarization Stokes $Q/U$ parameters and the coordinate-independent scalar $E$/$B$ modes that are widely used in Cosmic Microwave Background (CMB) and cosmic shear analysis. We also derive real space operators that decompose the measured Stokes parameters into those corresponding to $E$-modes and $B$-modes respectively, without ever evaluating the scalar fields themselves. We cast the standard CMB polarization analysis operators in a matrix-vector notation which elucidates these derivations.  For all these real space operators we show that the kernels split naturally into angular and radial parts and we show explicitly how the radial extent of these kernels depends on the targeted band-limit. We show that the kernels can be interpreted either as a complex convolving beam or as a Green's function when they are expressed in terms of the forward or inverse rotation Euler angles. We show that an arbitrary radial function can produce $E/B$-like maps, provided it vanishes at the origin and the  antipodal point.  These maps are simply filtered versions of the standard $E/B$ maps.  We can recover the standard power spectrum of the polarized CMB sky by correcting the power spectra of these maps with a simple window function, which we show how to derive for any radial dependence.  For these reasons we can compute $E/B$ maps in real space with a compactly-supported kernel, an approach that can guarantee the avoidance of known foreground regions and could be employed in a massively-parallel scheme at high-resolution.  We show that the spin raising  and lowering  operators $\eth^2$/$\bar{\eth}^2$ are special cases of these generalized radial functions, and present their band limited versions. 
    The spatial structure of the real space operators provides great intuition for the $E/B$ structure of polarized, filamentary galactic foregrounds.  We predict a non-zero $B$-mode signature that is expected from polarized filaments in the sky. This paper is the first part in a series of papers that explore real-space computation of polarization modes and their applications.} 
}
\maketitle
\begingroup
\let\clearpage\relax
\section{Introduction}
During recombination, the Cosmic Mircowave Background (CMB) undergoes Thomson scattering that leaves it with $\sim 5$ percent linear polarization.  The polarization signal contains information about the plasma velocity and provides cosmological constraints independent from the signal in temperature anisotropies \citep{1997NewA....2..323H}.  Standard CMB analysis techniques involve translating the Stokes parameters of polarization into scalar ($E$) and pseudo-scalar ($B$) modes, since the statistics of these coordinate independent scalar fields are predicted by theory.   

The dominant contribution to the $E$-modes of polarization is sourced by primordial scalar perturbations, but these do not generate $B$-modes of polarization at first order. Various phenomena may generate $B$-modes in polarization measurements of the microwave sky: primordial tensor perturbations (gravitational waves) \citep{1997PhRvD..56..596H,1997PhRvL..78.2054S}; weak gravitational lensing of $E$-modes induced by the potentials of intervening large scale structures;
Milky Way foregrounds (especially Galactic synchrotron and dust emission)
\citep{2016A&A...586A.133P}; uncorrected systematic problems in the data \citep{2003PhRvD..67d3004H,2008PhRvD..77h3003S}; and unknown, exotic phenemena like cosmic birefringence or primordial magnetic fields
\citep{1996ApJ...469....1K,1999PhRvL..83.1506L,2004ApJ...616....1C,2014MNRAS.438.2508P}.

The pseudo-scalar $B$-modes particularly on large angular scales ($\sim 2$ degrees) are expected to have a significant contribution from primordial tensor perturbations generated during inflation. Hence a measurement of these large angular scale B-modes will yield information about the statistical properties of tensor perturbations that will eventually lead to important constraints on models of inflation.  The measurement of lensing $B$-modes on small angular scales ($\sim$ few arcminute) will yield information on the clustering of matter across cosmic ages \citep{Abazajian2015, Kamionkowski2016,Abazajian2016,Hu2002c,Wehus2016}.
  
The primary aim of the current CMB experiments is to make precise measurements of CMB polarization. While the $E$-modes of polarization have been measured reasonably well by a number of experiments (\cite{2018RPPh...81d4901S} gives a recent review),  ongoing and future CMB experiments aim to measure the $B$-modes of CMB polarization with unprecedented accuracy. The instruments are swiftly approaching the desired sensitivities to enable us to in principle measure $B$-mode signals with $r\gtrsim 0.001$ \cite{Spider, CLASS,Litebird, BICEP22015, 2016arXiv161002743A, 2017arXiv170602464A,Delabrouille2017}. However $B$-modes generated by galactic foreground are expected to be a few orders of magnitude higher than $B$-mode amplitude measurable by these detectors.  Precise modelling and subtraction of this large foreground contribution poses a major challenge for a robust unravelling of the minuscule $B$-mode signal. 
 
The formalism for converting the Stokes parameters to scalar quantities is well established \citep{1997PhRvD..55.7368K,1997PhRvD..55.1830Z}. The spin-0 scalar $E$/$B$ modes relate to the spin-2 complex Stokes parameters via the spin-raising and -lowering operators ($\eth^2,\bar \eth^2$), which are derivatives evaluated locally and filtered.  In practice the $E$/$B$ modes are computed to some specified band limit, and the filtering and band limit make them non-local functions of the polarization field.  In other words, $E$/$B$ modes evaluated at a point receive contributions from all over the sky. In this work we aim to gain real space insights into the non-locality of the $E/B$ fields compared to the Stokes parameters. With renewed focus on foreground contamination to the $B$-mode signal we aim to gain an intuition for $E/B$ mode patterns resulting from physical polarized structures in the galaxy. These real space insights may yield new ideas for minimizing foreground contamination that are not obvious using conventional approaches. 

{Some of the ideas presented in this work bear resemblance to those in Zaldarriaga (2001) \citep{Zaldarriaga2001a}.  Here we dig deeper into the mathematical formulation of the real space operations on the curved sky.  This has consequently led to updates in interpretation and some substantial differences in detail.  We describe our approach in this article and will discuss applications of the codes and analysis tools we have developed in a subsequent publication.}
 
This paper is organized in the following manner: In \sec{sec:pol-primer} we present a primer on the description of CMB polarization on the sphere, beginning with a heuristic argument that makes transparent the real space construction of $E$/$B$ modes.  We discuss the standard harmonic-space procedures for this operation. Finally, we introduce a matrix-vector notation which yields a more concise description of the harmonic space procedures. In \sec{sec:qu2eb} and \sec{sec:eb2qu} we derive and discuss the real space operators that transform $Q$/$U$ to $E$/$B$ and vice versa. In \sec{sec:visualize_operator} we evaluate these real space operators and present visualizations of these functions. In \sec{sec:purify_stokes_qu} we derive a real space operator that decomposes the Stokes $Q$/$U$  parameters into components that correspond to $E$ and $B$ modes respectively and present its visualizations.  In \sec{sec:radial_locality} we study the locality of the real space operators and explore its band limit dependence. In \sec{sec:generalized_operators} we present a systematic method of generalizing the real space operators by controlling the non-locality while recovering the standard power spectra.  We discuss the connection to the standard spin raising and lower operators. In \sec{sec:pol_filaments} we discuss $E/B$ mode signatures of foreground filaments. In \sec{sec:discussion}, we conclude with a summary and discuss the prospects of this new method for analyzing CMB polarization maps.
\section{Polarization primer} \label{sec:pol-primer}
\subsection{Heuristic, real-space construction of E/B fields on the sphere} \label{sec:qu2eb_heuristic}

CMB polarization is measured in terms of Stokes parameters, time averages of the linear polarization of the electric field along cartesian axes perpendicular to the line of sight.\footnote{Throughout we use the conventions of HEALPix \cite{healpix_primer}, measuring the polarization angle East of South.} Thus Stokes $Q$ and $U$ depend on the choice of the local coordinate system, and a rotation by an angle $\psi$ around the line of sight transforms them as:
\beq \label{eq:qu-rot}
\fqu' = \begin{bmatrix} \cos{2 \psi} &  \sin{2 \psi} \\ -\sin{2\psi} & \cos{2 \psi} \end{bmatrix} \fqu \,.
\eeq
Equivalently, the object $_{\pm 2}{X}(\hat{n}) = Q(\hat{n}) \pm i U (\hat{n})$ transforms as ${}_{\pm 2}f' = e^{\mp 2i\psi} {}_{s}f$ and hence forms a spin $\pm$2 field \cite{Zaldarriaga1997}.

The standard construction of $E$ and $B$ fields arise from the desire to have a coordinate independent description of the polarization. This follows from operations that raise (or lower) the spin of the field ${}_{\pm2}{X}$ to construct scalar fields.  By understanding the transformation properties of the Stokes parameters and those of the Euler angles on the sphere, we can already construct a heuristic argument for what these operations must look like in real space. We consider the contribution to a scalar field at $\hat n_e$ from the polarization field at $\hat n_q$. 

\fig{fig:euler_angles} shows that the transformation of the local coordinate system between the two positions can be described by a counter-clockwise rotation around the local $\hat n_q$ (unit radial vector at $(\theta_q,\phi_q)$ pointing outward) by angle $\alpha$, parallel transport by angle $\beta$ along the shorter geodesic and a counter-clockwise rotation around $\hat n_e$ by $-\gamma$.  This corresponds to a rotation by Euler angles $(\alpha,\beta,-\gamma)$ in the $z-y_1-z_2$ convention.\footnote{The Euler angles in the more standard $z-y-z$ convention are related to those in the $z-y_1-z_2$ convection by the following rule: $(\alpha,\beta,\gamma)_{z-y-z} =(\gamma,\beta,\alpha)_{z-y_1-z_2}$ \cite{varshalovich}.}
\begin{figure}
\centering
\includegraphics[width=0.5\columnwidth]{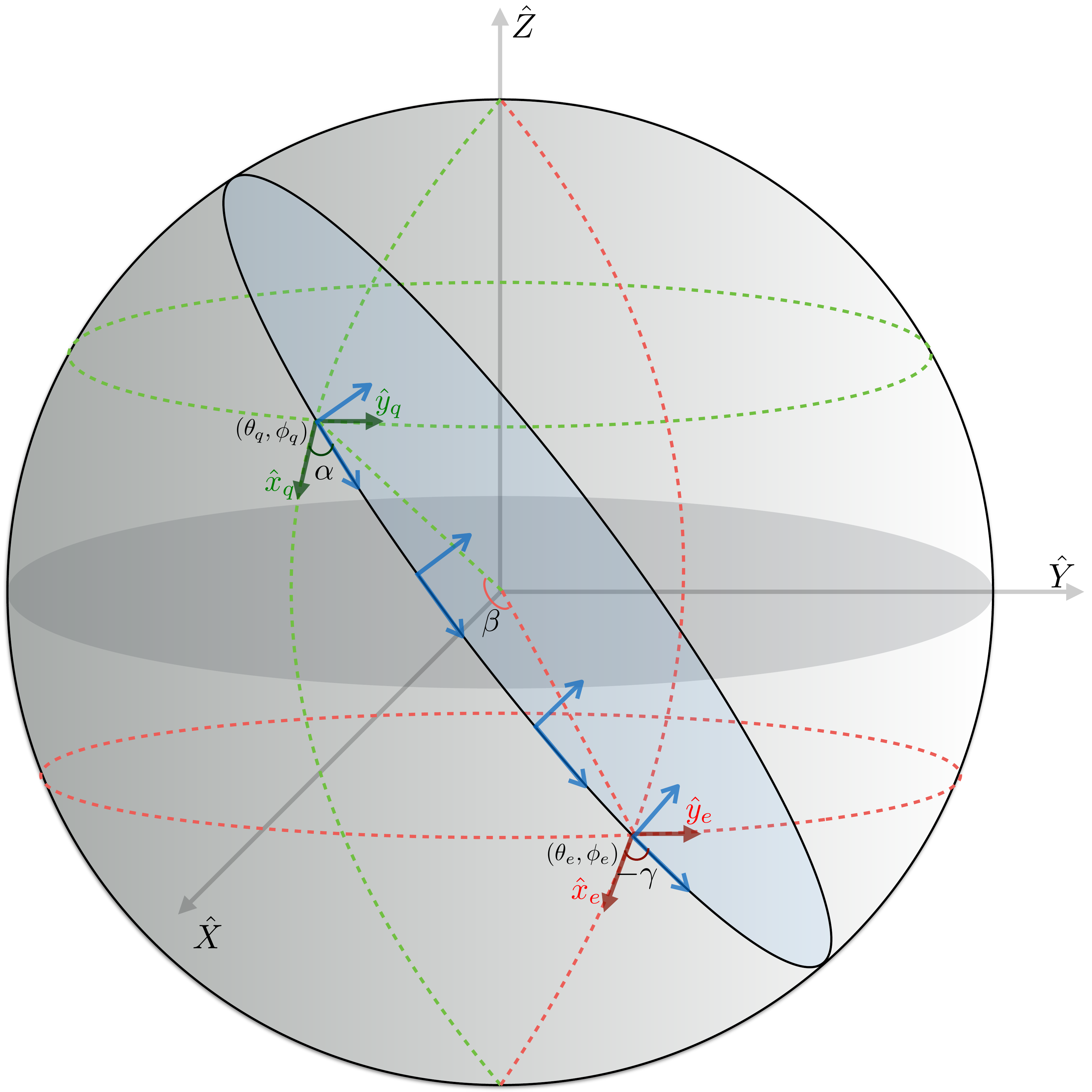}
\caption{This figure depicts the Euler angles in the $z-y_1-z_2$ convention. The cartesian coordinate axes shown in dark solid green are those that lie in the tangent plane at location $\hat{n}_q = (\theta_q, \phi_q)$ while those shown in dark solid red are the ones that lie in the tangent plane at location $\hat{n}_e = (\theta_e, \phi_e)$. The blue axes represent the parallel transport along the geodesic connection between the two locations $\hat{n}_q$ and $\hat{n}_e$ on the sphere.}
\label{fig:euler_angles}
\end{figure}

We consider the impact of local rotations on the Stokes parameters and on these Euler angles. Rotating the cartesian coordinates in the tangent plane at location $\hat{n}_q$ by an angle $\psi$ about the local $\hat{z}_q$ axis, the Stokes parameters in the new coordinate system relate to those in the original coordinate system as:
$\mathcal{R}_{\hat{z}_q}(\psi)[{}_{+2}X(\hat{n}_q)] =  {}_{+2}X(\hat{n}_q) e^{-i2\psi} $.
This same rotation by $\psi$ alters the Euler angle $\alpha_{qe}$, the angle that aligns the $x$-axis at $\hat{n}_q$ along the geodesic to the location $\hat{n}_e$, so that: $\mathcal{R}_{\hat{z}_q}(\psi)[\alpha_{qe}] = \alpha_{qe} - \psi$.  Therefore one can see that: $\mathcal{R}_{\hat{z}_q}(\psi)[e^{-i2\alpha_{qe}}] =  e^{-i2\alpha_{qe}} e^{i2\psi}$.

Given these transformation properties, the combination ${}_{+2}X(\hat{n}_q)e^{-i2\alpha_{qe}}$ is invariant under rotations and hence must be spin-0 or scalar quantity by definition:
\beq
\mathcal{R}_{\hat{z}_q}(\psi)[{}_{+2}X(\hat{n}_q)e^{-i2\alpha_{qe}}] = {}_{+2}X(\hat{n}_q) e^{-i2\alpha_{qe}} \,. \label{eq:invariant}
\eeq
Thus we can build a scalar polarization quantity out of such combinations.
Further note that both $Q$ and $\cos2 \alpha$ have even parity since they do not change sign when $\hat{x} \rightarrow -\hat{x}$ (or $\hat{y} \rightarrow -\hat{y}$).  Meanwhile $U$ and $\sin 2 \alpha$ change sign under this transformation and hence have odd parity. The real part of the function  ${}_{+2}X(\hat{n}_q)e^{-i2\alpha_{qe}}$ must have even parity, because it is composed of terms containing $Q\cos{2\alpha}$ and  $U\sin{2 \alpha}$ which are product of functions with the same parity. Similarly, the imaginary part of the function must have odd parity, because it is composed of $Q\sin{2 \alpha}$ and $U\cos{2\alpha}$ which are product of functions with opposite parity.  Therefore we can make the association that contributions to  $(E+iB)(\hat n_e)$ must be proportional to $ {}_{+2}X(\hat{n}_q) e^{-i2\alpha_{qe}}$.

The same rotation $\mathcal{R}_{\hat{z}_q}(\psi)$ leaves the Euler angle $|\beta_{qe}|$ unaltered (it measures the angular distance between the points).  Thus we further conclude that the contribution to $(E+iB)(\hat n_e)$ from the position $\hat{n}_q$ must have the generic form:
\beq
{}_{+2}X(\hat{n}_q) f(\beta_{qe})  e^{-i2\alpha_{qe}}
\eeq
for some real function $f$. The total $(E+iB)$ field results from summing over all  contributions  from Stokes parameters ${}_{+2}X(\hat{n}_q)$ across the sphere.

Geometry places immediate constraints on function $f$.  When the two locations coincide ($\beta_{qe}=0$) then  $\alpha_{qe}=0,2\pi,4\pi,\dots$, implying $E + iB \propto Q+iU$.  This is a contradiction because $Q+iU$ does not transform as a spin-0 field under local rotations, hence we must have $f(\beta_{qe} = 0 ) = 0$. This implies that the $E/B$ fields are necessarily defined non-locally.  A similar contradiction arises when the two locations are diametrically opposite, $\beta_{qe} = \pi$, and therefore we also require that $f(\beta_{qe} = \pi ) = 0$.  Any function that satisfies these constraints will let us construct $E/B$-like scalar fields.   Below we derive the particular $f$ that gives rise to our familiar $E/B$ modes.

Note that this type of real-space construction can be generalized to transform a field of any spin to a field of any other spin, not just two and zero, and so we can use a similar construction (in the opposite direction) to transform $E/B$ maps back to the Stokes parameters (i.e. transforming spin-0 fields to spin-2).

\subsection{Standard $E$/$B$ fields}

The standard construction of $E/B$ fields is expressed in terms of the spin-raising and spin-lowering operators and this operation is usually carried out in harmonic space. The spin-raising operator ($\eth$) applied to a field of spin-s $_{s}g$, results in a field with spin-$(s+1)$: $(\eth _{s}g)' = e^{-i(s+1)\psi}(\eth _{s}g)$  \cite{goldberg67}.  The complementary spin-lowering operator $(\bar{\eth})$  similarly results in a field with spin-$(s-1)$: $(\bar{\eth} _{s}g)' = e^{-i(s-1)\psi}(\bar{\eth} _{s}g)$.  The complex spin-0 scalar now arise from these spin lowering/raising operations of the spin-2 fields ${_{\pm 2}X}$ as follows:
\begin{subequations}\label{eq:ebdef}
\beqry
\mathcal{E}(\hat{n}) + i \mathcal{B}(\hat{n}) &=& -\bar{\eth}^2 _{+ 2}\bar{X}(\hat{n}) \,,\label{eq:ebdef_lower}\\
\mathcal{E}(\hat{n}) - i \mathcal{B}(\hat{n}) &=& -{\eth}^2 _{-2}\bar{X}(\hat{n}) \,.
\eeqry
\end{subequations}
The $\cal E/B$ fields are defined locally at point $\hat n$ in terms of the operators $\eth$ and $\bar \eth$. It is possible to decompose the complex field $_{\pm 2}\bar{X}$ into spin spherical harmonic functions: ${}_{\pm 2}\bar{X}(\hat{n}) = \sum_{\ell m} {}_{\pm 2} \tilde X_{\ell m} {}_{\pm 2}Y_{\ell m}(\hat{n})$. Applying the spin raising and lowering operators on the spin spherical harmonic functions leads to the following identities \cite{goldberg67}:
\begin{subequations}\label{eq:spinopylm} 
\beqry
\eth _s Y_{lm}(\hat{n}) &=& \sqrt{(\ell-s)(\ell+s+1)} _{s+1} Y_{lm}(\hat{n}) \,, \\
\bar{\eth} _s Y_{lm}(\hat{n}) &=& -\sqrt{(\ell+s)(\ell-s+1)} _{s-1} Y_{lm}(\hat{n}) \,, 
\eeqry
\end{subequations}
where $_s Y_{lm}(\hat{n}) $ denote the spin-s spherical harmonics.

From the definition of $\mathcal{E/B}$, the spin spherical harmonic decomposition of ${}_{\pm2}\bar{X}$, and the identities given in \eq{eq:spinopylm}, it follows that the scalar fields $\mathcal{E}/\mathcal{B}$ are given by the equations:
\beq \label{eq:pseudo}
\mathcal{E}(\hat{n}) = \sum_{\ell m} a^{E}_{\ell m} \sqrt{\frac{(\ell+2)!}{(\ell-2)!}} Y_{\ell m} (\hat{n})\qquad;\qquad
\mathcal{B}(\hat{n})  =\sum_{\ell m} a^{B}_{\ell m} \sqrt{\frac{(\ell+2)!}{(\ell-2)!}} Y_{\ell m} (\hat{n}) \,,
\eeq
where the harmonic coefficients $a^{E}_{\ell m}$ and  $a^{B}_{\ell m}$ relate to the harmonic coefficients of the spin-2 polarization field via the following equations:
\beq\label{eq:x2eb}
a^{E}_{\ell m} = -\frac{1}{2} \Big[ {}_{+2}\tilde{X}_{\ell m} + {}_{-2}\tilde{X}_{\ell m} \Big]\qquad;\qquad a^{B}_{\ell m} = -\frac{1}{2i} \Big[ {}_{+2}\tilde{X}_{\ell m} - {}_{-2}\tilde{X}_{\ell m} \Big] \,.
\eeq
In the remainder of this article, we will work with the scalar $E$ and pseudo scalar $B$ fields, defined by: 
\beq \label{eq:realeb}
E(\hat{n}) = \sum_{\ell m} a^{E}_{\ell m} Y_{\ell m} (\hat{n})\qquad;\qquad B(\hat{n})  =\sum_{\ell m} a^{B}_{\ell m} Y_{\ell m} (\hat{n}) \,.
\eeq
These $E/B$ fields are merely versions of $\mathcal{E}/\mathcal{B}$ that downweight higher-$\ell$ modes  (i.e. their spherical harmonic coefficients of expansion are reduced by the factor $[{(\ell-2)!}/{(\ell+2)!}]^{1/2}$).

\subsection{Matrix notation} \label{sec:mat_pol_intro}
Our derivations of the real space operators are more transparent in a matrix-vector notation\footnote{While we work with the matrix and vector sizes given in terms of some pixelization parameter $\rm N_{\rm pix}$, all the relations are equally valid in the continuum limit attained by allowing $\rm N_{\rm pix}\rightarrow \infty$}. We introduce a matrix that encodes spin spherical harmonic basis vectors:
\beq
{}_{|s|}\mathcal{Y}= \bmat _{+s}Y & 0 \\ 0 & _{-s}Y \emat _{2 \rm N_{\rm pix} \times 2 \rm N_{\rm alms}} ;\qquad  {}_{|s|}\mathcal{Y}^{\ddagger}= \Delta \Omega \bmat _{+s}Y^{\dagger} & 0 \\ 0 & _{-s}Y^{\dagger} \emat _{2 \rm N_{\rm alms} \times 2 \rm N_{\rm pix}} \,,
\eeq
where $s$ denotes the spin of the basis functions and our definition of ${}_{|s|}\mathcal{Y}^{\ddagger}$ differs from the conventional conjugate transpose operation by the factor $\Delta \Omega$.  We introduce this to ensure the orthonormality of these operations on the discretized sphere when the pixel size is sufficiently small, ${}_{|s|}\mathcal{Y}^{\ddagger} {}_{|s|}\mathcal{Y} = I_{2 \rm N_{\rm alms} \times2 \rm N_{\rm alms}}$, and also maintain the standard definition of spherical harmonics. 

We will be working with cases $s \in [0,2]$. Each column of ${}_{|s|}\mathcal{Y}$ maps to a specific harmonic basis function (i.e. indexed by $\ell m$) and each row maps to a pixel on the sphere. This matrix is not square in general: the number of rows is determined by the pixelization and the number of columns is set by the number of basis functions (e.g. determined by the band limit).

We now define the different polarization data vectors and their representation in real and harmonic space as follows\footnote{We adopt a convention in which real space quantities are denoted by bar-ed variable while those in harmonic space are denoted by tilde-ed variables.}:
\beqrys
\bar{S} &=& \bmat E \\ B  \emat_{2 \rm N_{\rm pix} \times 1};\qquad \bar{X} = \bmat _{+2}X \\ _{-2}X \emat_{2 \rm N_{\rm pix} \times 1};\qquad \bar{P} =\fqu_{\tiny {2 \rm N_{\rm pix} \times 1}} \,, \\
\tilde{S} &=& \bmat a^{E} \\ a^{B} \emat _{2 \rm N_{\rm alms} \times 1};\qquad \tilde{X} = \bmat _{+2} \tilde{X} \\ _{-2} \tilde{X} \emat_{2 \rm N_{\rm alms} \times 1} \,.
\eeqrys
The symbols have the same meaning as in \sec{sec:pol-primer}, except that the subscript ${\ell m}$ for the spherical harmonic coefficients is suppressed for cleaner notation.

We define an operator that transforms between different representations of the polarization field (i.e. from $Q,U$ to $_{\pm2}\bar{X}$ and back):
\beqrys
\bar T &=& \qutox_{2 \rm N_{\rm pix} \times 2 \rm N_{\rm pix}} ;\qquad \bar T^{-1} = \frac{1}{2} \bar T^{\dagger} \,, \\
\tilde T &=& -\qutox_{2 \rm N_{\rm alms} \times 2 \rm N_{\rm alms}};\qquad \tilde T^{-1} = \frac{1}{2} \tilde T^{\dagger} \,,
\eeqrys
The sign conventions we have chosen matches that of HEALPix.
Using the data vectors and the matrix operators defined above we can now express, in compact notation, the forward and inverse relations between different representations of the polarization data vectors via the following equations:
\begin{subequations} \label{eq:pol_data_relns}
  \beqry
  \bar{X} &= \bar T  \bar{P} ; &\qquad \bar{P} = \frac{1}{2} \bar T^{\dagger}  \bar{X} \,, \\
  \tilde{X} &= \tilde T \tilde{S}; &\qquad \tilde{S} = \frac{1}{2}\tilde T^{\dagger} \tilde{X} \,.
  \eeqry
  Meanwhile the spherical harmonic transforms are written as:
  \beqry
  \bar X &=  {{}_2\mathcal{Y}}  \tilde X; &\qquad \tilde X ={{}_2\mathcal{Y}}^{\ddagger}  \bar X  ; \\
  \bar S &=  {{}_0\mathcal{Y}} \tilde S; &\qquad  \tilde S =  {{}_0\mathcal{Y}}^{\ddagger} \bar S \,.
  \eeqry
\end{subequations}
Finally we introduce the operators that project harmonic space data vector to the $E$ or $B$ subspace:
\begin{subequations} \label{eq:har_eb_op}
\beqry
\tilde O_E &=& \bmat \mathbb{1} & \mathbb{0} \\ \mathbb{0} & \mathbb{0} \emat _{2 \rm N_{\rm alms} \times 2 \rm N_{\rm alms} }; \qquad \tilde S_E = \tilde O_E  \tilde S ,\\
\tilde O_B &=& \bmat \mathbb{0} & \mathbb{0} \\ \mathbb{0} & \mathbb{1} \emat _{2 \rm N_{\rm alms} \times 2 \rm N_{\rm alms} }; \qquad \tilde S_B = \tilde O_B  \tilde S .
\eeqry
\end{subequations}
Note that these harmonic space matrices are idempotent ($\tilde O_E  \tilde O_E = \tilde O_E;  \tilde O_B  \tilde O_B= \tilde O_B$), orthogonal ($\tilde O_E  \tilde O_B = \mathbb{0}$) and sum to the identity matrix ($\tilde O_E + \tilde O_B = \mathbb{1}$).

The above relations for these harmonic space operators are exactly valid.  In the following sections we derive the real space analogues ($O_E,O_B$) of these harmonic space operators.

\section{Real space polarization operators} \label{sec:real_space_operators}
\subsection{Evaluating scalars $E/B$ from Stokes $Q/U$}\label{sec:qu2eb}
In \sec{sec:pol-primer} we described the conventional procedure of computing the scalar fields $E/B$ from the Stokes parameters $Q/U$. 
In this section we derive the real space operators which can be used to directly evaluate the scalar fields $E$/$B$ on the sphere.  We use the vector-matrix notation introduced in \sec{sec:mat_pol_intro} to write down an operator equation relating the real space vector of scalars \vs to the Stokes polarization vector \vp{}:
\beqrys
\bar{S} &=& {{}_0\mathcal{Y}} \, \tilde T^{-1}  \, {{}_2\mathcal{Y}^{\ddagger}} \, \bar T  \bar{P}
= \frac{1}{2} {{}_0\mathcal{Y}} \, \tilde T^{\dagger} {{}_2\mathcal{Y}^{\ddagger}} \, \bar T \bar{P} \,,   \\
&=&  \bar O \bar{P}. \label{eq:qu2eb_op}
\eeqrys
The explicit form of the real space operator $\bar O$ can be derived by contracting over all the matrix operators. This procedure is explicitly worked out in the following set of equations:
\beqrys
\bar{O} &=& \frac{1}{2} {{}_0\mathcal{Y}}\, \tilde T^{\dagger} {{}_2\mathcal{Y}^{\ddagger}} \, \bar T \,, \\
&=& -0.5 \Delta \Omega \yzmat{e} \qutoxd \ymatc{q} \qutox   \,, \\
&=& -0.5 \Delta \Omega \begin{bmatrix} \sum ({}_{0}Y_e ~{}_{2}Y^{\dag}_q  +  {}_{0}Y_e~ {}_{-2}Y^{\dag}_q) & {\rm i}  \sum ({}_{0}Y_e ~ {}_{2}Y^{\dag}_q - {}_{0}Y_e ~{}_{-2}Y^{\dag}_q)  \\  - {\rm i} \sum  ({}_{0}Y_e ~ {}_{2}Y^{\dag}_q - {}_{0}Y_e~ {}_{-2}Y^{\dag}_q) & \sum ({}_{0}Y_e~ {}_{2}Y^{\dag}_q + {}_{0}Y_e ~{}_{-2}Y^{\dag}_q)  \end{bmatrix} \,, \label{eq:qu2eb_ker_1}
\eeqrys
where the symbol ${}_{0}Y_e$ is used to denote the sub-matrix ${}_{0}Y_{\hat{n}_e \times \ell m} \equiv {}_{0}Y_{\ell m}(\hat{n}_e)$, the symbol ${}_{\pm 2}Y^{\dag}_q$ is used to denote the transposed conjugated matrix ${}_{\pm 2}Y^*_{\ell m \times \hat{n}_q} \equiv {}_{\pm 2}Y^*_{\ell m}(\hat{n}_q)$ and the summation is over the multipole indices $\ell,m$. As before, we use the notation that the index $e$ denotes the location where the scalar fields are being evaluated, and the index $q$ denotes the location from which  the Stokes parameters are being accessed. Using the conjugation properties of the spin spherical harmonic functions it can be shown that the following identity holds true:
\beq
 \left [\sum_{\ell m} {}_{0}Y_{\ell m}(\hat{n}_e){}_{+2}Y^*_{\ell m}(\hat{n}_q)\right]^* = \sum_{\ell m} {}_{0}Y_{\ell m}(\hat{n}_e){}_{-2}Y^*_{\ell m}(\hat{n}_q) \,,
 \eeq
where the terms on either side of the equation are those that appear in \eq{eq:qu2eb_ker_1}. Note that the operator $\bar{O}$ is real as one expects, since each sub-matrix in \eq{eq:qu2eb_ker_1} is formed by summing a complex number and its conjugate. 

\eq{eq:qu2eb_ker_1} already presents a real space operator, but it is not in a form which can be practically implemented. To proceed, we use the fact that the $m$ sum over the product of two spin spherical harmonic functions can be expressed as a function of the Euler angles \cite{varshalovich}:
\beq \label{eq:sum_spin_shf}
 \sum_{m}{{}_{s_1}Y}^*_{\ell m}(\hat{n}_i)\,{{}_{s_2}Y}_{\ell m}(\hat{n}_j) = \sqrt{\frac{2\ell+1}{4 \pi}} {{}_{s_2}}Y_{\ell \,-s_1}(\beta_{ij},\alpha_{ij}) e^{- i s_2 \gamma_{ij}} \,,
\eeq
where $(\alpha_{ij}, ~\beta_{ij}, \gamma_{ij})$ denote the Euler angles that specifically transform $(i \rightarrow j)$ so that the coordinate system at $\hat{n}_i$ aligns with the coordinate system at $\hat{n}_j$\footnote{The sense of the rotation becomes more obvious when this equation is written in terms of the Wigner-$D$ functions.}. Using this identity, the different parts of the real space operator $\bar{O}$  (from eq.~\ref{eq:qu2eb_ker_1}) are completely specified by the following complex function:
\begin{subequations}\label{eq:qu2eb_gen_kernel}
\beqry
\mathcal{M}( \hat{n}_e, \hat{n}_q)  &=& \mathcal{M}_{r} + i \mathcal{M}_{i}  \,,\nonumber \\ 
 &=&\sum_{\ell m} {{_0}Y}_{\ell m}(\hat n_e) \, {{_{-2}}Y}^*_{\ell m}(\hat n_q) = \sum_{\ell} \sqrt{\frac{2\ell+1}{ 4 \pi}}{{_0Y}_{\ell 2}}(\beta_{qe},\alpha_{qe})\,,\\
&=&  \Big [ \cos(2 \alpha_{qe}) + i \sin(2 \alpha_{qe} ) \Big]   \sum_{\ell=\ell_{\rm min}}^{\ell_{\rm max}} {\frac{2\ell+1}{ 4 \pi}} \sqrt{\frac{(\ell-2)!}{(\ell+2)!}}P_{\ell}^2 (\cos\beta_{qe}) \,, \label{eq:rad_ker_queb} \\
\mathcal{M}(\beta_{qe}, \alpha_{qe})  &=&  \Big [ \cos(2 \alpha_{qe}) + i \sin(2 \alpha_{qe} ) \Big] \quad {{}_{\mm}f}(\beta_{qe},\ell_{\rm min},\ell_{\rm max}) \,, 
\eeqry
\end{subequations}
where we have used the identity in \eq{eq:sum_spin_shf} to simplify the product of the spherical harmonic functions. Note that the function depends only on two out of the three Euler angles.  The azimuthal part depends only on the Euler angle $\alpha_{qe}$ and its harmonic transform has no multipole $\ell$ dependence.  The azimuthal part is the crucial operation that translates between different spin representation of CMB polarization. The radial part $f(\beta_{qe})$ depends only on the angular separation between locations and completely incorporates all the multipole $\ell$ dependence. Recall that we had guessed the general form of the kernel using simple heuristic arguments in \sec{sec:qu2eb_heuristic}. Here we have rigorously derived the exact form of the function $f(\beta)$. Studying the $P_{\ell}^2$ functions in the limits $\beta \rightarrow 0,\pi$ it can be shown that $f(\beta)$ vanishes at $\beta=0,\pi$, which we had argued is a crucial property to yield a field of correct spin.

Employing \eq{eq:qu2eb_gen_kernel} to simplify the product of spherical harmonic functions in \eq{eq:qu2eb_ker_1}, the real space operator $\bar{O}$ can now be cast in this more useful form:
\beq\label{eq:op_qu2eb_rad}
\bar O =- \Delta \Omega \bmat  \mathcal{M}_{r} & \mathcal{M}_{i} \\  -\mathcal{M}_{i}  & \mathcal{M}_{r} \emat = - \Delta \Omega {{}_{\mm}f}(\beta_{qe},\ell_{\rm min},\ell_{\rm max})\bmat \cos(2 \alpha_{qe}) & \sin(2\alpha_{qe})\\  -\sin(2 \alpha_{qe})  & \cos(2 \alpha_{qe}) \emat \,,
\eeq
where $(\alpha_{qe}, ~\beta_{qe}, \gamma_{qe})$ denote the Euler angles which rotate the local cartesian system at $\hat{n}_q$ (location where Stokes parameters are accessed) to the cartesian system at  $\hat{n}_e$ (location where the scalar fields are evaluated).

It turns out that the real space operations can \emph{also} be expressed in terms of the inverse rotation.  These two expressions lead to two sets of kernels that are conceptually different, although they both ultimately yield the same mathematical result. The first set act like Green's functions, where, for example, a pixel of the Stokes parameters broadcasts or radiates an $E$/$B$ field.  The second set act like convolving beams, gathering Stokes contributions to the $E/B$ fields at a point. On the equator, or in the flat sky approximation, these kernels are identical and this distinction is immaterial.  On the curved sky this distinction is important, and the kernels are especially different near the poles.

\paragraph{Radiation kernel.} The expression above, based on the $\hat n_q \rightarrow \hat n_e$ rotation,  we call the \textit{radiation kernel}.  It allows us, like a Green's function, to evaluate the $E/B$ field contribution due to a single Stoke parameter ``charge'' at a fixed location. The total $E/B$ maps can then be thought of as the superposed radiation emerging from Stokes charges across the sphere. In this picture, we are effectively in the frame of the Stokes charge ${}_{\pm2}X$ and evaluating its contribution to the complex spin-0 scalar field $E+iB$ across the sphere. This one-to-many mapping from a point in the spin-2 Stokes field to the complex spin-0 (scalar) field across the sphere is graphically represented by the blue circle in \fig{fig:planar_euler_angles}.

The $E$/$B$ contribution from the Stokes parameters at some location $\hat{n}_q$ is given by the following expression (\eq{eq:op_qu2eb_rad} and \eq{eq:qu2eb_op}):
\beq  \label{eq:qu2eb_radiation_explicit}
\bar{S}_q(\hat n_e) = \bmat E_e \\ B_e  \emat_{q} =- {{}_{\mm}f}(\beta_{qe},\ell_{\rm min},\ell_{\rm max})\bmat \cos(2 \alpha_{q e}) & \sin(2\alpha_{q e})\\  -\sin(2 \alpha_{q e})  & \cos(2 \alpha_{q e}) \emat  \bmat Q_{q} \\ U_{q}  \emat \Delta \Omega \,.
\eeq
The total map can be simply evaluated by summing over the contribution from the Stokes parameters at each location $\hat{n}_q$: $\bar{S} = \sum_{q=1}^{N_{\rm pix}} \bar{S}_q$. This operation can be cast concisely as:
\begin{subequations} \label{eq:qu2eb_radiation_concise}
\beqry 
\left[E + iB\right](\hat{n}_e) &=& -\Delta \Omega  \sum_{q=1}^{N_{\rm pix}} \Big[ {}_{+2}X(\hat{n}_{q}) e^{-i2\alpha_{q e}} \Big]  {{}_{\mm}f}(\beta_{q e}) \,, \\
&=& \Delta \Omega \sum_{q=1}^{N_{\rm pix}} {}_{+2}X(\hat{n}_{q}) \,  \mathcal{M}_{G}(\hat n_q) \,,
\eeqry
\end{subequations}
where $\Delta \Omega$ denotes the pixel area and the last line is a simple scalar multiplication between complex numbers.  The radiation kernel is then: $ \mathcal{M}_{G} =- \mm(\beta_{qe},\alpha_{qe})^*$ which can be thought of as the Green's function of the operator, since $[E +iB]= \mathcal{M}_{G}$ is the spin-0 scalar field generated from the delta-function Stokes field $[Q+iU] = [\delta_{\hat{n},\hat{n}_q}/{\Delta \Omega} + i0]$. We display the kernel later in \fig{fig:vis_kernel}.

\begin{figure}[!t]
\centering
\includegraphics[width=0.5\columnwidth]{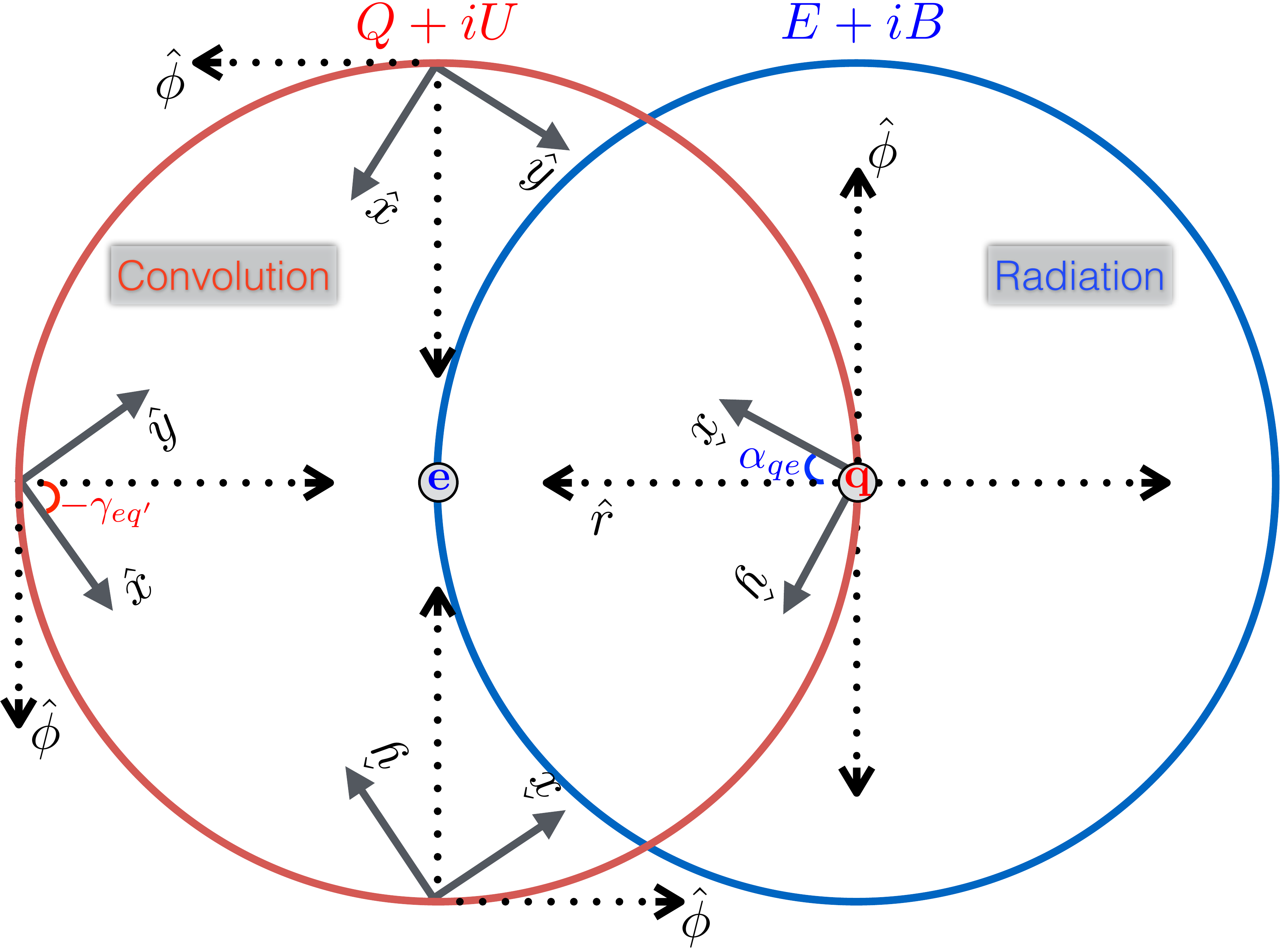}
\caption{The local cartesian coordinates $(\hat{x},\hat{y})$ are drawn on the red circle(sphere), representative of the coordinate dependence of the Stokes parameters. The two sets of dotted lines drawn at representative points denote great circles, one which passes through the central point labelled `$e$' and the other chosen such that the two have locally orthogonal tangent vectors $(\hat{r},\hat{\phi})$. The angle $\alpha_{qe}$ defines a rotation operator that aligns the local $\hat{x}$ with $\hat{r}$.   \textit{Radiation kernel:} We can compute the contribution from the Stokes parameter at `$q$' to all the points on the blue circle and this is a function of the Euler angle $\alpha_{qe}$. \textit{Convolution kernel:} The resultant scalar field at `$e$' can also be evaluated by summing over the contribution from all the Stokes parameters on the red circle. This convolution is performed with kernels which are defined in term of the Euler angle $\gamma_{eq}$.}
\label{fig:planar_euler_angles}
\end{figure}

\paragraph{Convolution kernel.} We can also formulate the real space operator as a convolution operation, where the scalar field at $\hat n_e$ gathers contributions from the Stokes fields.  This is based around the inverse rotation from the previous section (to align the coordinate system at $\hat{n}_e$ with that at $\hat{n}_q$).  The inverse rotation Euler angles relates to the forward rotation Euler angles by the following relations: $\alpha_{eq}=-\gamma_{qe}$, $\beta_{eq} = -\beta_{qe}$ and  $\gamma_{eq} =-\alpha_{qe}$. Since the kernel depends on the cosine of the Euler angle $\beta$, it is immune to changes in its sign. The operator equation can be expressed as a function of the Euler angle $\gamma_{eq}$ as follows:
\beq \label{eq:qu2eb_convolution_explicit}
\bmat E_e \\ B_e  \emat =- \Delta \Omega\sum_{q=1}^{N_{\rm pix}}{{}_{\mm}f}(\beta_{eq},\ell_{\rm min},\ell_{\rm max})\bmat \cos(2 \gamma_{eq}) & -\sin(2\gamma_{eq})\\  \sin(2 \gamma_{eq})  & \cos(2 \gamma_{eq}) \emat  \bmat Q_q \\ U_q  \emat \,,
\eeq
This formulation of the real space operator can be interpreted as integrating at some fixed location $\hat{n}_e$ the $E/B$ mode contribution arising from the Stokes parameters at all location $\hat{n}_q$ on the sphere. This operation can be expressed more concisely as follows:
\begin{subequations} \label{q:qu2eb_convolution_concise}
\beqry 
[E + iB](\hat{n}_e) &=& - \Delta \Omega \sum_{q=1}^{N_{\rm pix}}{{}_{\mm}f}(\beta_{eq},\ell_{\rm min},\ell_{\rm max}) {\Bigg( e^{i2 \gamma_{eq}}   {}_{+2}X (\hat{n}_q) \Bigg)}, \label{eq:qu2eb_physical}\\
&=& \Bigg\lbrace \mathcal{M}_{B} \star {}_{+2}X \Bigg\rbrace(\hat{n}_e) \,, \label{eq:qu2eb_convolution} 
\eeqry
\end{subequations}
where $\star$ denotes a convolution and $\mathcal{M}_{B} = -\mm(\beta_{eq},\gamma_{eq})$.  When $\mm$ is expressed as a function of the Euler angle $\gamma_{eq}$ it can be thought of as an effective instrument beam pointing to the direction $\hat{n}_e$. This many-to-one mapping from the spin-2 Stokes field on the sphere to the complex spin-0 (scalar) field at a point on the sphere is graphically represented in \fig{fig:planar_euler_angles}. We display the complex conjugate of this kernel in \fig{fig:vis_kernel}.  (Later we see that the complex conjugate of this kernel is the radiation kernel for the inverse transform.)

\subsection{Evaluating Stokes $Q$/$U$ from scalar $E$/$B$}\label{sec:eb2qu}
The real space operator which translates $E$/$B$ fields to Stokes parameters $Q$/$U$ can be derived using a similar procedure. Expressed in the matrix-vector notation, the inverse operator is given by the following equation:
\begin{subequations}
\beqry
\bar{P} &=& \bar{T}^{-1} {{}_2\mathcal{Y}}\, \tilde T {{}_0\mathcal{Y}^{\ddagger}}\bar{S} = \frac{1}{2} \bar{T}^{\dagger} {{}_2\mathcal{Y}} \,\tilde T {{}_0\mathcal{Y}^{\ddagger}}\bar{S}\,,  \\
&=&  \bar O^{-1} \bar{S}\,.
\eeqry
\end{subequations}

The inverse operator expressed in terms of the function $\mm$ given in \eq{eq:qu2eb_gen_kernel} is given by the following equation:
\beq
{\bar O}^{-1}=- \Delta \Omega\bmat \mathcal{M}_{r} & -\mathcal{M}_{i} \\  \mathcal{M}_{i}  & \mathcal{M}_{r} \emat=- \Delta \Omega{{}_{\mm}f}(\beta_{eq},\ell_{\rm min},\ell_{\rm max})\bmat \cos(2 \alpha_{qe}) & -\sin(2\alpha_{qe})\\  \sin(2 \alpha_{qe})  & \cos(2 \alpha_{qe}) \emat \,,
\eeq
where all the symbols have the same meaning as discussed in \sec{sec:qu2eb}. Note that the kernel in the above equation differs from the one in \eq{eq:op_qu2eb_rad} by a change in sign on the off-diagonals of the block matrix. When expressed in terms of the same set of Euler angles used to define the operator $\bar{O}$, it can be shown that the different forms of the real space operator are given by the following equations:
\beqry
    {}_{+2}X(\hat{n}_q) &&=  \Delta \Omega \sum_{e=1}^{N_{\rm pix}} [E+iB](\hat{n}_{e})\   \mathcal{M}^*_{B}(\hat{n}_e) \hspace{0.8cm}\textrm{\emph{Radiation kernel}},\label{eq:eb2qu_radiation} \\
    {}_{+2}X(\hat{n}_q) &&= \Bigg\lbrace \mathcal{M}^*_{G} \star [E+iB] \Bigg\rbrace(\hat{n}_q) \hspace{1.4cm}\textrm{\emph {Convolution kernel}}, \label{eq:eb2qu_convolution}
\eeqry
where all the symbols have the same meaning as defined in \sec{sec:qu2eb}. Note that $M_B$ (the convolution kernel before for $Q/U \rightarrow E/B$) is here the radiation kernel (as $M_B^*$) for $E/B \rightarrow Q/U$.  Similarly $M_G$ (the radiation kernel before for $Q/U \rightarrow E/B$) is here the convolution kernel (as $M_G^*$) for $E/B \rightarrow Q/U$.
Thus the conjugated forms of the radiation kernel (Green's function) and the convolution kernel (effective beam) for the operator $\bar{O}$ reverse roles for the inverse operator $\bar{O}^{-1}$.

\subsection{Visualizing the real space kernels} \label{sec:visualize_operator}
\newlength{\kernelfigwidth}
\setlength{\kernelfigwidth}{0.2\columnwidth}
\newlength{\kernelfigspace}
\setlength{\kernelfigspace}{-1.8mm}

\begin{figure}[t] 
\begin{center}
\begin{tabular}{m{8ex}m{\kernelfigwidth}m{\kernelfigwidth}|m{\kernelfigwidth}m{\kernelfigwidth}}
$b=90^\circ$ &
\hspace{\kernelfigspace}\includegraphics[width=\kernelfigwidth]{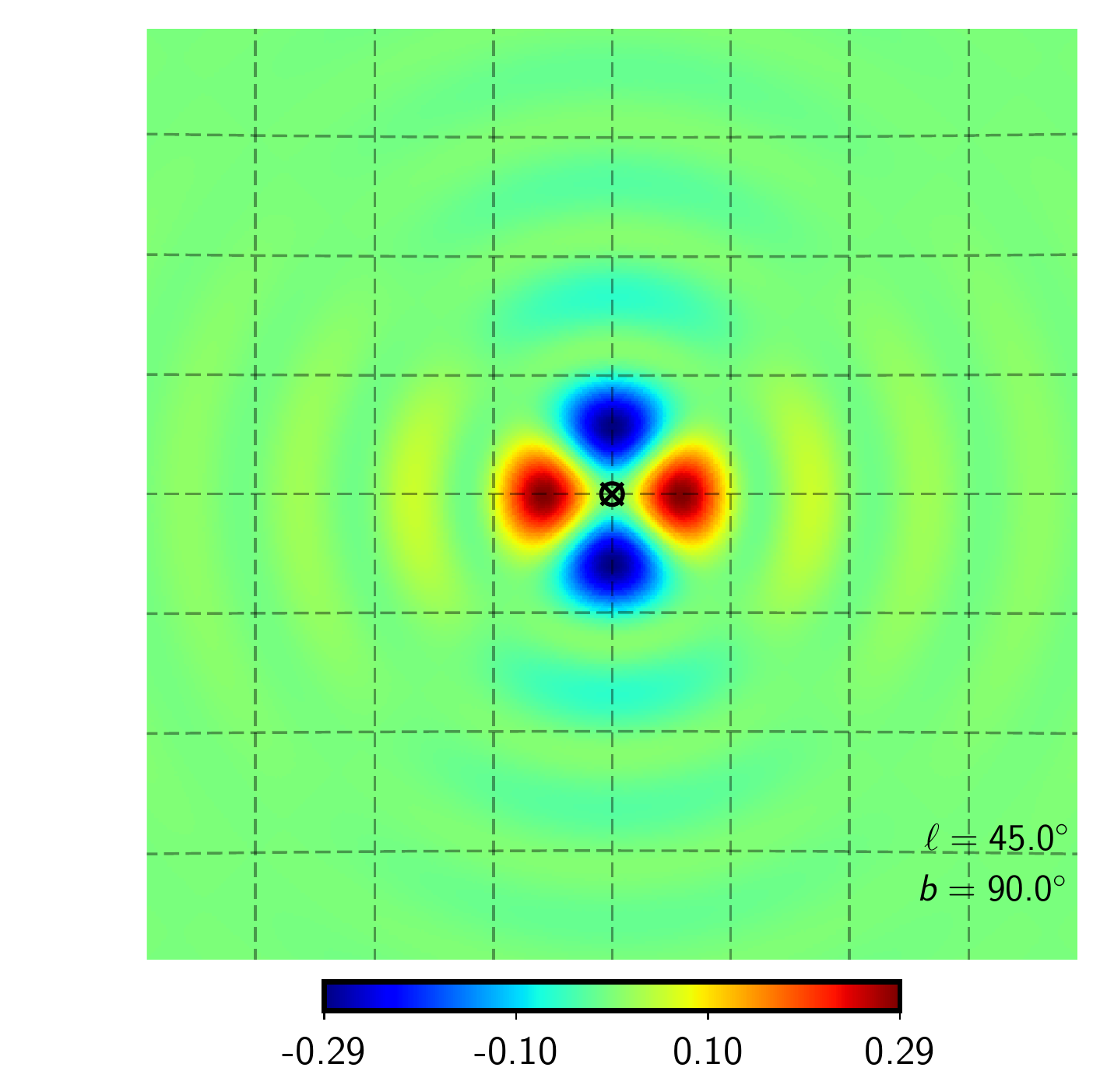} &
\hspace{\kernelfigspace}\includegraphics[width=\kernelfigwidth]{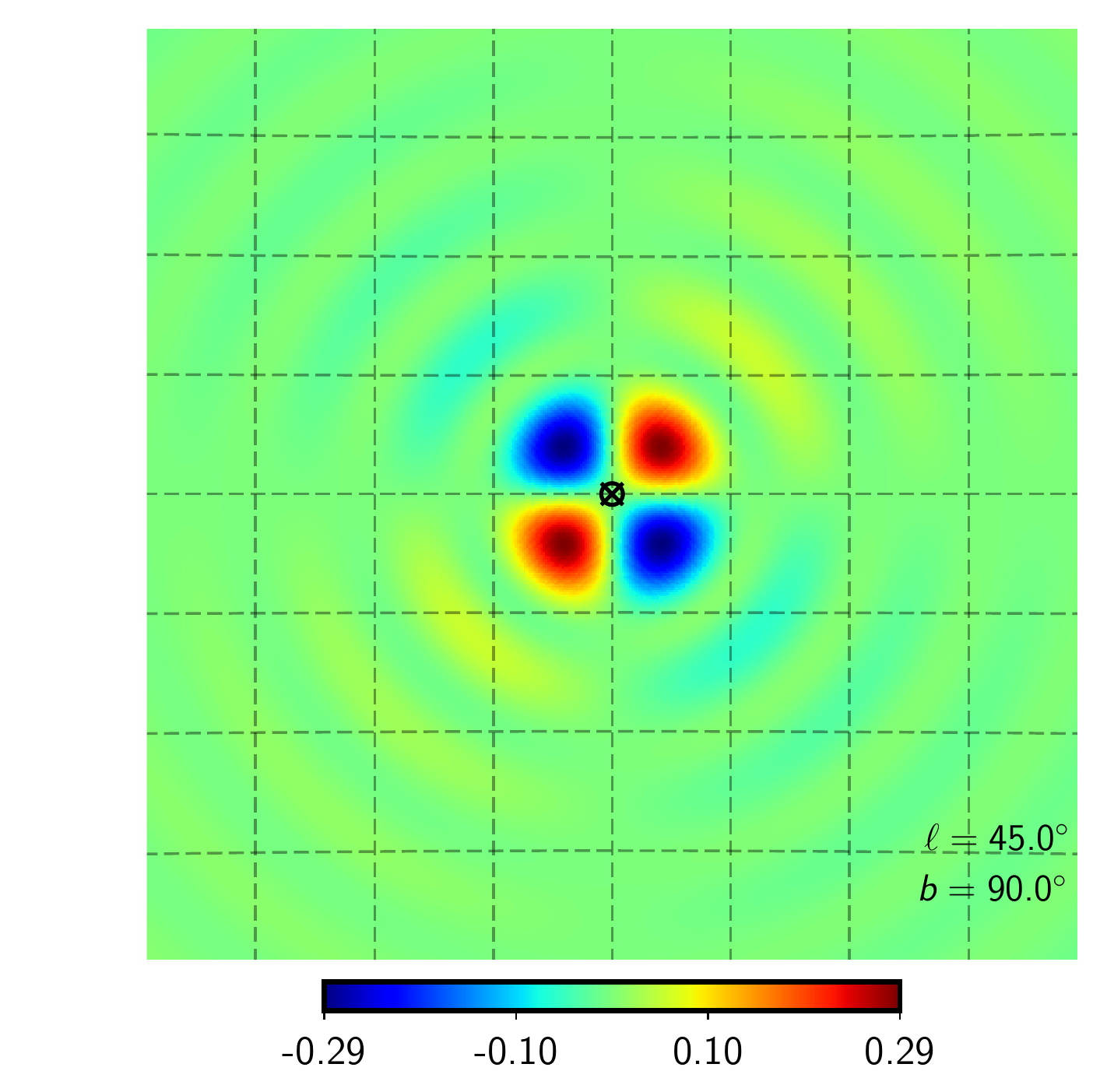} &
\hspace{\kernelfigspace}\includegraphics[width=\kernelfigwidth]{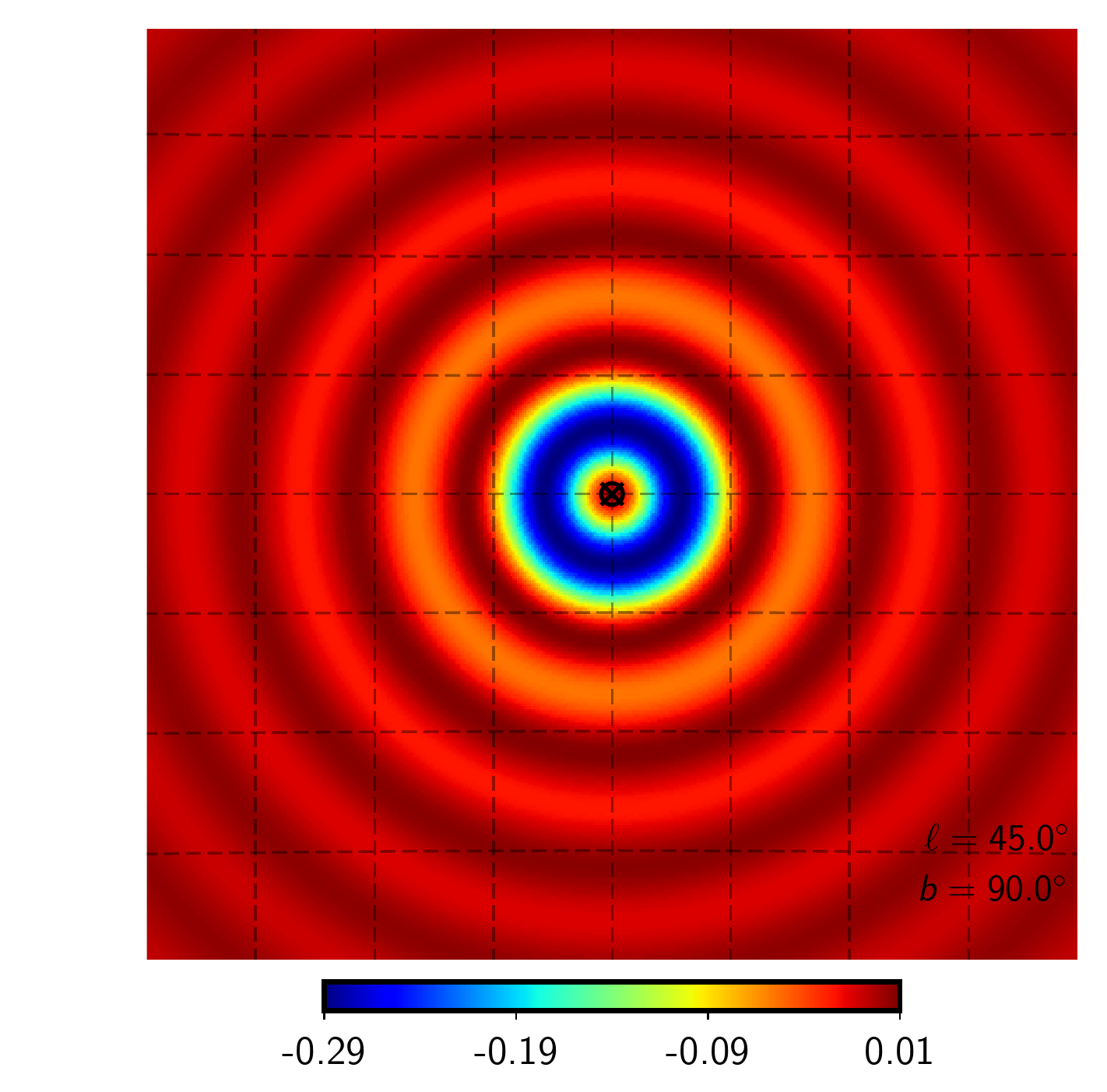} &
\hspace{\kernelfigspace}\includegraphics[width=\kernelfigwidth]{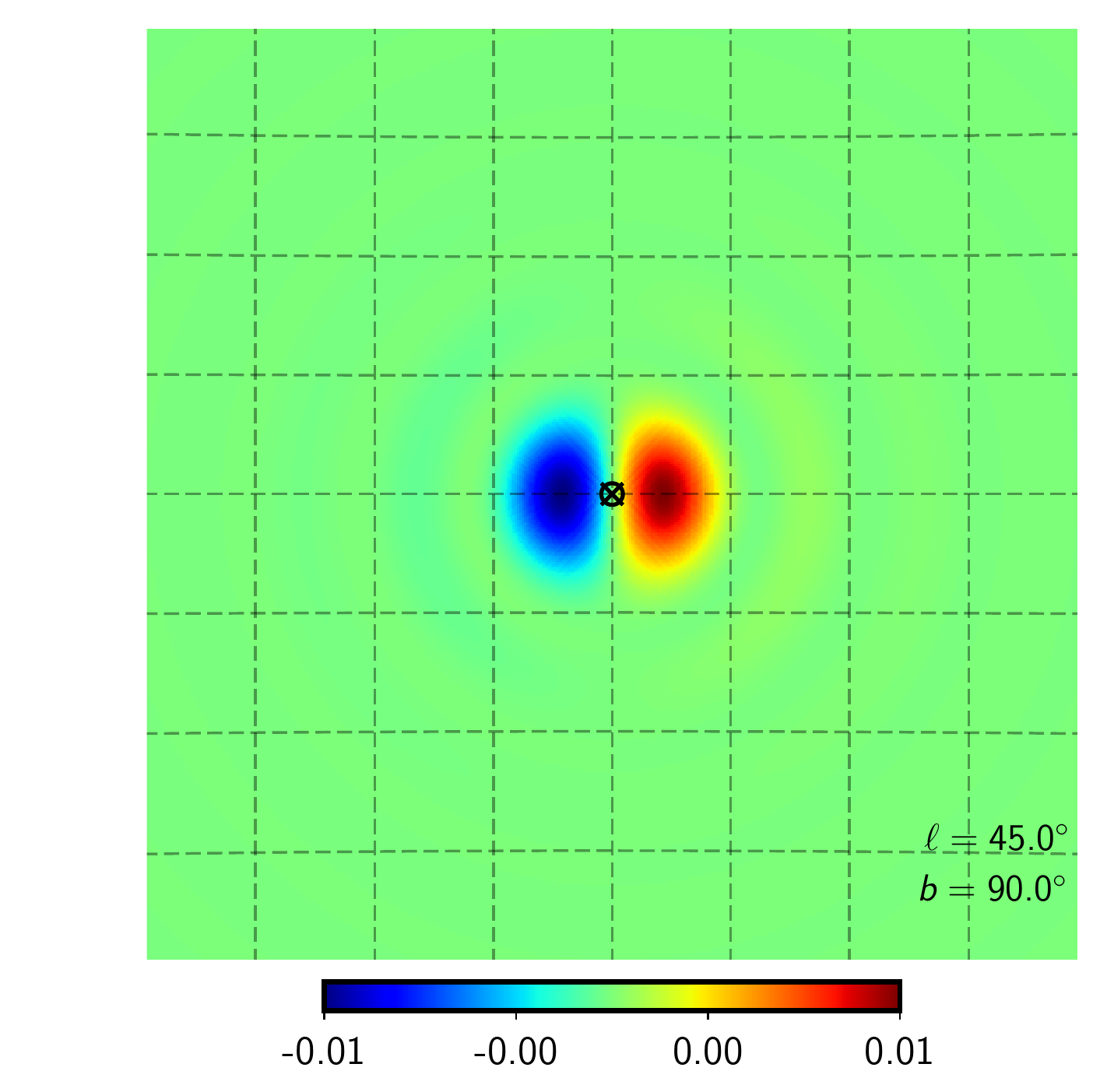} \\
$b=87^\circ$&
\hspace{\kernelfigspace}\includegraphics[width=\kernelfigwidth]{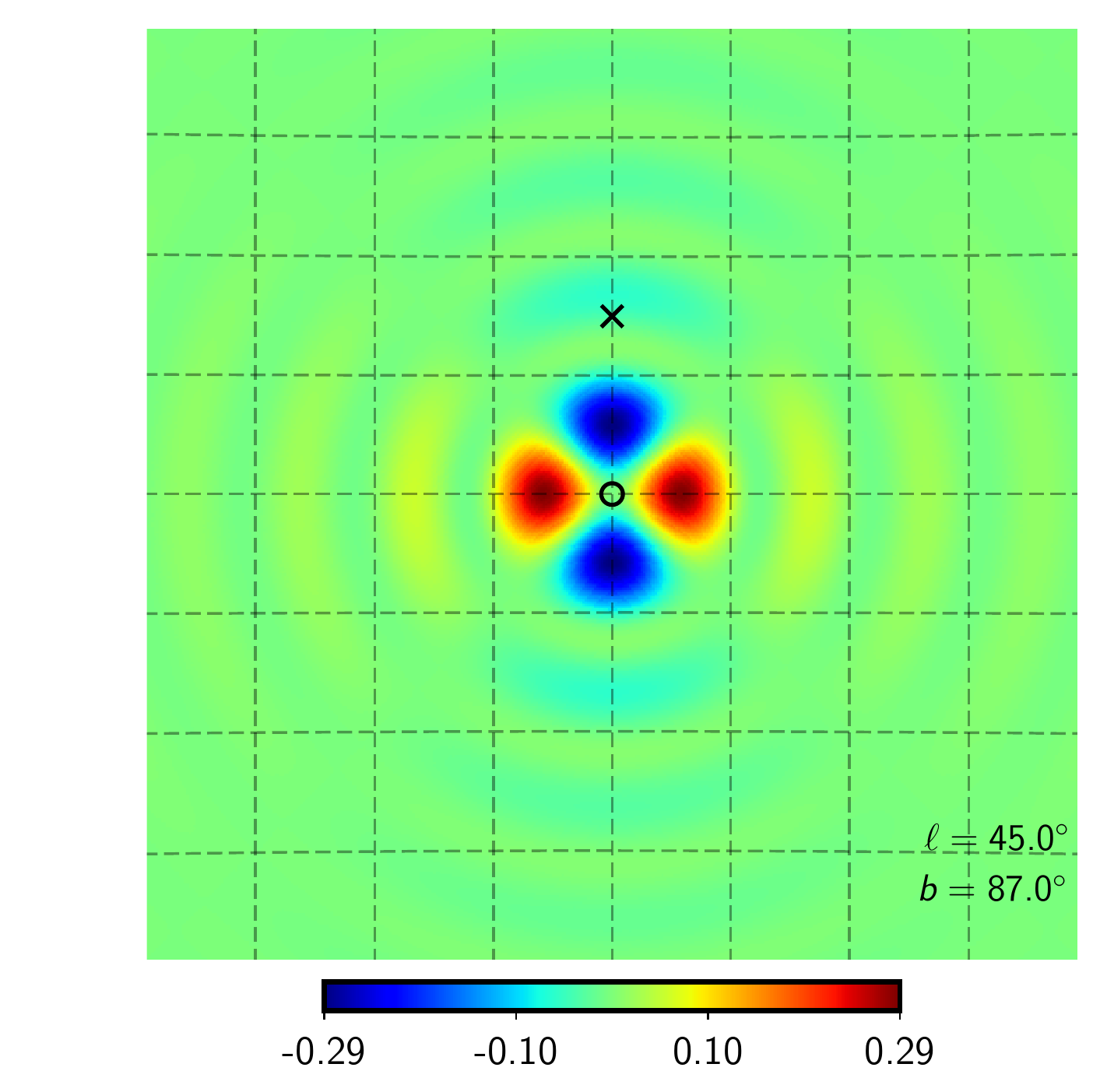} &
\hspace{\kernelfigspace}\includegraphics[width=\kernelfigwidth]{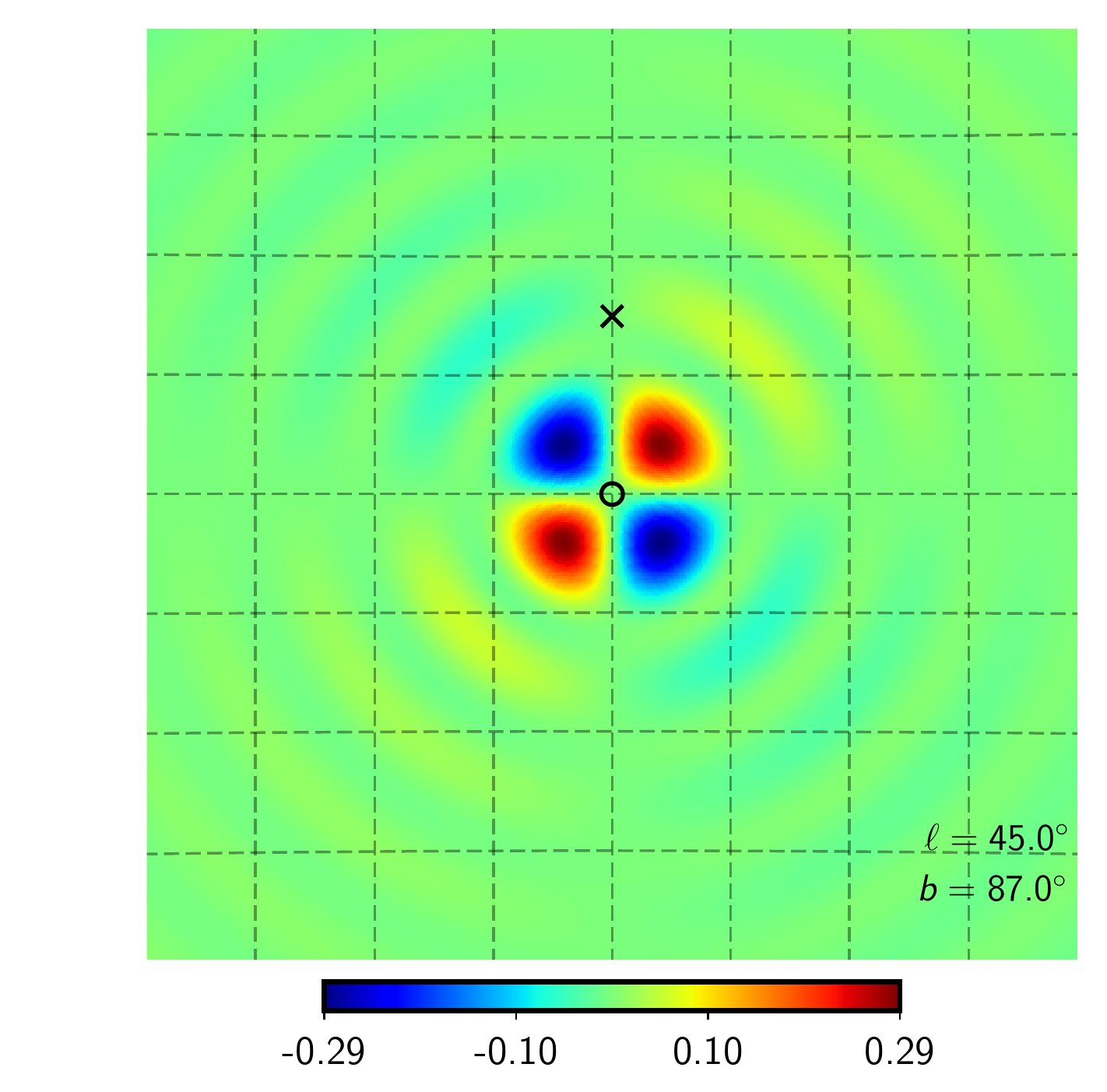} &
\hspace{\kernelfigspace}\includegraphics[width=\kernelfigwidth]{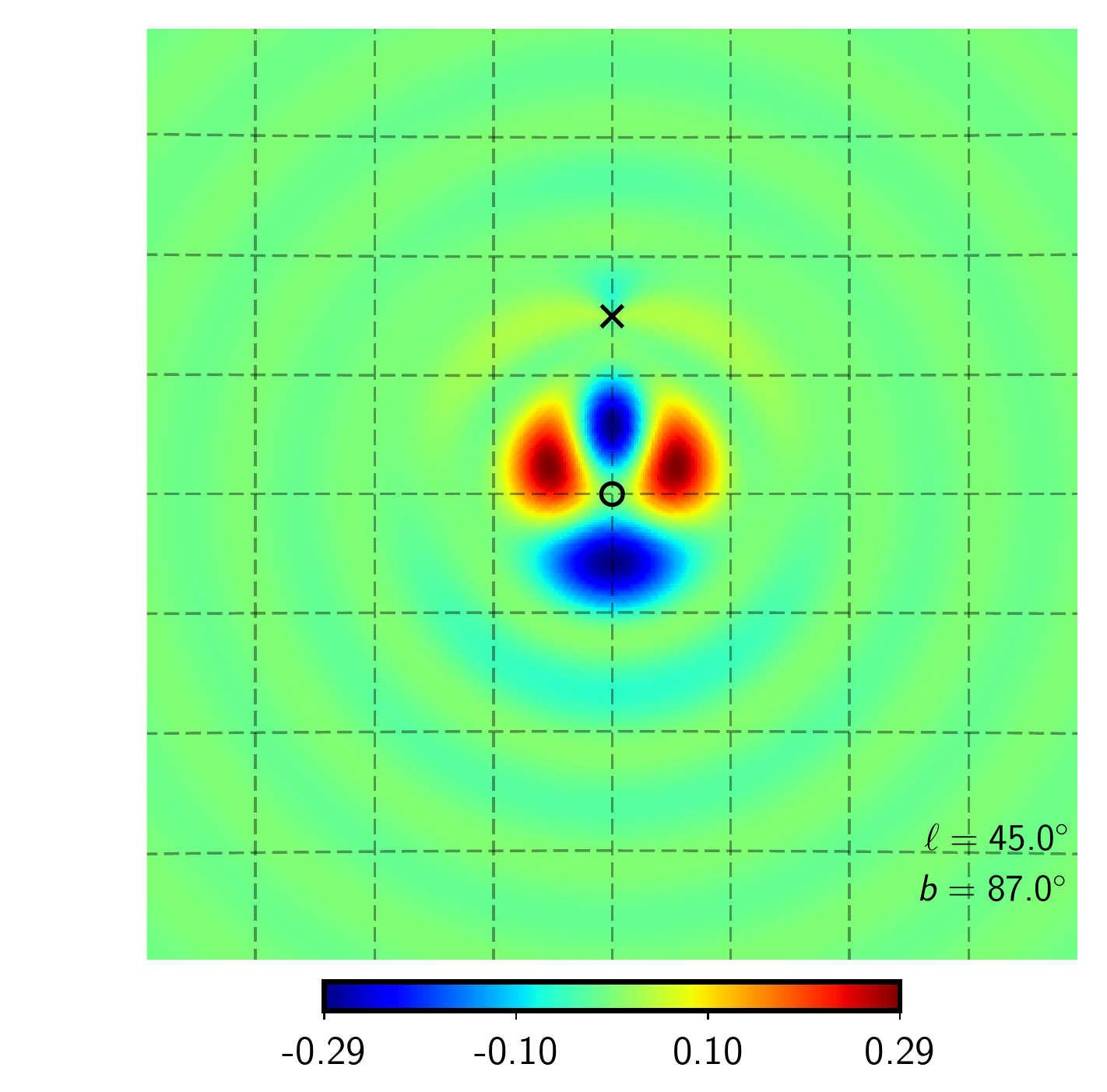} &
\hspace{\kernelfigspace}\includegraphics[width=\kernelfigwidth]{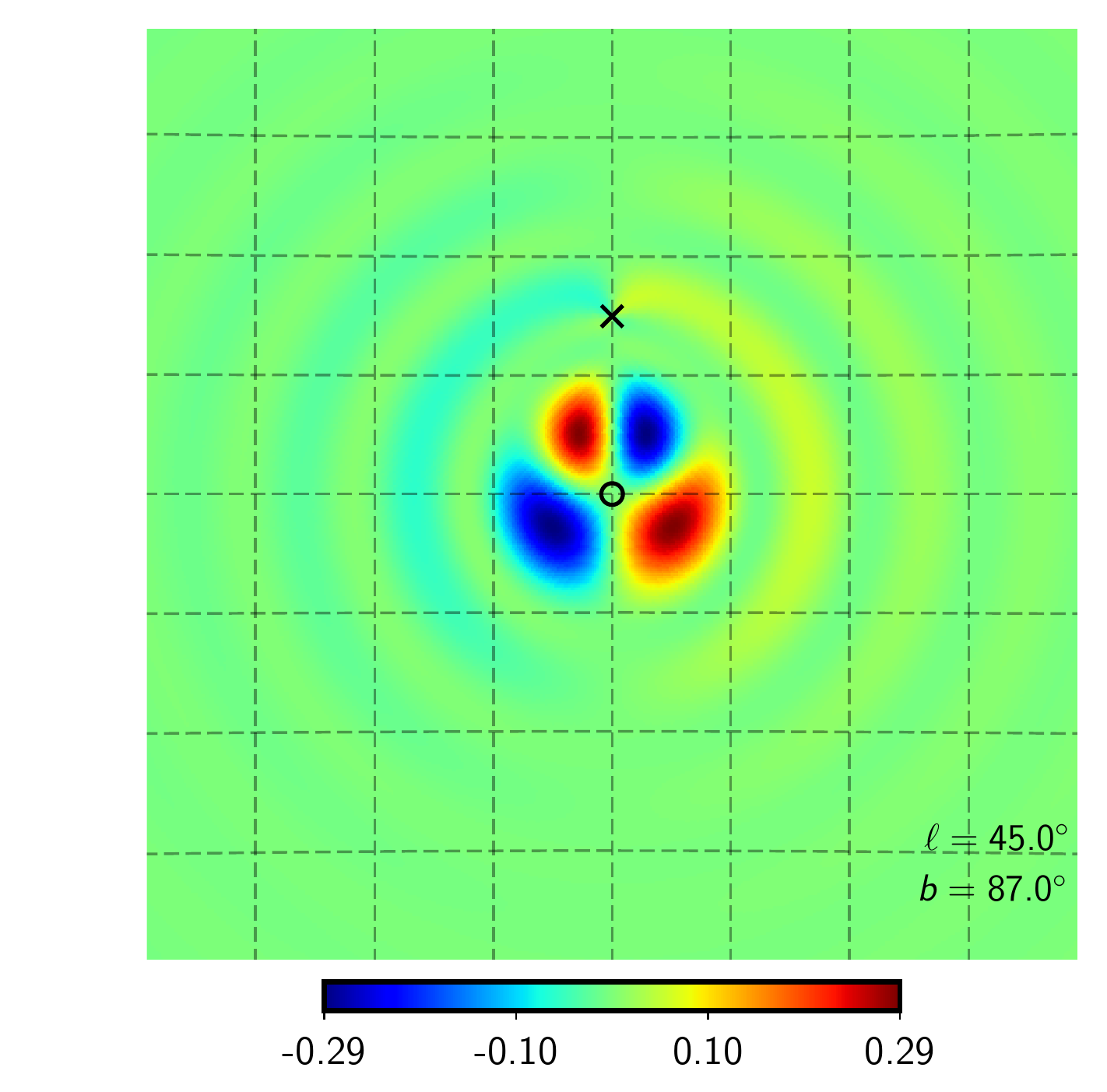} \\
$b=80^\circ$&
\hspace{\kernelfigspace}\includegraphics[width=\kernelfigwidth]{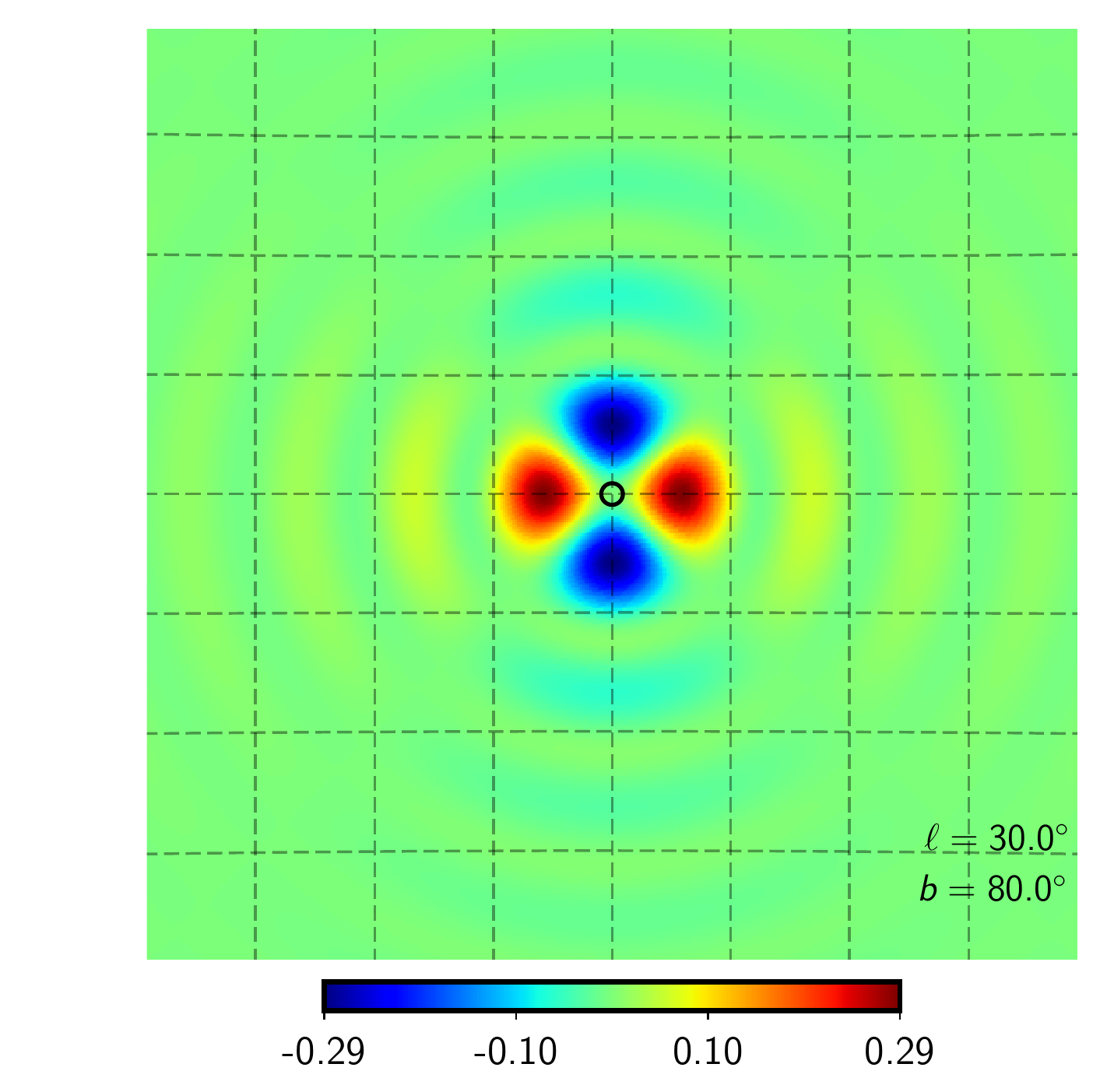} &
\hspace{\kernelfigspace}\includegraphics[width=\kernelfigwidth]{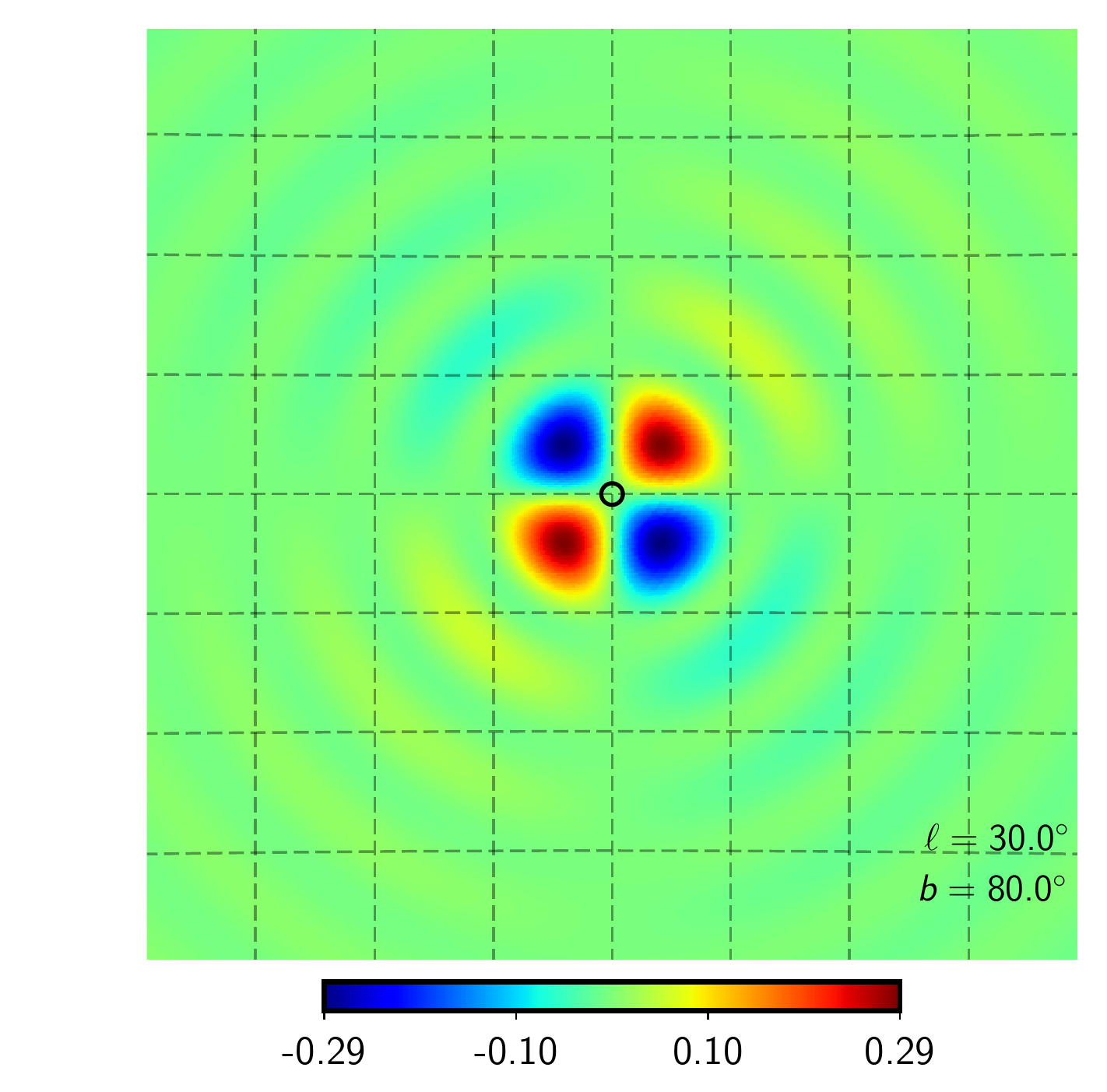} &
\hspace{\kernelfigspace}\includegraphics[width=\kernelfigwidth]{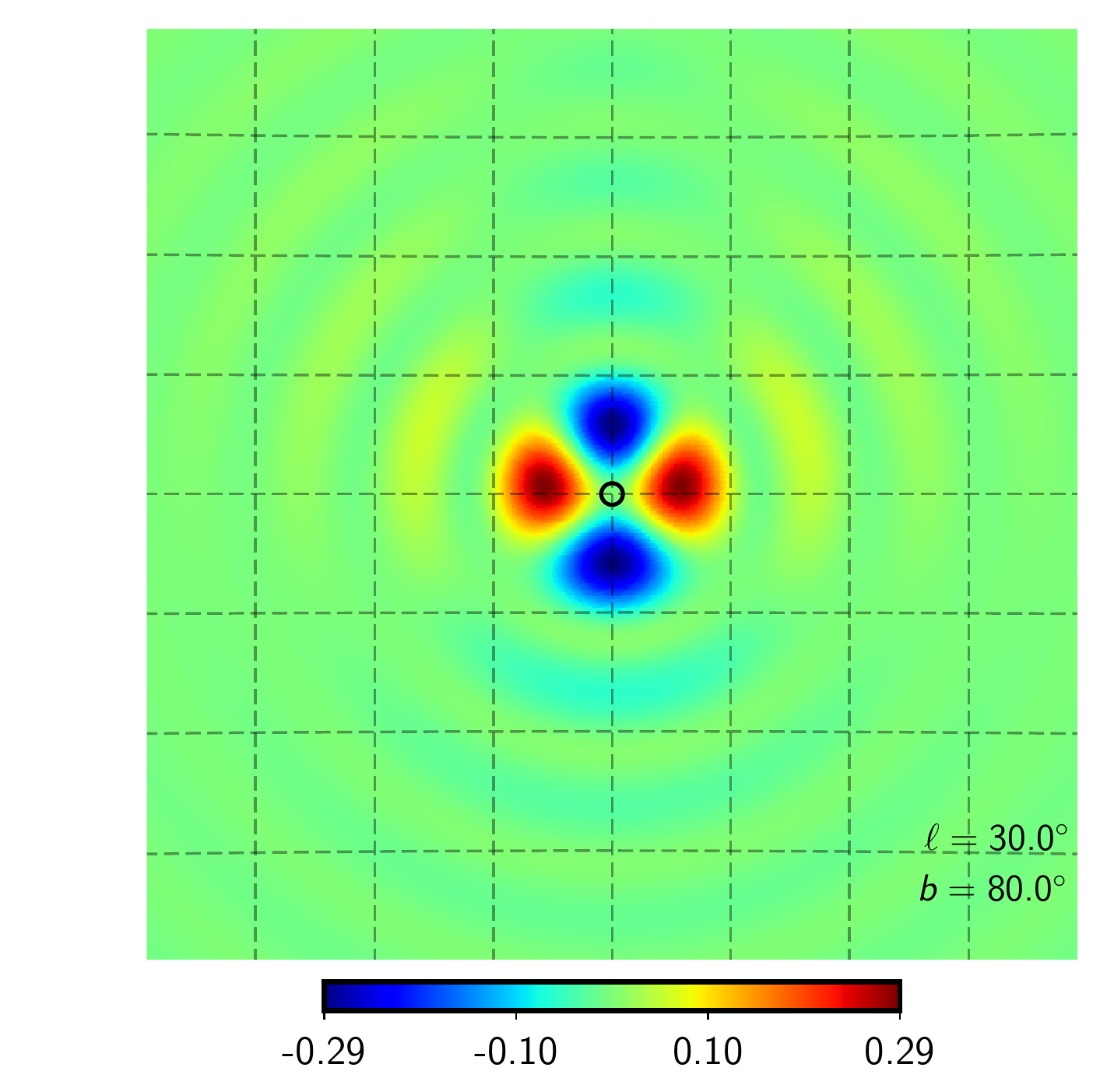} &
\hspace{\kernelfigspace}\includegraphics[width=\kernelfigwidth]{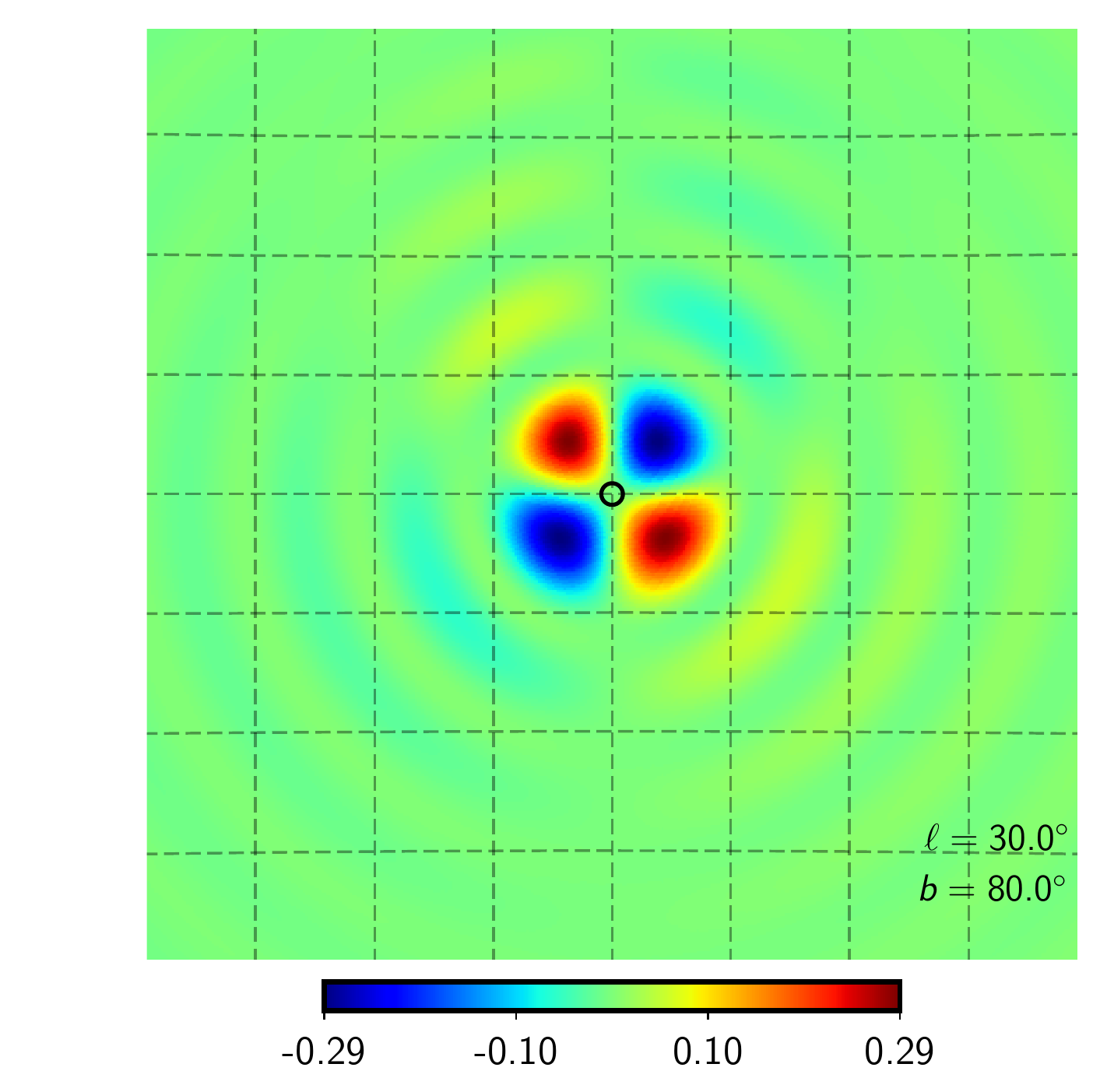} \\
$b=0^\circ$&
\hspace{\kernelfigspace}\includegraphics[width=\kernelfigwidth]{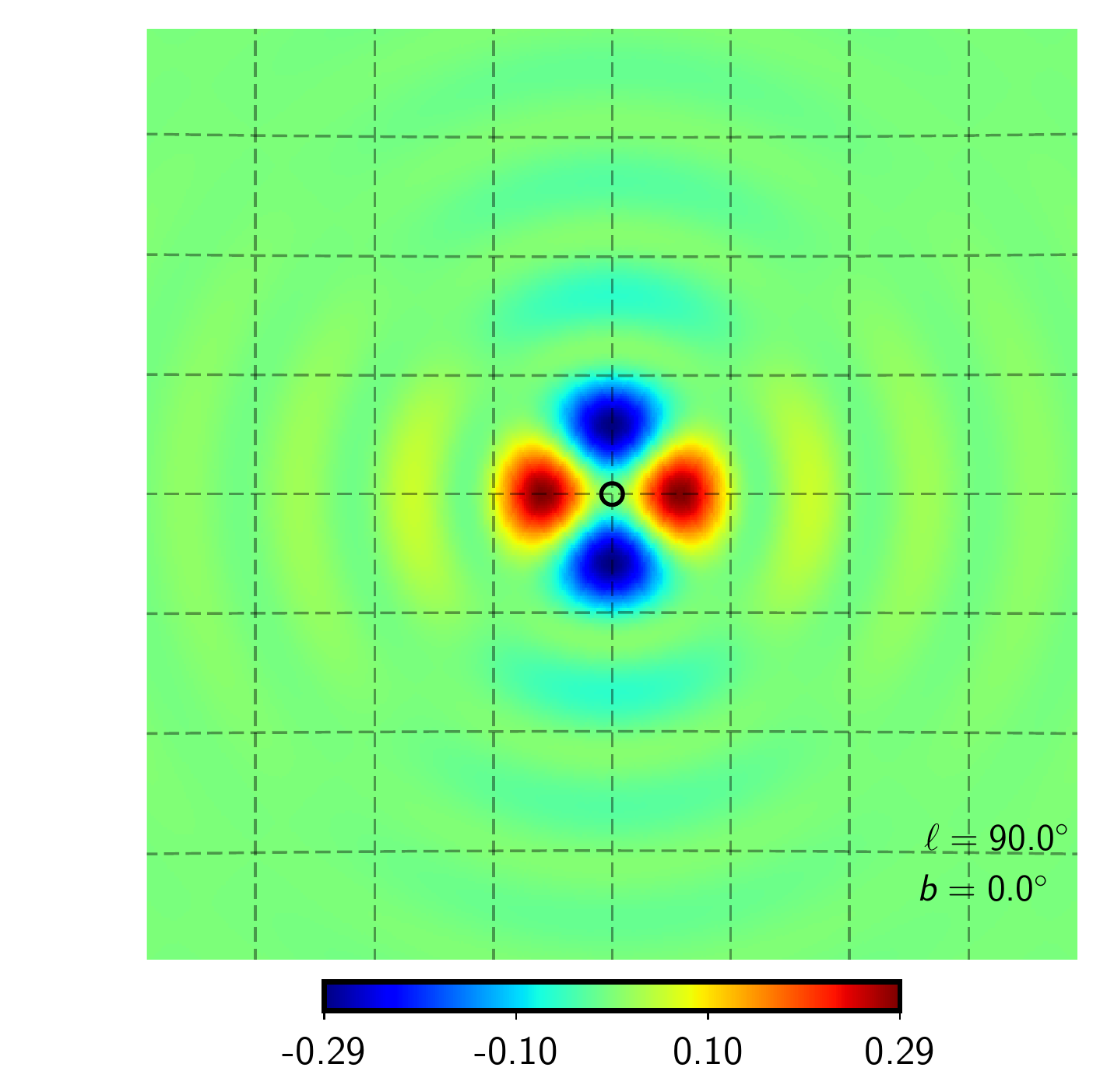} &
\hspace{\kernelfigspace}\includegraphics[width=\kernelfigwidth]{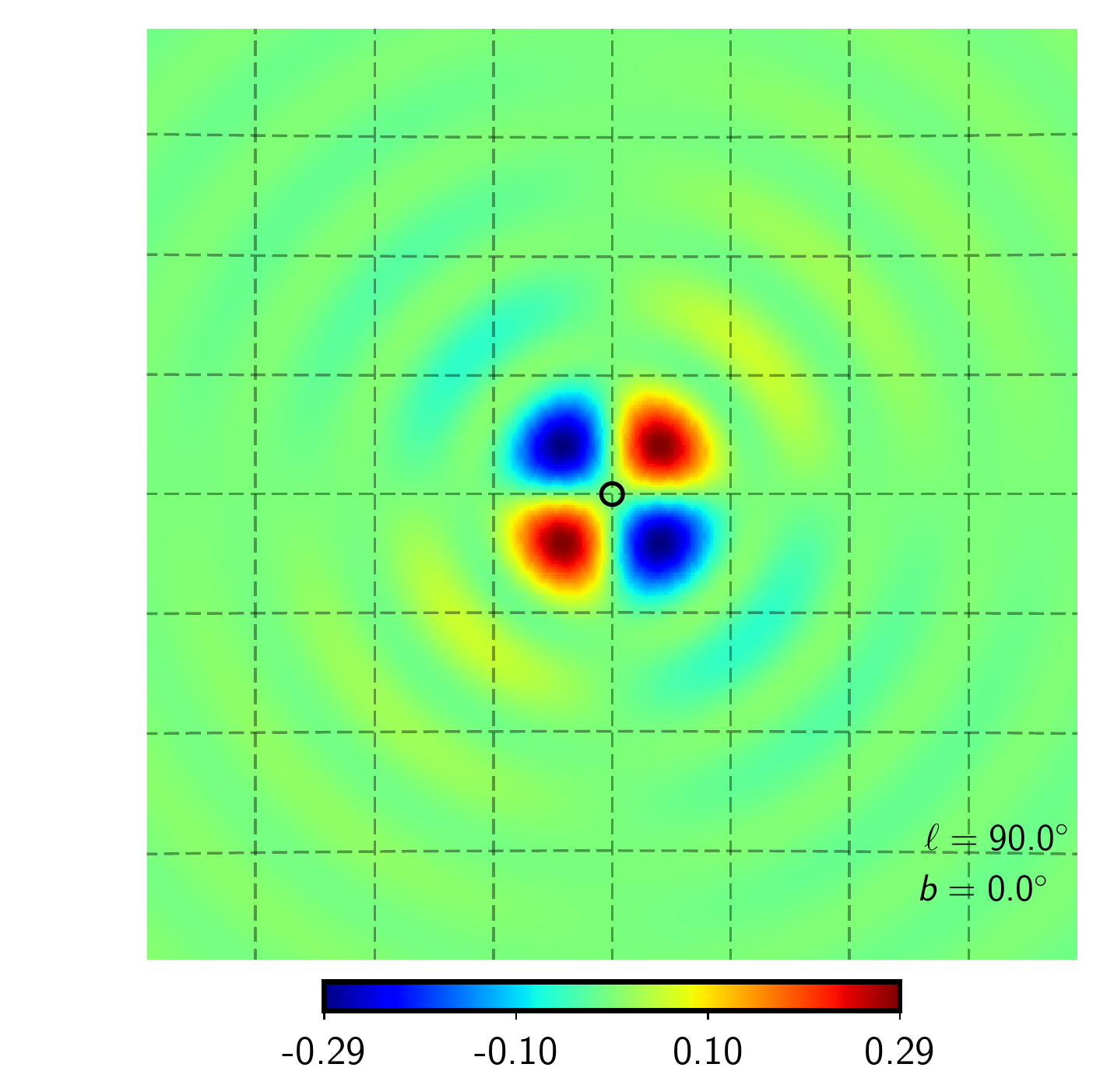} &
\hspace{\kernelfigspace}\includegraphics[width=\kernelfigwidth]{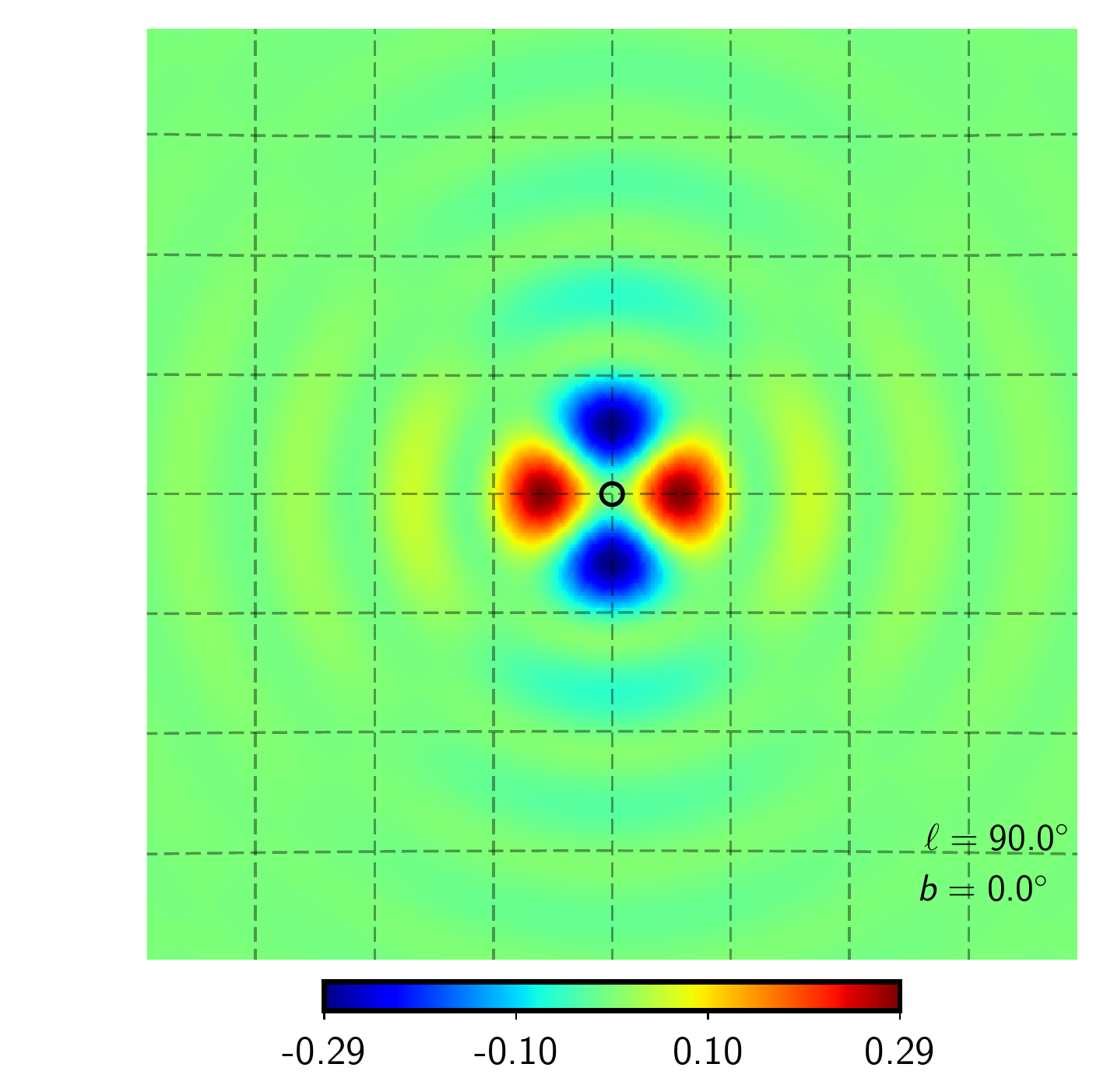} &
\hspace{\kernelfigspace}\includegraphics[width=\kernelfigwidth]{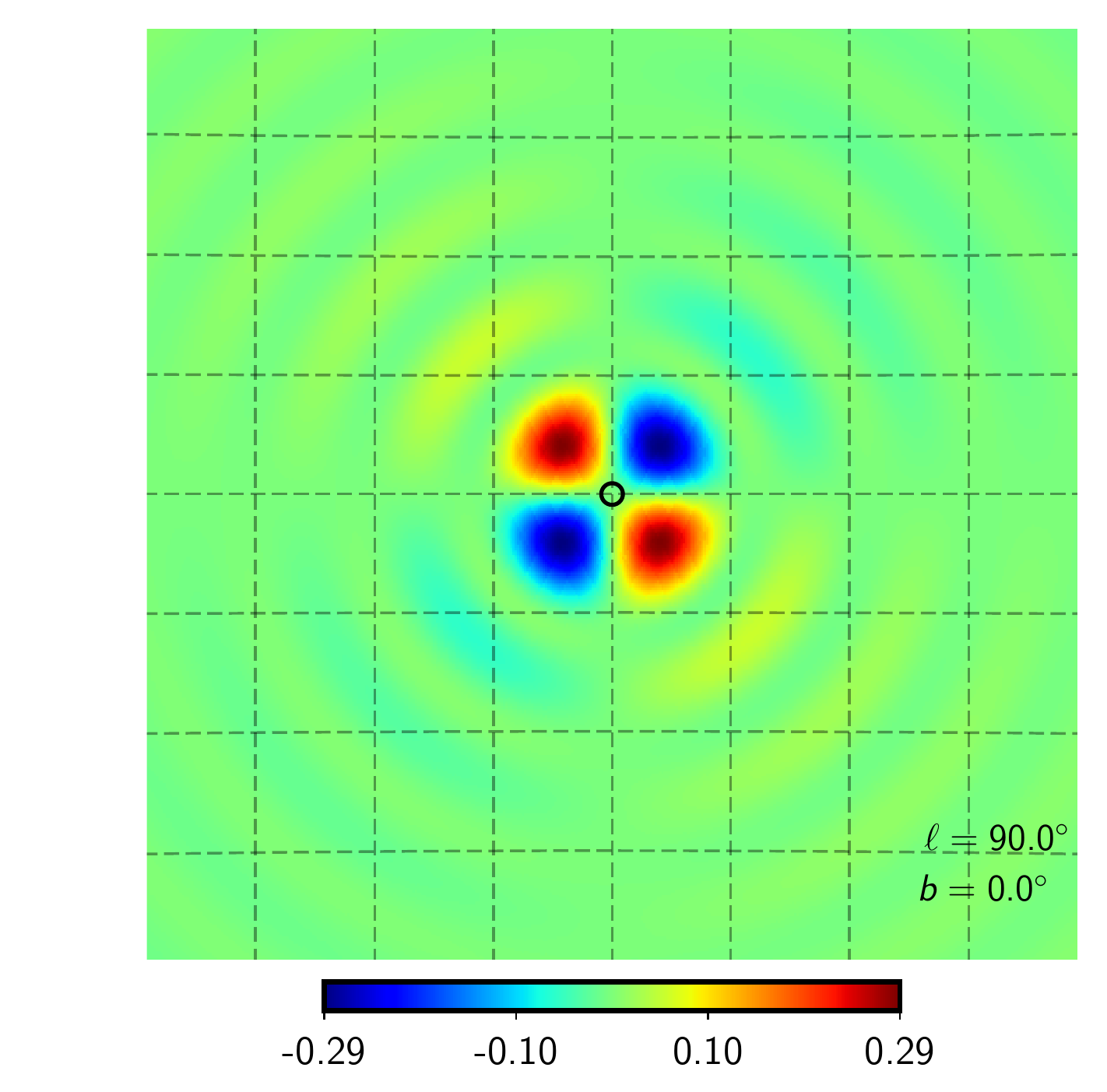} \\
&
\centering $ \textrm{Re} \left(\mathcal{M}_{G} \right) $ &
\centering $\textrm{Im} \left(\mathcal{M}_{G} \right) $ &
\centering $\textrm{Re}  \left(\mathcal{M}_{B}^* \right) $ &
\centering $\textrm{Im} \left(\mathcal{M}_{B}^* \right) $
\end{tabular}
\end{center}
\caption{Real and imaginary parts of real space kernels for $Q/U$ to $E/B$ translation (and vice versa).  The function $\mathcal{M}_{G}$ is the Green's function (radiation kernel) that gives $E+iB$ for a delta function Stokes input $_{+2}X = Q+iU$ at the center.  On the other hand, $\mathcal{M}^*_{B}$ is the Green's function (radiation kernel) that gives $_{+2}X$ for a delta function scalar input $E+iB$ at the center.  The black circles denotes the position of the center around which the kernels have been evaluated while the black cross marks the location of the North Pole. The four rows depict the kernels at different latitudes on the sphere.   The kernels are unchanged for center points at constant latitude.  $\mathcal{M}_G$ is invariant over the sphere because $E/B$ fields are coordinate independent. The \textit{convolution kernels} for the corresponding translations are $\mathcal{M}_B$ and $\mathcal{M}_G^*$ respectively: the functions are conjugated and switch roles between forward and inverse translation. The kernels have been evaluated with the band limit $\ell \in [2,192]$ and sampled at the HEALPix resolution parameter $N_{\rm side}=2048$. Each panel is approximately $16^{\circ} \times 16^{\circ}$ in size with grid lines every 2 degrees. East is to the left, as in sky map convention.} \label{fig:vis_kernel} 
\end{figure}
We compute the Euler angles $(\alpha, \beta, \gamma)$ given the angular coordinates of any two HEALPix pixels and use these to evaluate the convolution and radiation kernels. To provide an intuition for how these kernels vary as a function of position of the central pixel we depict in \fig{fig:vis_kernel} the respective kernels at a few different locations on the sphere.
While the kernels are evaluated in the band limit $\ell \in [2,192]$, for illustration these functions are sampled at a very high Healpix resolution parameter of $N_{\rm side}=2048$. All the plots have been rotated such that the central location marked by the black circle are in the centre of the figure. The grid spacing is 2 degrees.

The kernel $\mathcal{M}_G$ is the Green's function of the operator that transforms Stoke parameters to coordinate independent $E/B$.  The coordinate independence implies that the real and imaginary parts of the kernel do not vary with changes in the galactic latitude and longitude of the central pixel. In particular these functions are not distorted when a part of the domain overlaps with the poles, as can be seen in the first two rows of \fig{fig:vis_kernel}. From \eq{eq:qu2eb_radiation_concise}, we can see the $E/B$ patterns that Kronecker $\delta$-functions in the Stokes parameter pixels produce:
\beq
\bmat E= +\textrm{Re}(\mathcal{M}_G) \\ B =+\textrm{Im}(\mathcal{M}_G)  \emat  \leftarrow \bmat Q=\frac{\delta_{\hat{n},\hat{n}_q}}{\Delta \Omega}\\ U=0 \emat ;\qquad
\bmat E= -\textrm{Im}(\mathcal{M}_G) \\ B = + \textrm{Re}(\mathcal{M}_G)  \emat  \leftarrow \bmat Q=0 \\ U=\frac{\delta_{\hat{n},\hat{n}_q}}{\Delta \Omega} \emat.
\eeq
On the other hand, $\mathcal{M}^*_B$ is the Green's function of the operator that transforms $E/B$ to coordinate-dependent Stokes parameters $Q/U$.   The kernel $\mathcal{M}^*_B$ does not change with the central longitude, but varies as a function of galactic latitude. This latitude-dependent shape carries the  coordinate dependence of the Stokes parameters. From \eq{eq:eb2qu_radiation}, we can see the Q/U patterns that arise from $\delta$-functions in $E/B$:
\beq
\bmat Q= +\textrm{Re}(\mathcal{M}^*_B) \\ U = +\textrm{Im}(\mathcal{M}^*_B)  \emat  \leftarrow \bmat E=\frac{\delta_{\hat{n},\hat{n}_e}}{\Delta \Omega}\\ B=0 \emat;\qquad
\bmat Q= -\textrm{Im}(\mathcal{M}^*_B) \\ U = +\textrm{Re}(\mathcal{M}^*_B)  \emat  \leftarrow \bmat E=0 \\ B=\frac{\delta_{\hat{n},\hat{n}_e}}{\Delta \Omega} \emat\,.
\eeq
Recall that the complex conjugates of these functions switch roles to form the convolution kernels (e.g., $\mathcal{M}_G^*$ is the convolution kernel for $E/B \rightarrow Q/U$.)  It is also interesting to note that the radiation kernel ($\mm_{G}$) and the convolution kernel ($\mm_{B}$) become progressively identical as one approaches the equator, which is a consequence of $\gamma \simeq -\alpha$ in the vicinity of the equator.  This is visible in \fig{fig:vis_kernel} (although it displays $\mm_B^*$ and not $\mm_B$). The equator is the place on the sphere most akin to the flat sky case, where the radiation and convolution kernels are expected to be identical.

\subsection{Purifying Stokes parameters $Q$/$U$ for $E$/$B$ modes} \label{sec:purify_stokes_qu}
We can only measure the total Stokes vector, a sum of the part that corresponds to scalar $E$ and the part that corresponds to $B$.  The $E$/$B$ modes are orthogonal to each other in the sense that their respective operators are orthogonal to each other as discussed in \sec{sec:mat_pol_intro}. It is possible to decompose the Stokes vector \vp{} into one \vp{\rm E} that purely contributes to $E$ modes and another \vp{\rm B} that purely contribute to the $B$ modes of polarization. In this section we derive the real space operators which operate on the total Stokes vector and yield this decomposition, without ever having to explicitly evaluate the scalar $E/B$ modes \footnote{While this manuscript was in preparation, work appeared in \cite{2018JCAP...05..059L} that discussed a similar idea of real-space purification of Stokes $Q/U$ parameters to their $E/B$ counterparts, but we treat it here in much more detail.}. The algebra is more involved, but the derivation is similar to that discussed in \sec{sec:qu2eb}, so we refrain from presenting the detailed calculations here and outline only the key points. We use the harmonic space projection operators $\tilde O_{E/B}$, defined in \eq{eq:har_eb_op}, to derive the respective real space operators. The Stokes parameters corresponding to each scalar mode are given by the following expressions:
\beqry
\bar{P}_E &=&  [\bar T^{-1}  {{}_2\mathcal{Y}} \, \tilde T  \tilde O_E \tilde T^{-1} {{}_2\mathcal{Y}^{\ddagger}}\, \bar T] \bar{P}  \,, \\
&=& [\frac{1}{4} \bar T^{\dagger }  {{}_2\mathcal{Y}} \, \tilde T  \tilde O_E  \tilde T^{\dagger}  {{}_2\mathcal{Y}^{\ddagger}}\, \bar T ]\bar{P}  \,, \nonumber \\
&=&  \bar O_{E} \bar{P} \,,\nonumber \\
\bar{P}_B &=&  [\bar T^{-1}  {{}_2\mathcal{Y}} \, \tilde T  \tilde O_B \tilde T^{-1} {{}_2\mathcal{Y}^{\ddagger}} \bar T]\bar{P}  \,, \\
&=& [\frac{1}{4} \bar T^{\dagger }  {{}_2\mathcal{Y}}\, \tilde T  \tilde O_B \tilde T^{\dagger} {{}_2\mathcal{Y}^{\ddagger}}\, \bar T] \bar{P}   \,, \nonumber\\
&=&  \bar O_{B} \bar{P} \,. \nonumber
\eeqry
We contract over all the matrix operators to arrive at the the real space operators. On working through the algebra it can be shown that the real space operators have the following form:
\beq
\bar O_{E/B} = 0.5 \Delta \Omega \Bigg\lbrace \bmat \mathcal{I}_{r} & \mathcal{I}_{i} \\  -\mathcal{I}_{i}  & \mathcal{I}_{r} \emat \pm \bmat \mathcal{D}_{r} & \mathcal{D}_{i} \\  \mathcal{D}_{i}  & - \mathcal{D}_{r} \emat \Bigg\rbrace \,,\\
\eeq
where $\mathcal{I}_{r}, \mathcal{D}_{r}$ and $\mathcal{I}_{i}, \mathcal{D}_{i}$ are the real and imaginary parts of the following complex functions:
\begin{subequations}
\beqry
\mathcal{I} (\hat{n}_e,\hat{n}_q) &=& \mathcal{I}_{r} + i \mathcal{I}_{i} = \sum_{\ell m} {_{-2}Y}_{\ell m}(\hat n_e) {_{-2}Y}^*_{\ell m}(\hat n_q) \,, \\
\mathcal{D}(\hat{n}_e,\hat{n}_q)  &=& \mathcal{D}_{r} + i\mathcal{D}_{i} = \sum_{\ell m} {_2Y}_{\ell m}(\hat n_e) {_{-2}Y}^*_{\ell m}(\hat n_q) \,.
\eeqry
\end{subequations}
These functions can be further simplified using the identity of spin spherical harmonics given in \eq{eq:sum_spin_shf}. Specifically, it can be shown that these functions reduce to the following mathematical forms:
\beqrys \label{eq:fn_i}
\mathcal{I}(\hat{n}_e, \hat{n}_q) &=& \sum_{\ell} \sqrt{\frac{2\ell+1}{ 4 \pi}}{_{-2}Y}_{\ell2}(\beta_{qe}, \alpha_{qe}) ~ \rm{e}^{i2 \gamma_{qe}} \label{eq:healpix-compatible-i} = \mathcal{I}_r + i \mathcal{I}_i \,, \\
\mathcal{I}_r + i \mathcal{I}_i &=& \Big [ \cos(2 \alpha_{qe} +  2\gamma_{qe}) + i \sin(2 \alpha_{qe} +  2 \gamma_{qe}) \Big]   {{}_{\mi}f}(\beta_{qe},\ell_{\rm min},\ell_{\rm max}) \,,
\eeqrys
\beqrys \label{eq:fn_d}
\mathcal{D}(\hat{n}_q, \hat{n}_e) &=& \sum_{\ell} \sqrt{\frac{2\ell+1}{ 4 \pi}}{_2Y}_{\ell 2}(\beta_{qe}, \alpha_{qe}) ~ \rm{e}^{- i2 \gamma_{qe}} \label{eq:healpix-compatible-m} =\mathcal{D}_r + i \mathcal{D}_i \,, \\
\mathcal{D}_r + i \mathcal{D}_i &=&  \Big [ \cos(2 \alpha_{qe} - 2\gamma_{qe}) + i \sin(2 \alpha_{qe} -  2 \gamma_{qe}) \Big]   {{}_{\md}f}(\beta_{qe},\ell_{\rm min},\ell_{\rm max}) \,,
\eeqrys
where the radial functions are given by:
\beq
{{}_{\mdi}f}(\beta,\ell_{\rm min},\ell_{\rm max}) = \sum_{\ell=\ell_{\rm min}}^{\ell_{\rm max}} \sqrt{\frac{2\ell+1}{ 4 \pi}} {{}_{ \mdi}f}_{\ell}(\beta) \label{eq:f2_rad_ker}\,,
\eeq
where the functions ${{}_{\mdi}f}_{\ell}(\beta)$ are expressed in terms of $P_{\ell}^2$ Legendre polynomials and are given by the following explicit mathematical forms:
 \beqry
 _{\mdi}f_{\ell}(\beta) &=& 2 \frac{(\ell-2)!}{(\ell+2)!}  \sqrt{\frac{2\ell +1 }{4 \pi}} \Bigg[ - P_{\ell}^{2} (\cos  \beta) \left( \frac{\ell-4}{\sin^2 \beta} + \frac{1}{2}\ell(\ell-1) \pm \frac{2 (\ell-1) \cos \beta}{\sin^2 \beta} \right) \nonumber \\ 
&+& P_{\ell-1}^2 (\cos \beta) \left( (\ell+2) \frac{\cos \beta}{\sin^2 \beta} \pm \frac{2 (\ell+2)}{ \sin^2 \beta } \right) \Bigg] \,. \label{eq:rad_ker_quequbqu}
 \eeqry
Finally the Stokes parameters corresponding to the respective scalar fields can be computed by evaluating the following expressions:
\beqry \label{eq:op_qu2equbqu}
\bmat Q_e \\ U_e  \emat_{E/B} &=& \sum_{q=1}^{N_{\rm pix}} \Bigg\lbrace {{}_{\mi}f}(\beta_{qe},\ell_{\rm min},\ell_{\rm max}) \bmat \cos(2 \alpha_{qe} + 2\gamma_{qe}) & \sin(2\alpha_{qe} +2 \gamma_{qe}) \\  -\sin(2\alpha_{qe} +2 \gamma_{qe})  & \cos(2 \alpha_{qe} + 2 \gamma_{qe}) \emat  \bmat Q_q \\ U_q  \emat  \\ &\pm& {}_{\md}f(\beta_{qe},\ell_{\rm min},\ell_{\rm max}) \bmat \cos(2 \alpha_{qe} - 2\gamma_{qe}) &  \sin(2\alpha_{qe} - 2 \gamma_{qe}) \\  \sin(2\alpha_{qe} - 2 \gamma_{qe})  & - \cos(2 \alpha_{qe} - 2 \gamma_{qe}) \emat  \bmat Q_q \\ U_q  \emat \Bigg\rbrace  \frac{\Delta\Omega}{2}  \,, \nonumber 
\eeqry
where all the symbols have their usual meaning. The above expression can be cast in the further simplified form,
\begin{subequations}
\beqry
{}_{+2}X_{E/B}(\hat{n}_e) &=& 0.5 \Delta \Omega\sum_{q=1}^{N_{\rm pix}}  {{}_{\mi}f}(\beta_{qe}) e^{-i2 (\alpha_{qe} + \gamma_{qe})} {}_{+2}X(\hat{n}_q)  \pm {{}_{\md}f}(\beta_{qe}) e^{i2 (\alpha_{qe} - \gamma_{qe})} {}_{+2}X(\hat{n}_q)^* \,, \nonumber \\
&=& 0.5 \Bigg\lbrace \Delta \Omega \sum_{q=1}^{N_{\rm pix}}  {}_{+2}X(\hat{n}_q)  \, \mathcal{I}_G(\hat{n}_q) \pm {}_{+2}X(\hat{n}_q)^* \, \mathcal{D}_G(\hat{n}_q) \Bigg\rbrace \hspace{.5cm } \textrm{\emph{  Radiation kernel}} \nonumber \,, \\ \vspace{-1cm} \\
&=& 0.5 \Bigg\lbrace \mathcal{I}_B \star {}_{+2}X\pm \mathcal{D}_B \star {}_{+2}X^* \Bigg\rbrace(\hat{n}_e)   \hspace{.5cm } \textrm{\emph {  Convolution kernel}} \,, \label{eq:qu2equbqu_convolution}
\eeqry
\end{subequations}
where all the symbols have their usual meaning and the explicit multipole dependence of the real space operators has been suppressed for brevity. Note that when the operators are expressed in terms of the Euler angles $(\alpha_{qe},\beta_{qe},\gamma_{qe})$ they can be interpreted as the Greens functions and  we denote them by $\mathcal{I}_G=\mathcal{I}^*$ and $\mathcal{D}_G=\mathcal{D}$. When expressed as function of Euler angles $(\alpha_{eq},\beta_{eq},\gamma_{eq})$ corresponding to the inverse rotations they can be interpreted as some convolving beam and we denote them by $\mathcal{I}_B=\mathcal{I}$ and $\mathcal{D}_B=\mathcal{D}$. Note that unlike in the case of the operators $\mm_G$ and $\mm_B$ which have different shapes owing to their dependence on Euler angles $\alpha$ and $\gamma$ respectively, the operators $D_G$ and $D_B$ are identical since $(\alpha_{qe}-\gamma_{qe}) = (\alpha_{eq}-\gamma_{eq})$, while $\mathcal{I}_{G}$ and $\mathcal{I}_B$ are related by conjugation since  $(\alpha_{qe}+\gamma_{qe}) = -(\alpha_{eq}+\gamma_{eq})$.

The operator $\mathcal{I}$ is Hermitian and is a band limited version of the delta function owing to the identity: $\lim_{\ell_{\rm max} \rightarrow \infty} \mathcal{I} = \delta_{\hat{n}_i, \hat{n}_j}$. For all practical purposes $\mathcal{I}$ acts like an identity operator as ascertained by the following set of identities: (i) $\mathcal{I} \mathcal{I}=\mathcal{I}$ ; (ii) $\mathcal{D} \mathcal{I}=\mathcal{D}$. $\mathcal{D}$ is a complex but symmetric matrix and $\mathcal{D}^*$ is its inverse in this band limited sense: $\mathcal{D}^* \mathcal{D}=\mathcal{I}$. Using these properties\footnote{While testing the real space operator identities one encounters terms like $\mathcal{D} \mathcal{I}^*,\mathcal{I}^*\mathcal{I}$ and $\mathcal{I}\mathcal{I}^*$ which cannot be simply interpreted but they always occur in pairs with opposite signs that exactly cancel each other.} of the operators $\mathcal{I}$ and $\mathcal{D}$ , one can verify that the real space operators satisfy the following identities:
\begin{subequations}
\beqry
\bar O_E \, \bar O_E &=& \bar O_E;\qquad \bar O_B \, \bar O_B = \bar O_B \,, \\
\bar O_E \, \bar O_B &=& 0 \,,\label{eq:real_ortho}\\
\bar O_E + \bar O_B &=& \mathcal{I} \,,
\eeqry
\end{subequations}
which are the real space analogues of their harmonic space counterparts discussed in \sec{sec:mat_pol_intro}. Thus they are exactly orthogonal and idempotent. Note that unlike in the harmonic case, the sum of the operators is the band limited identity operator $\mathcal{I}$. This non-exactness is representative of the loss of information resulting from making this transformation on measured data with some imposed band limit. If we were to force the sum of the operators to be exactly an identity matrix, we would compromise the orthogonality property of $\bar{O}_E$ and $\bar{O}_B$, which is exact (\eq{eq:real_ortho}) and a more crucial property of the operators.
\begin{figure}[!t]
  \begin{center}
  \begin{tabular}{m{8ex}m{\kernelfigwidth}m{\kernelfigwidth}|m{\kernelfigwidth}m{\kernelfigwidth}}
$b=90^\circ$&
\hspace{\kernelfigspace}\includegraphics[width=\kernelfigwidth]{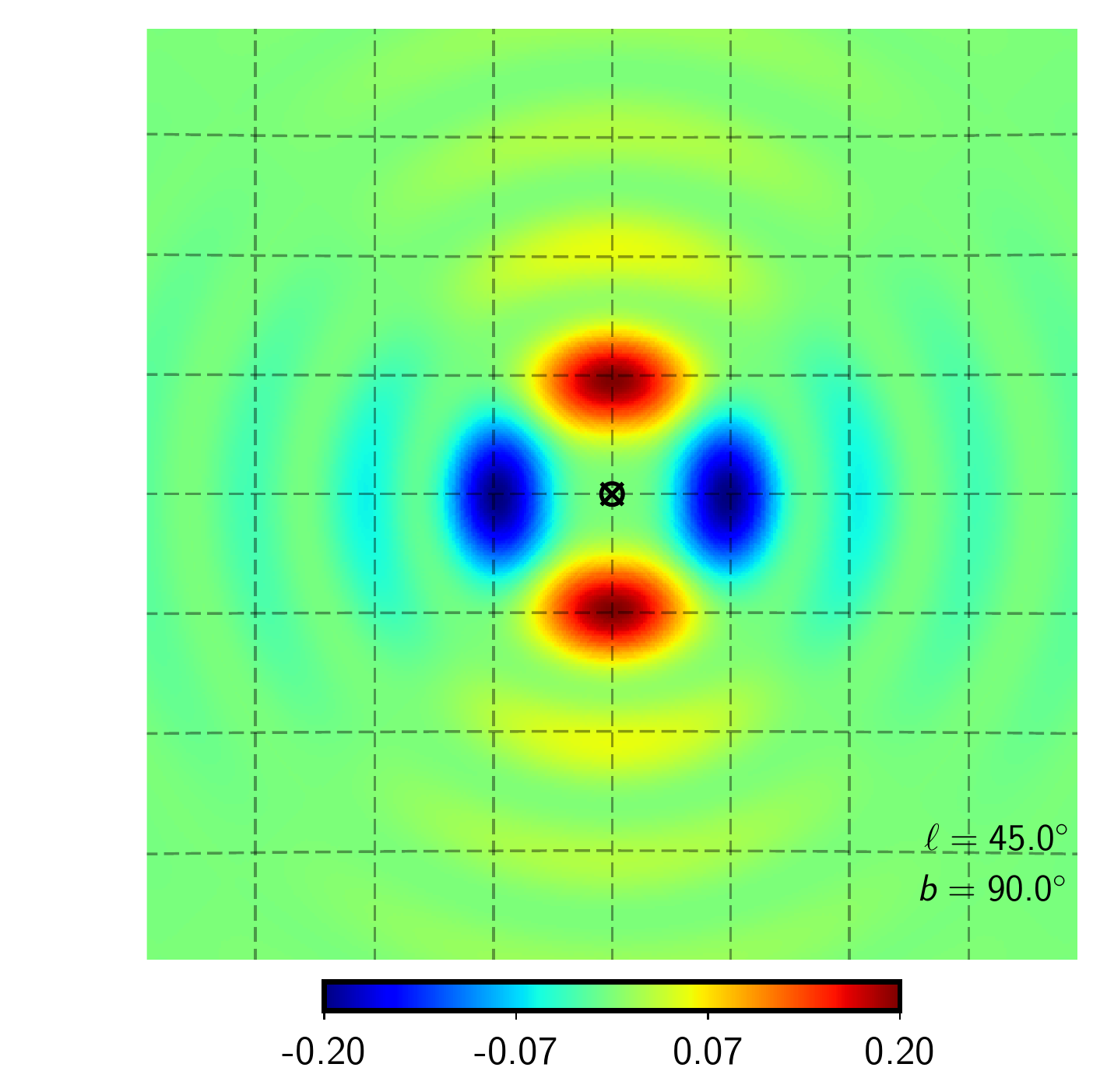} &
\hspace{\kernelfigspace}\includegraphics[width=\kernelfigwidth]{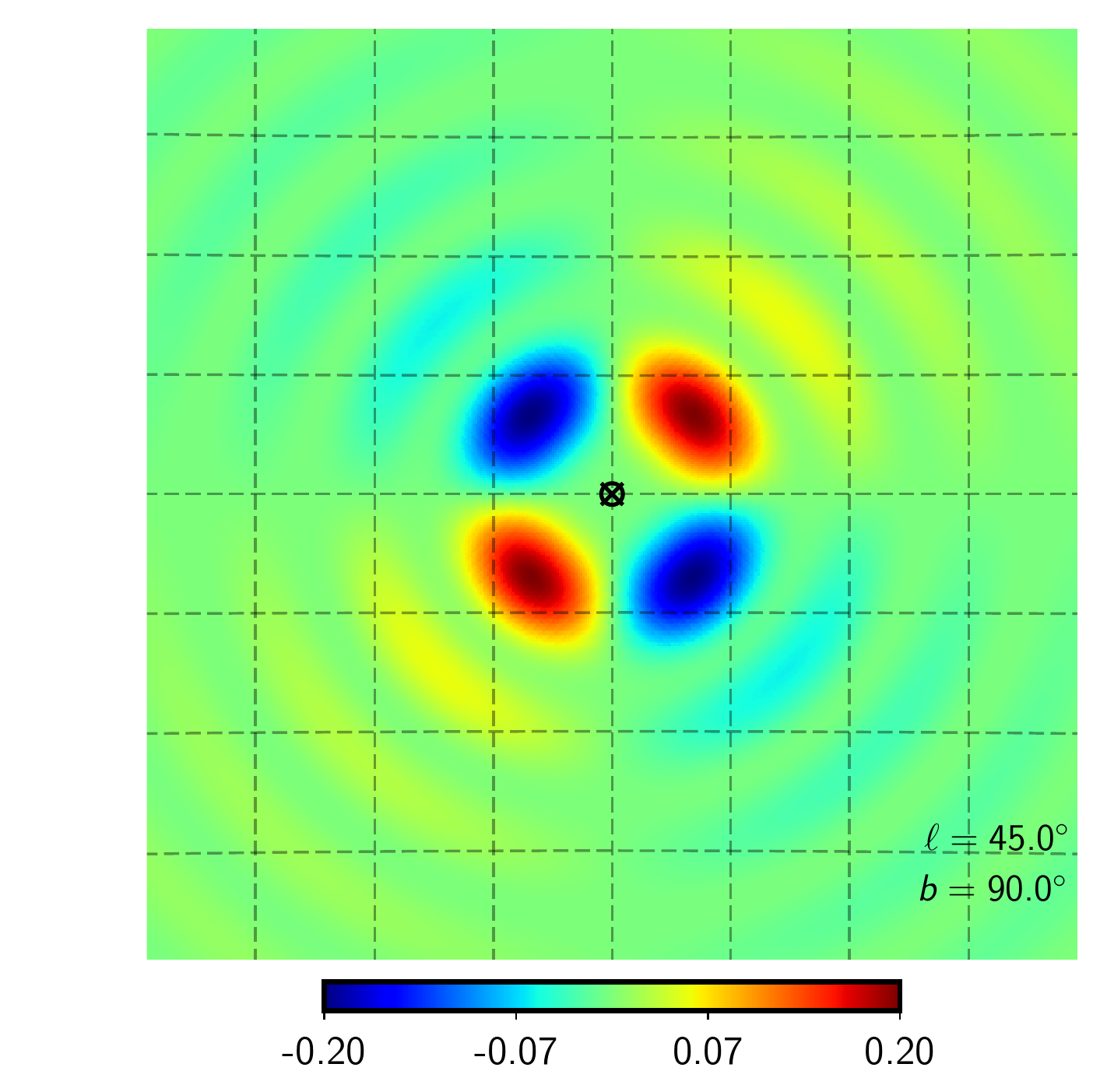} &
\hspace{\kernelfigspace}\includegraphics[width=\kernelfigwidth]{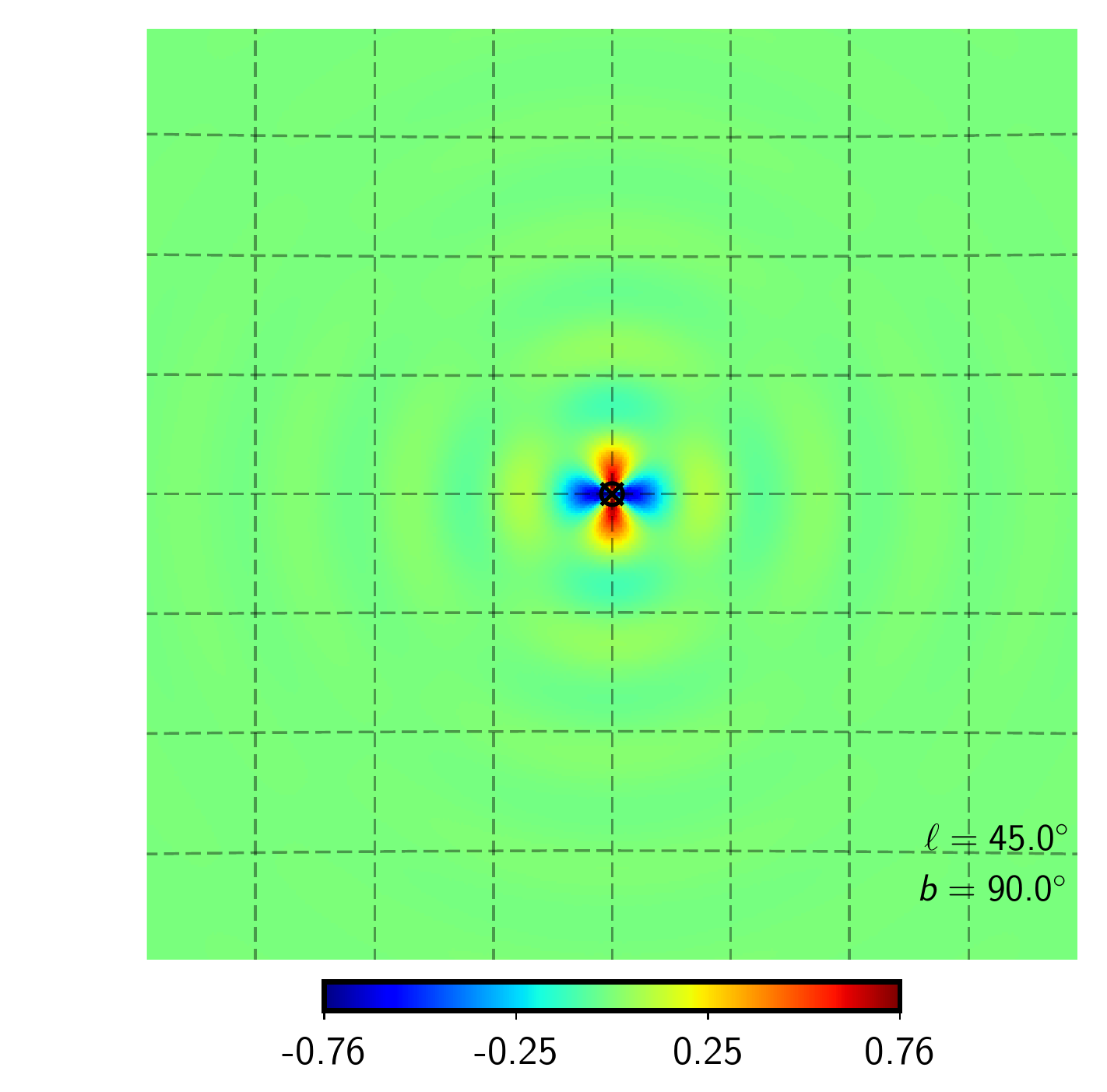} &
\hspace{\kernelfigspace}\includegraphics[width=\kernelfigwidth]{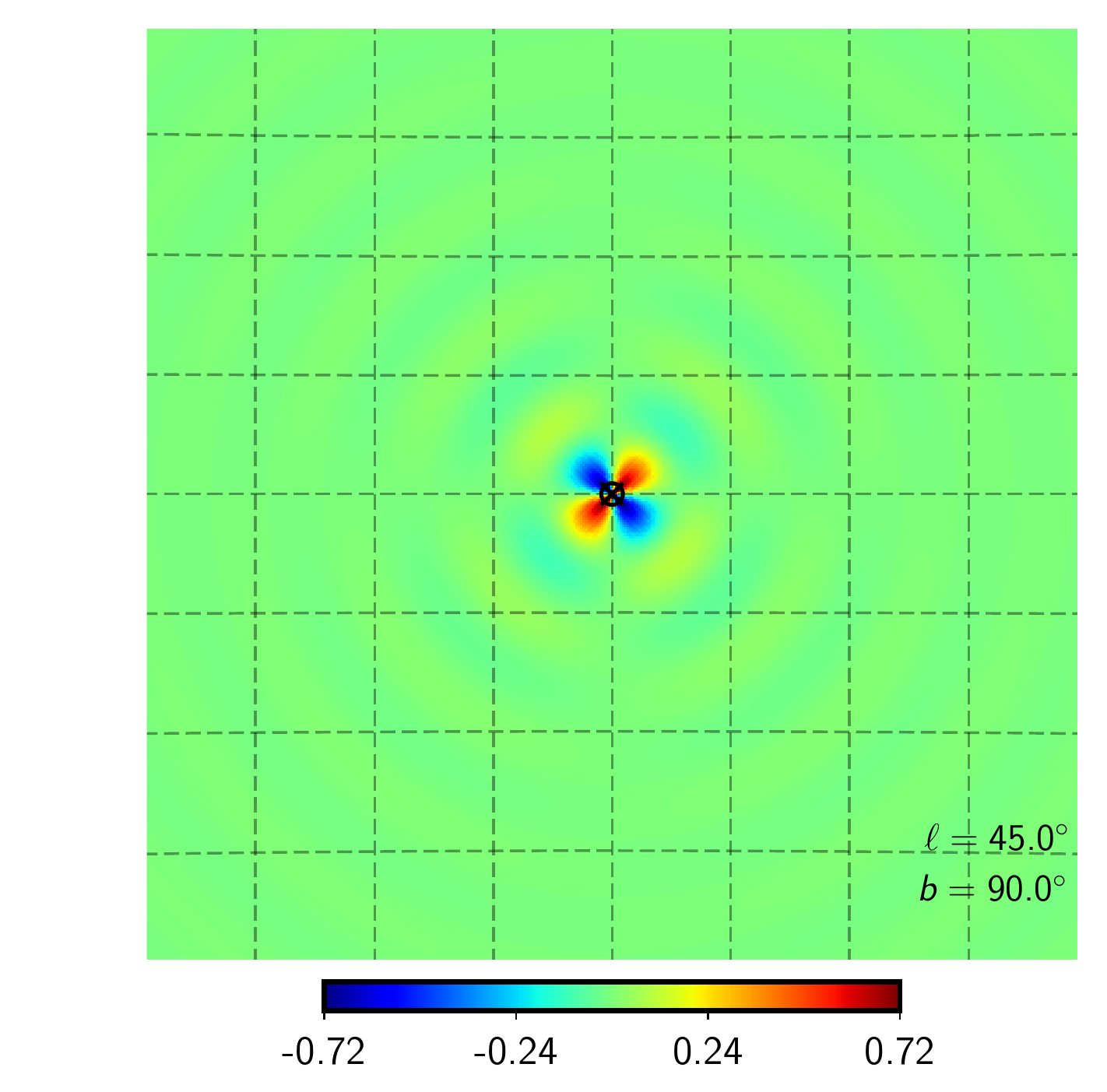} \\
$b=87^\circ$&
\hspace{\kernelfigspace}\includegraphics[width=\kernelfigwidth]{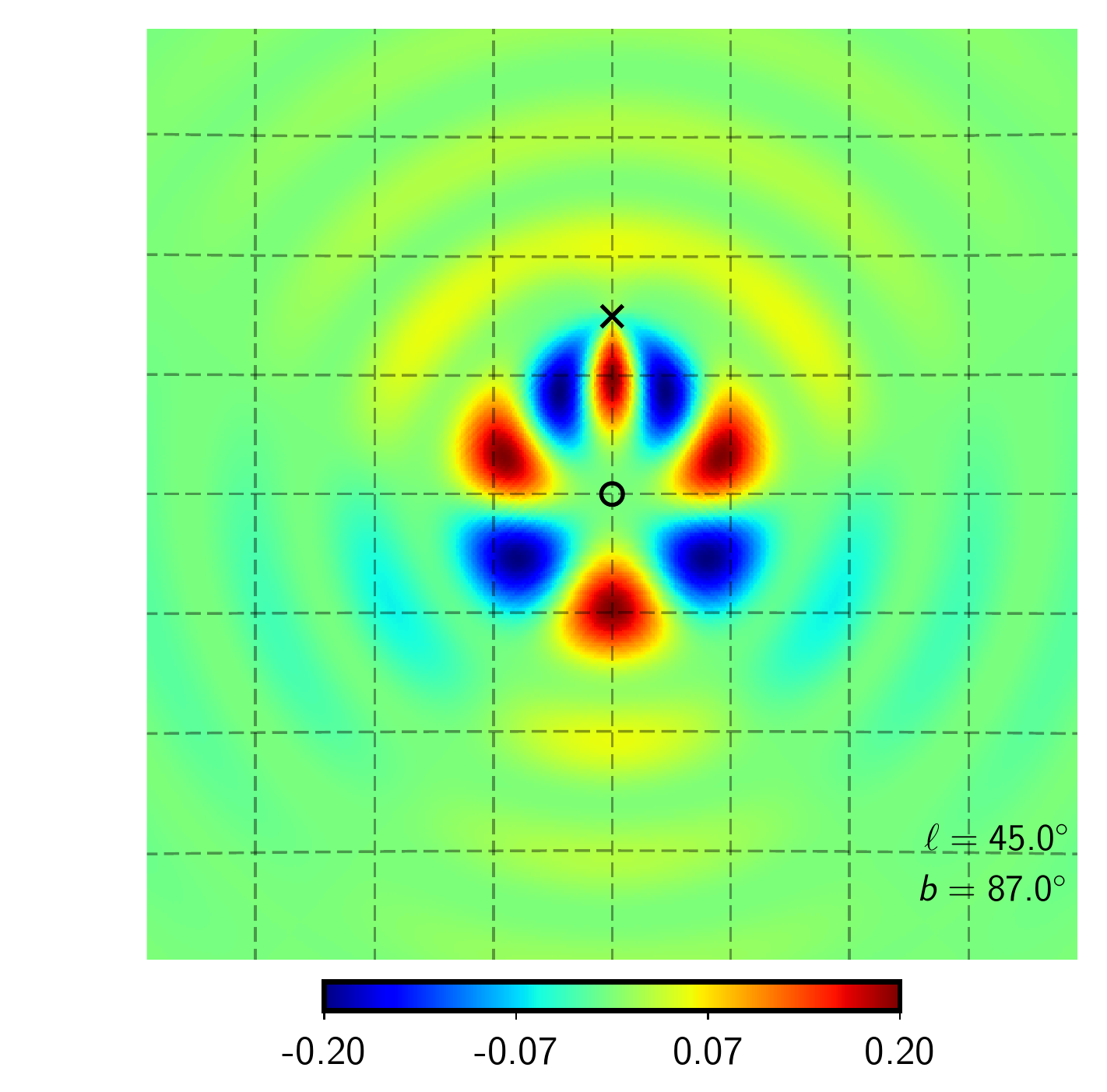} &
\hspace{\kernelfigspace}\includegraphics[width=\kernelfigwidth]{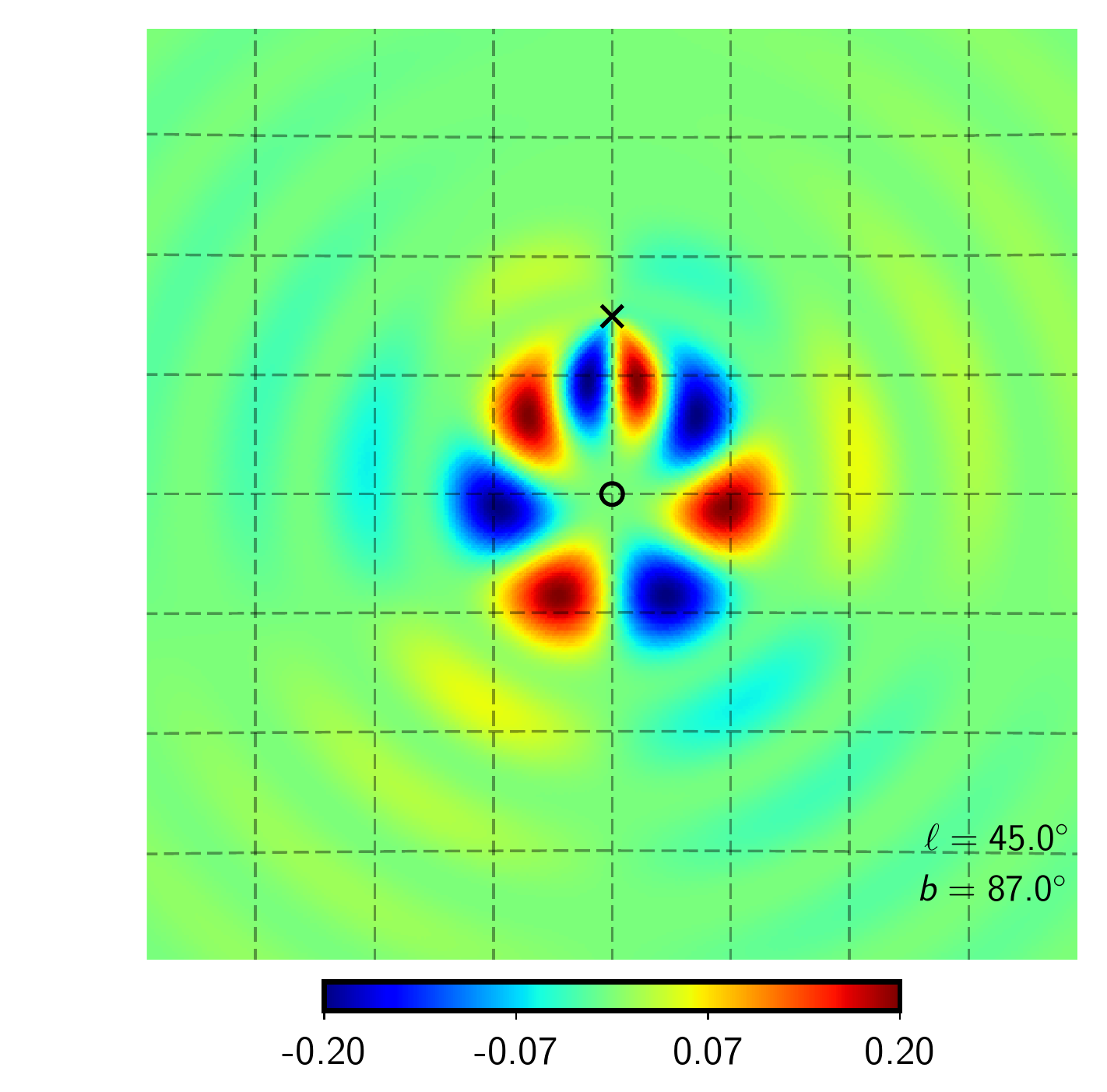} &
\hspace{\kernelfigspace}\includegraphics[width=\kernelfigwidth]{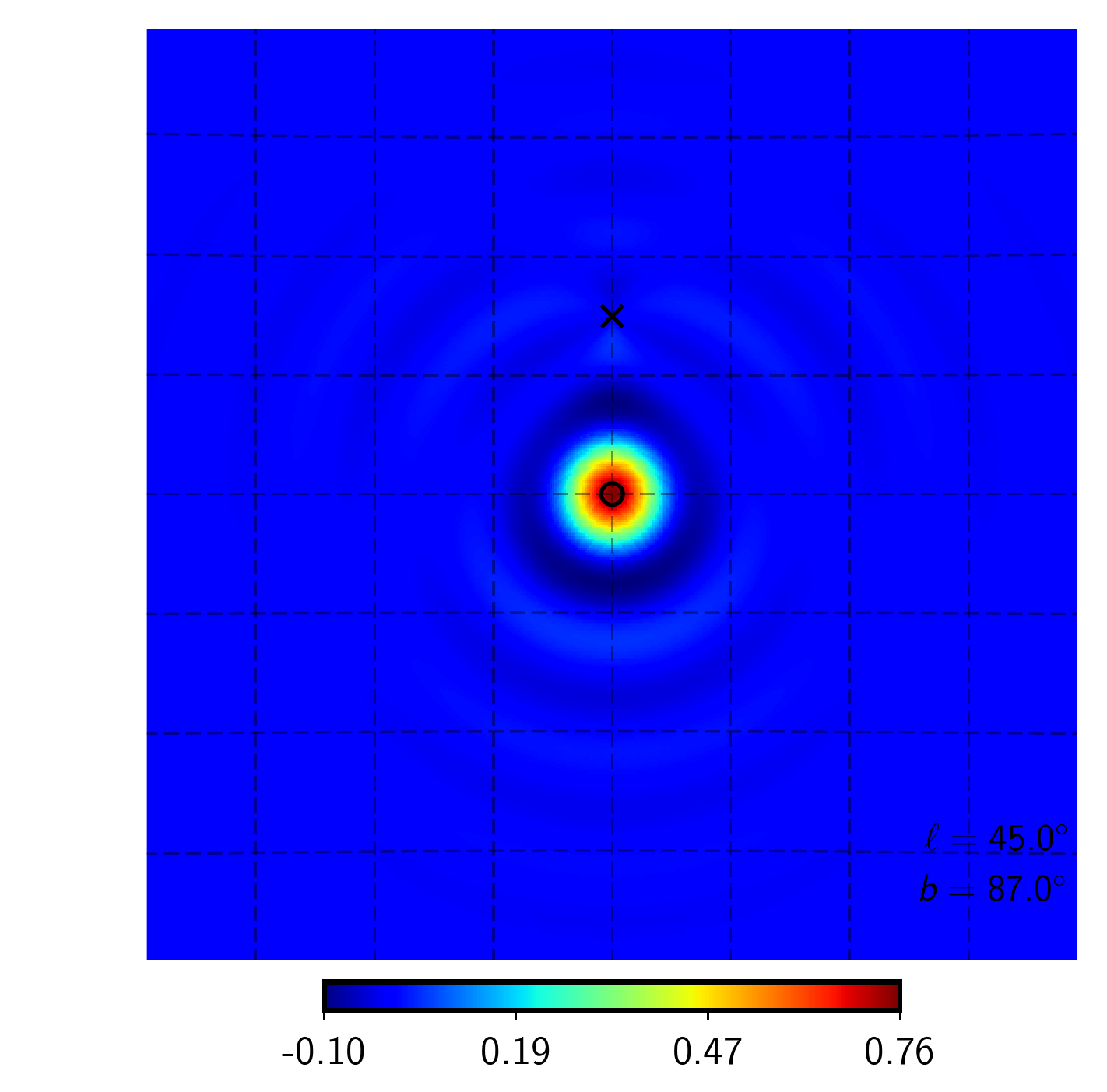} &
\hspace{\kernelfigspace}\includegraphics[width=\kernelfigwidth]{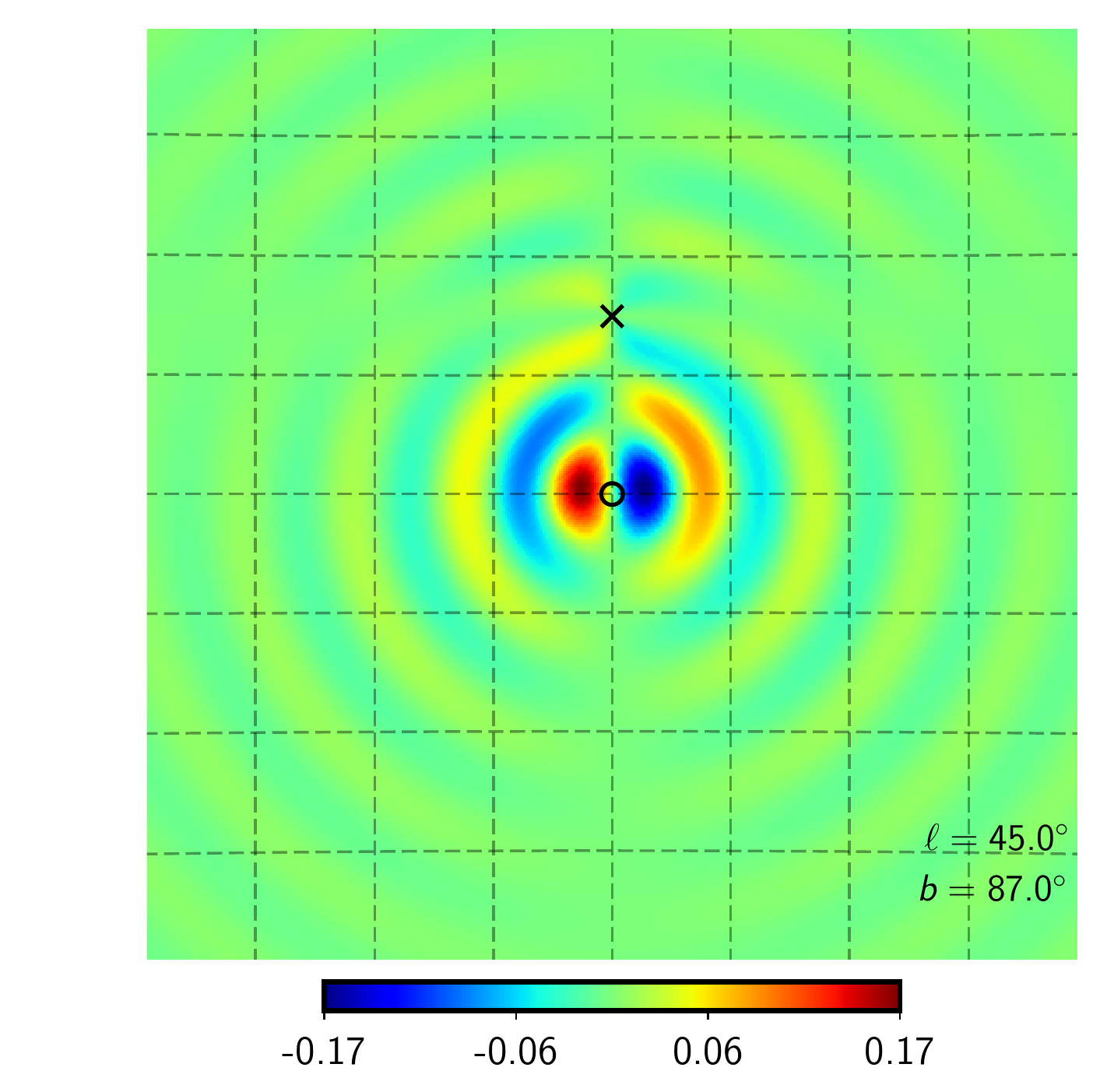} \\
$b=80^\circ$&
\hspace{\kernelfigspace}\includegraphics[width=\kernelfigwidth]{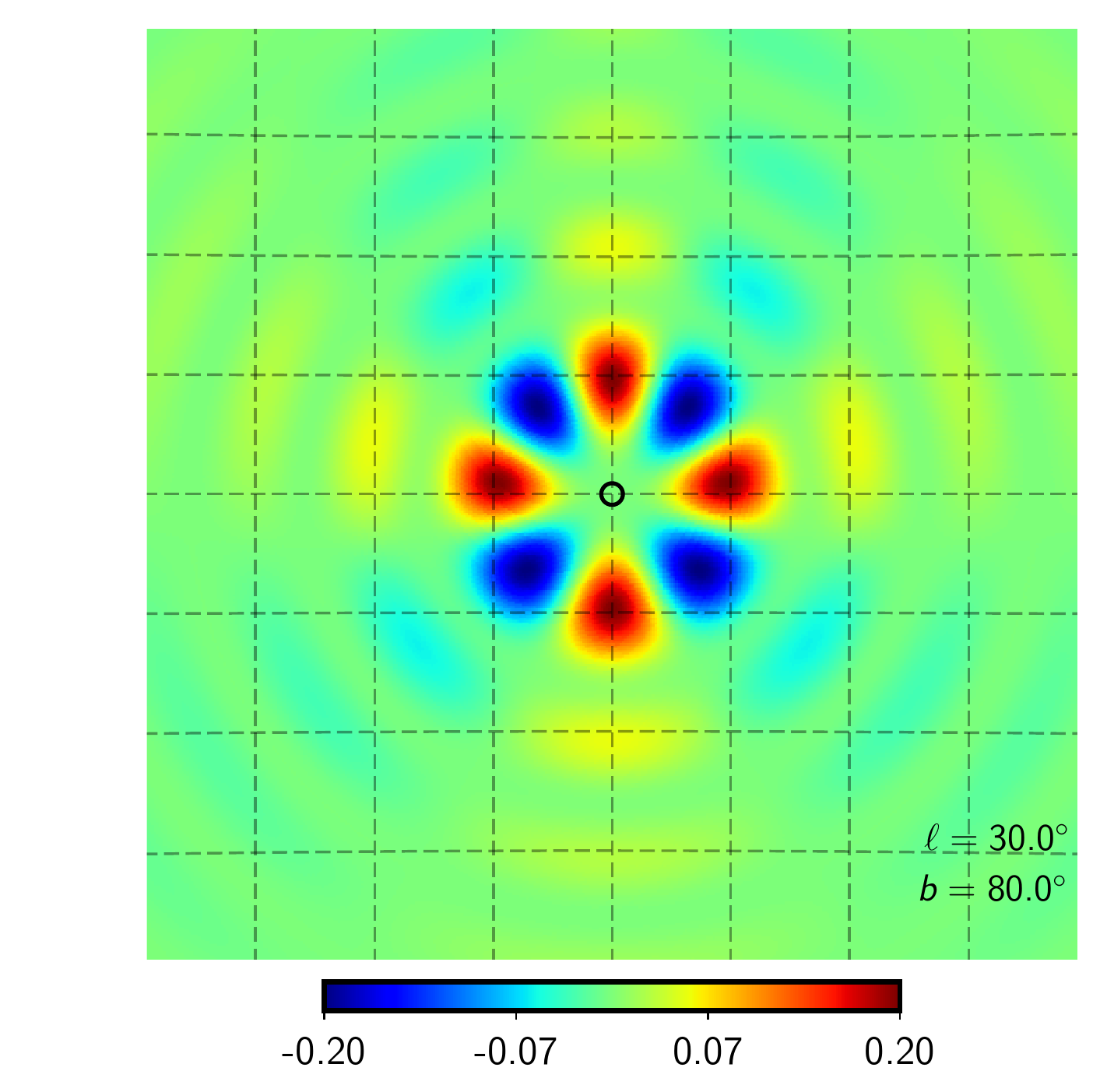} &
\hspace{\kernelfigspace}\includegraphics[width=\kernelfigwidth]{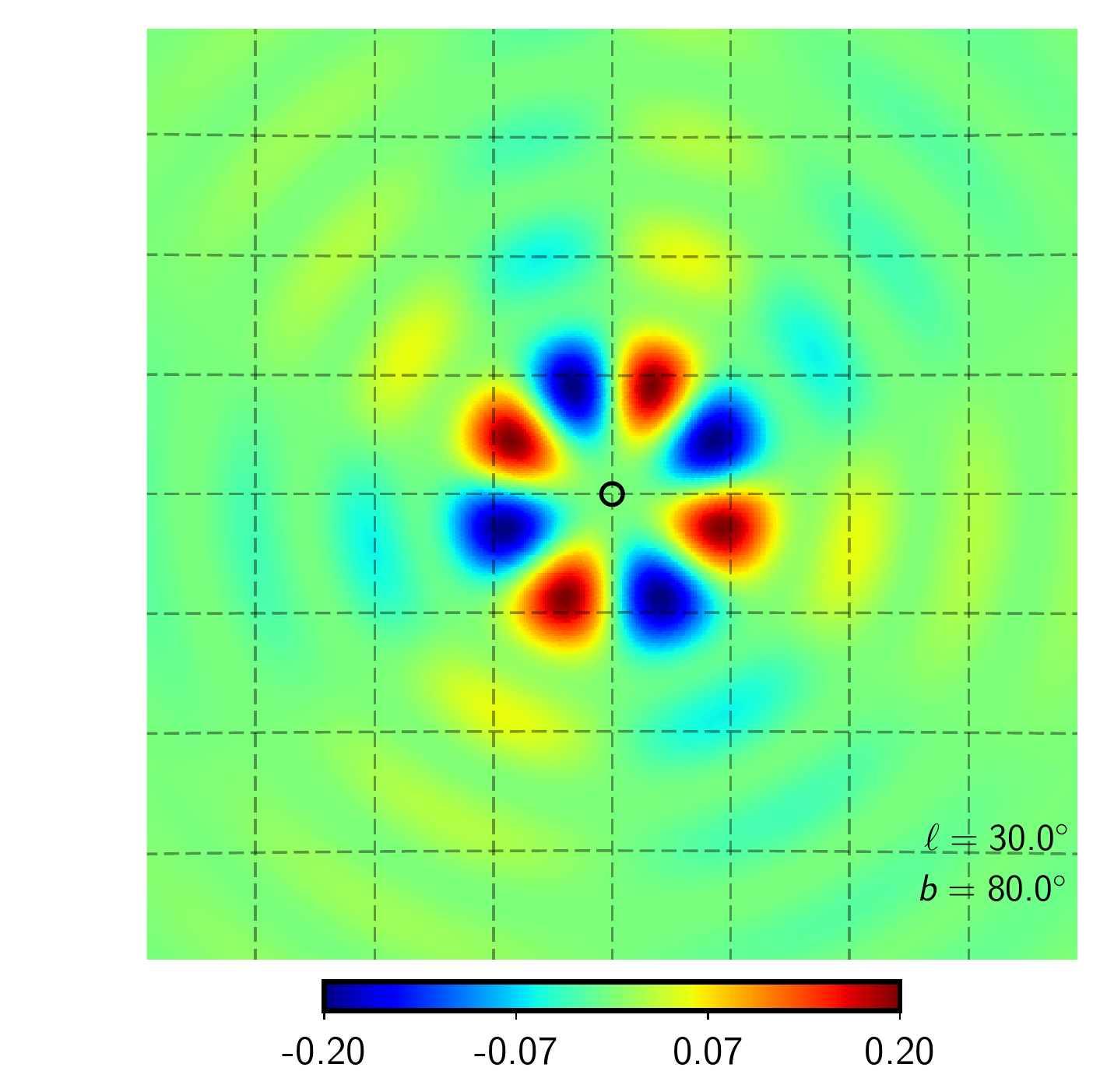} &
\hspace{\kernelfigspace}\includegraphics[width=\kernelfigwidth]{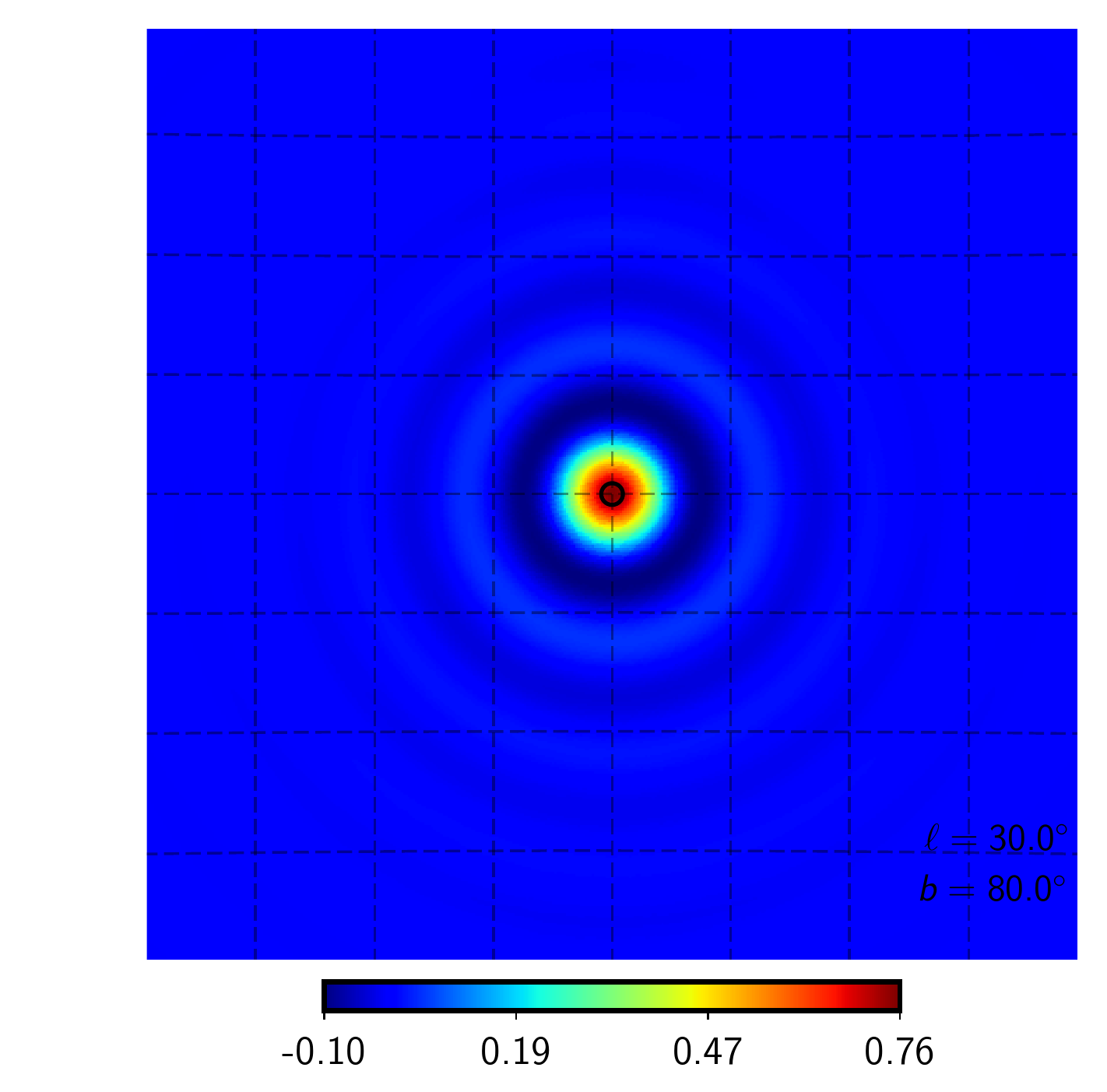} &
\hspace{\kernelfigspace}\includegraphics[width=\kernelfigwidth]{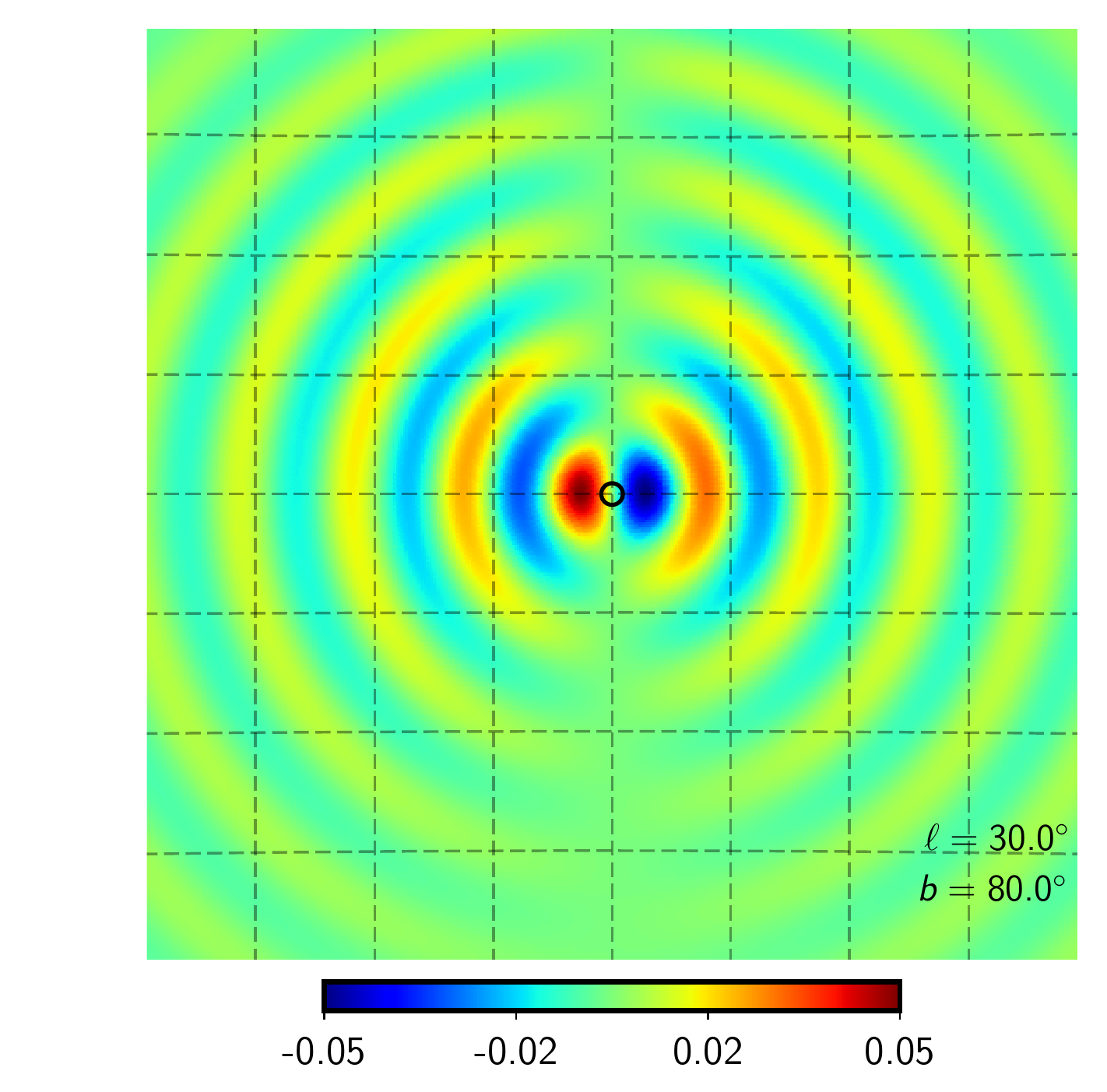} \\
$b=0^\circ$&
\hspace{\kernelfigspace}\includegraphics[width=\kernelfigwidth]{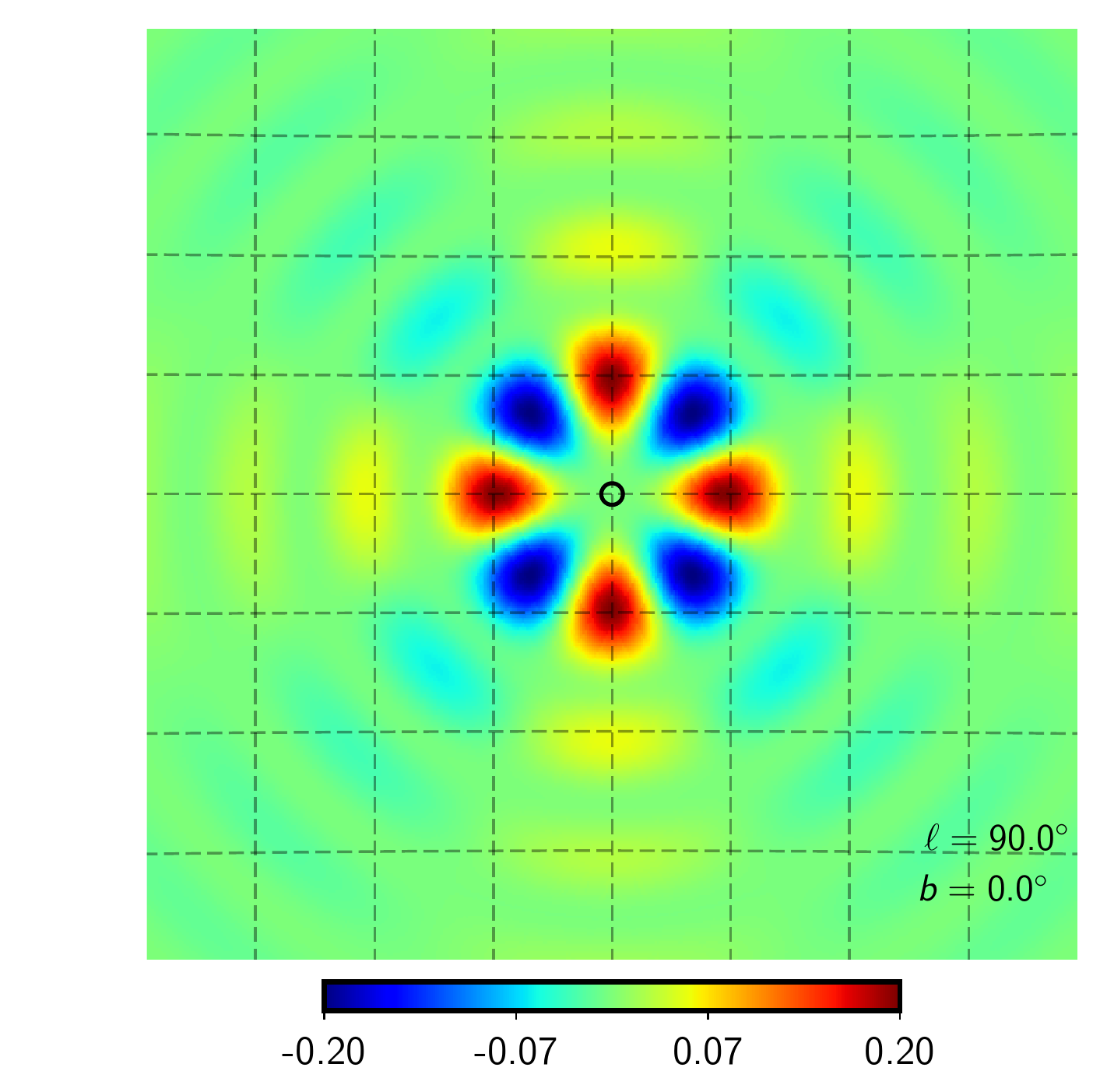} &
\hspace{\kernelfigspace}\includegraphics[width=\kernelfigwidth]{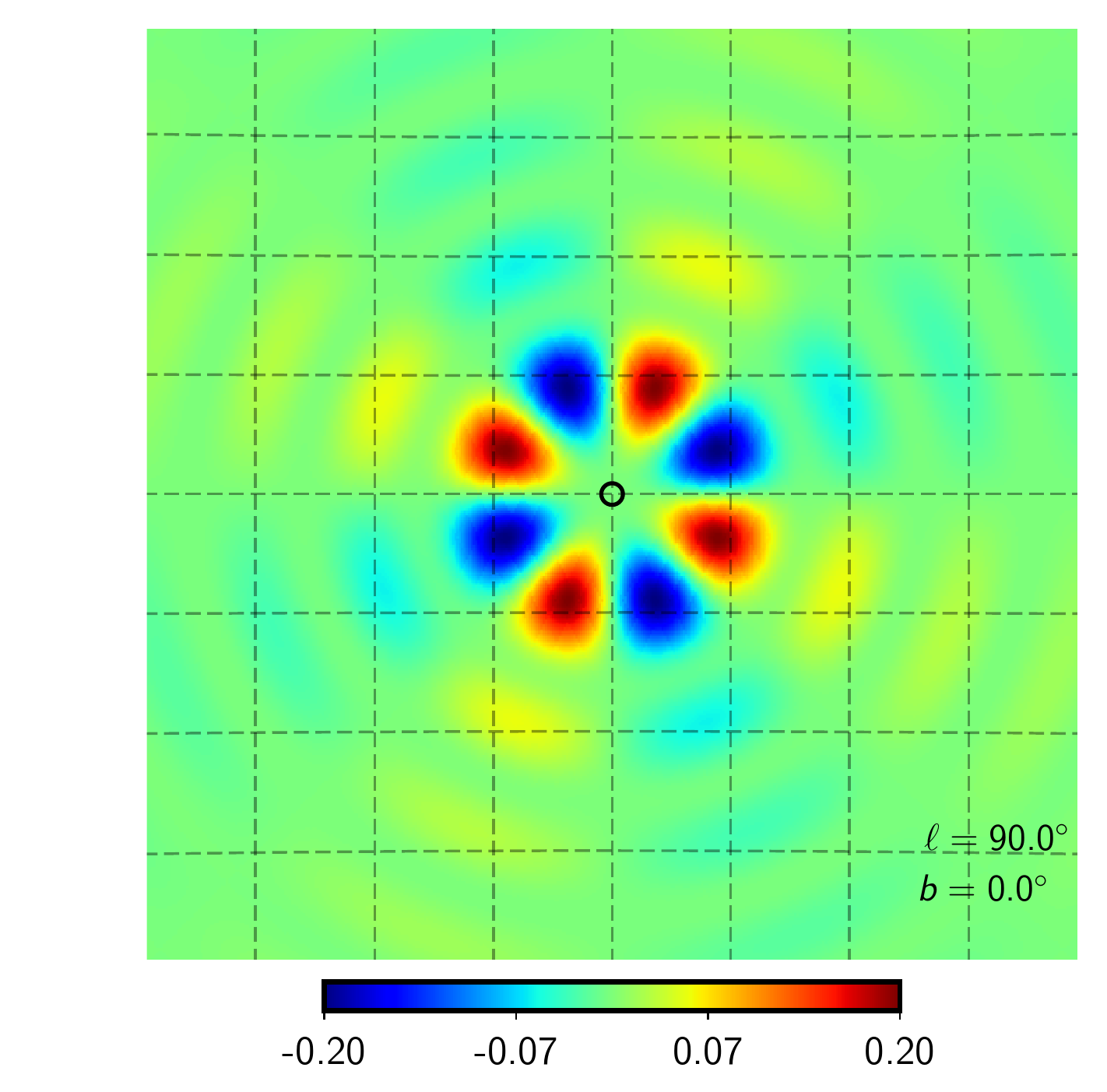} &
\hspace{\kernelfigspace}\includegraphics[width=\kernelfigwidth]{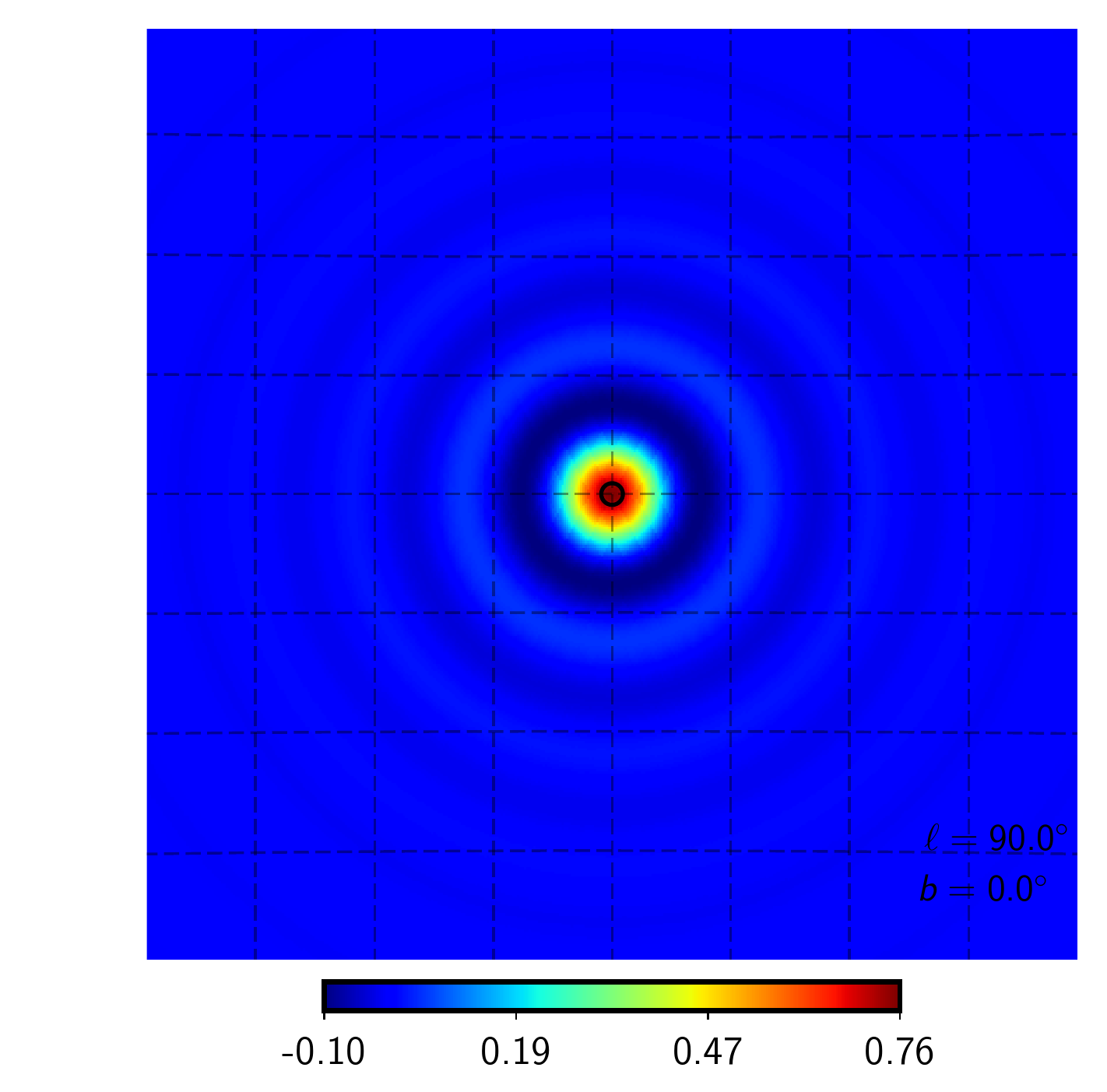} &
\hspace{\kernelfigspace}\includegraphics[width=\kernelfigwidth]{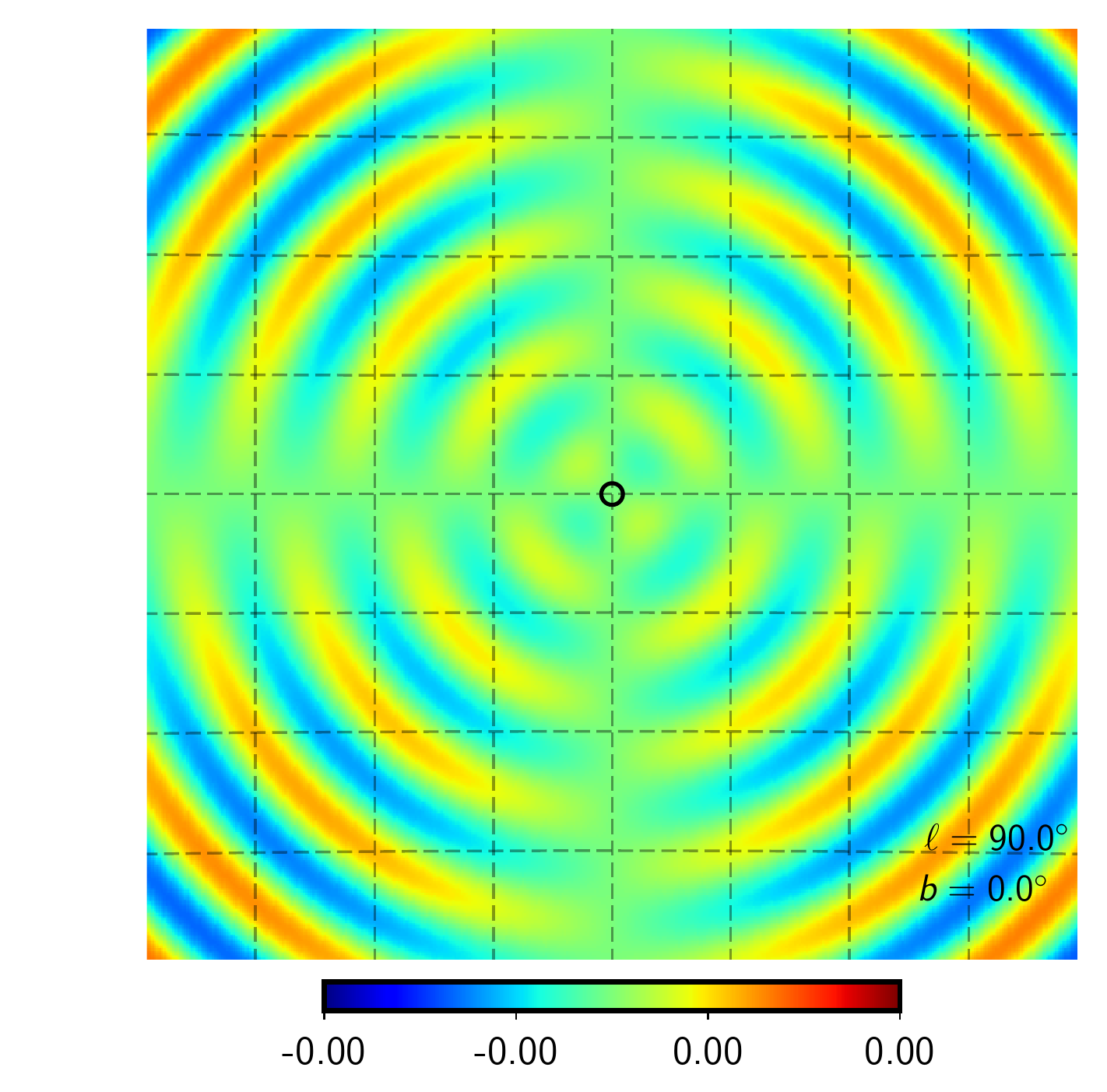} \\
&
\centering $\textrm{Re} \left(\mathcal{D} \right)$ &
\centering $\textrm{Im} \left(\mathcal{D} \right)$ &
\centering $\textrm{Re} \left(\mathcal{I} \right)$ &
\centering $\textrm{Im} \left(\mathcal{I} \right)$
  \end{tabular}
  \end{center}
  \caption{Like Figure \ref{fig:vis_kernel}, but for the kernels that purify Stokes parameter into their $E/B$ parts.} \label{fig:vis_kernel_DI}
\end{figure}

The kernels $\mathcal{D}$ and $\mathcal{I}$ vary significantly as a function of galactic latitude of the central pixel, as seen in \fig{fig:vis_kernel_DI}. These kernels show a two fold symmetry in the vicinity of the poles, which can be understood as a consequence of the Euler angle $\gamma \approx 0$ here and therefore $e^{i2(\alpha \pm \gamma)} \approx e^{i2\alpha}$. Note that in this region, the azimuthal profile of the real and imaginary part of these kernels is identical to $-\mathcal{M}^*_G$.  The imaginary part of the band limited delta function $\mathcal{I}$ contributes just as much as the real part in these regions. On transiting to lower latitudes, however, $\mathcal{D}$ quickly transitions to having a four fold symmetry while $\mathcal{I}$ transitions to being dominated by the real part and behaves more like the conventional delta function. This transition can be most easily understood in the flat sky limit where $\gamma \approx -\alpha$ which leads to the resultant 4 fold symmetry seen for $\mathcal{D}$ owing to $e^{i2(\alpha - \gamma)} \approx e^{i4\alpha}$ and $\mathcal{I}$ being dominated by the real part owing to $e^{-i2(\alpha + \gamma)} \approx 1 + i0$. Since the flat sky approximation has most validity in the proximity of the equator these limiting tendencies of the respective kernels are seen in the bottom row of \fig{fig:vis_kernel_DI} which depict the kernels evaluated at the equator $b=0^{\circ}$. The middle two row depict the kernels evaluated at intermediate latitudes: $b=87^{\circ}$ and $b=80^{\circ}$ and serve to indicate the rate of this transition. As before, these kernels are invariant under changes in longitude of the central pixel with the latitude fixed.
\subsection{The non-locality of the real space operators} \label{sec:radial_locality}
\begin{figure}[t]
\centering
\includegraphics[width=0.8\columnwidth]{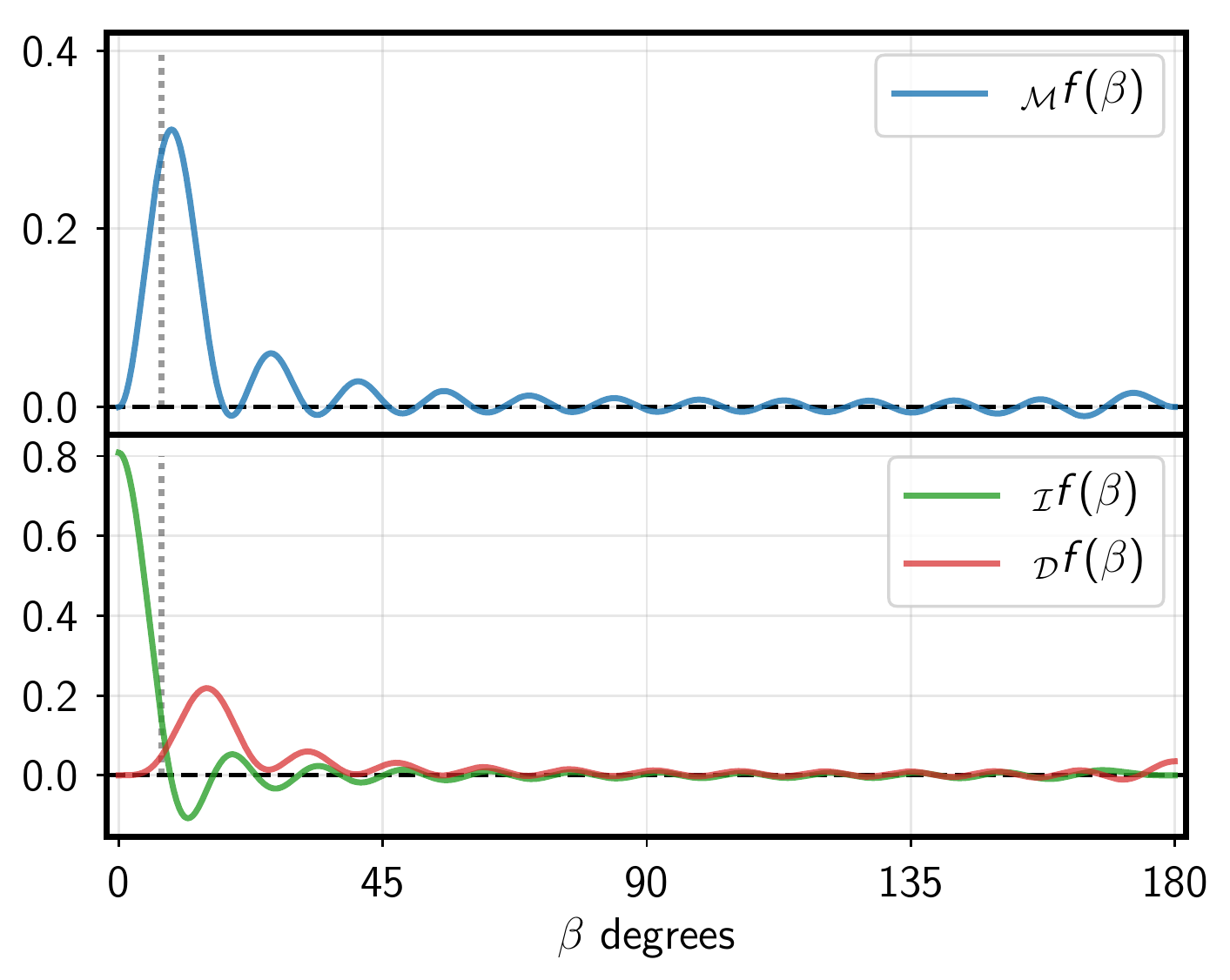}
\caption{The figure depicts the radial part of the convolution kernels. These radial function have been evaluated with the band limit fixed at $\ell \in [2,24]$. The vertical dashed line marks the approximate Healpix pixel size of a $N_{\rm side}=8$, which is the lowest resolution that allows access to $\ell_{\rm max}=24$.}
\label{fig:beta_kernel}
\end{figure}
Above, we have explored in detail the  azimuthal dependence of the real space kernels.  Here we probe the radial dependence, which both  determines the non-locality of the operators and encodes all their multipole dependencies. For illustration, \fig{fig:beta_kernel} shows the radial kernels ${_{\mm}f}, {_{\md}f},{_{\mi}f}$, evaluated using the respective multipole sums given in \eq{eq:rad_ker_queb} and \eq{eq:f2_rad_ker} in the band limit $\ell \in [2,24]$. We choose such a low band limit to highlight some key features of their radial profile.
\begin{figure}[t]
\subfigure[\label{fig:fbeta}]{\includegraphics[width=0.325\columnwidth]{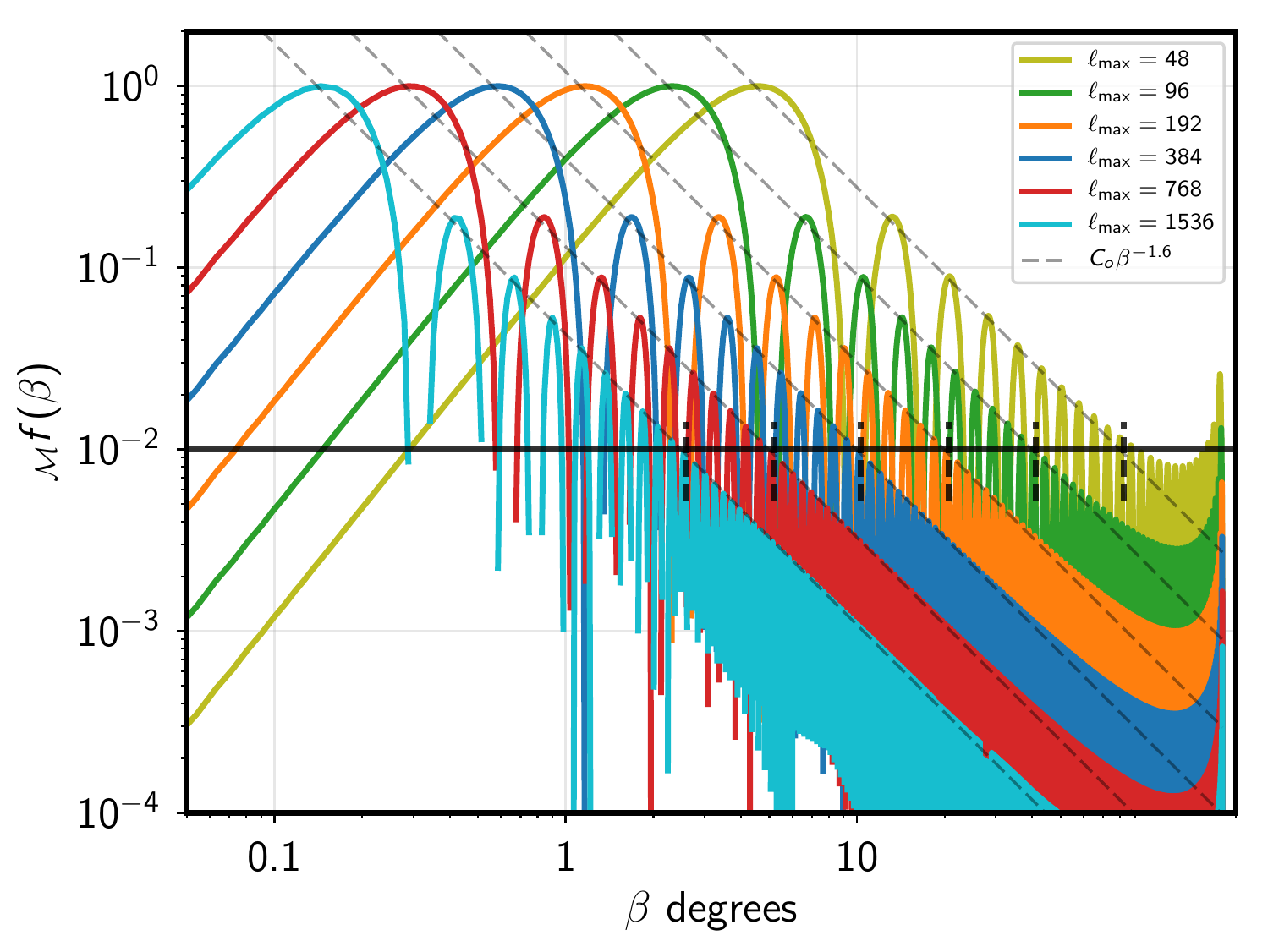}}\hfill
\subfigure[\label{fig:dfbeta}]{\includegraphics[width=0.325\columnwidth]{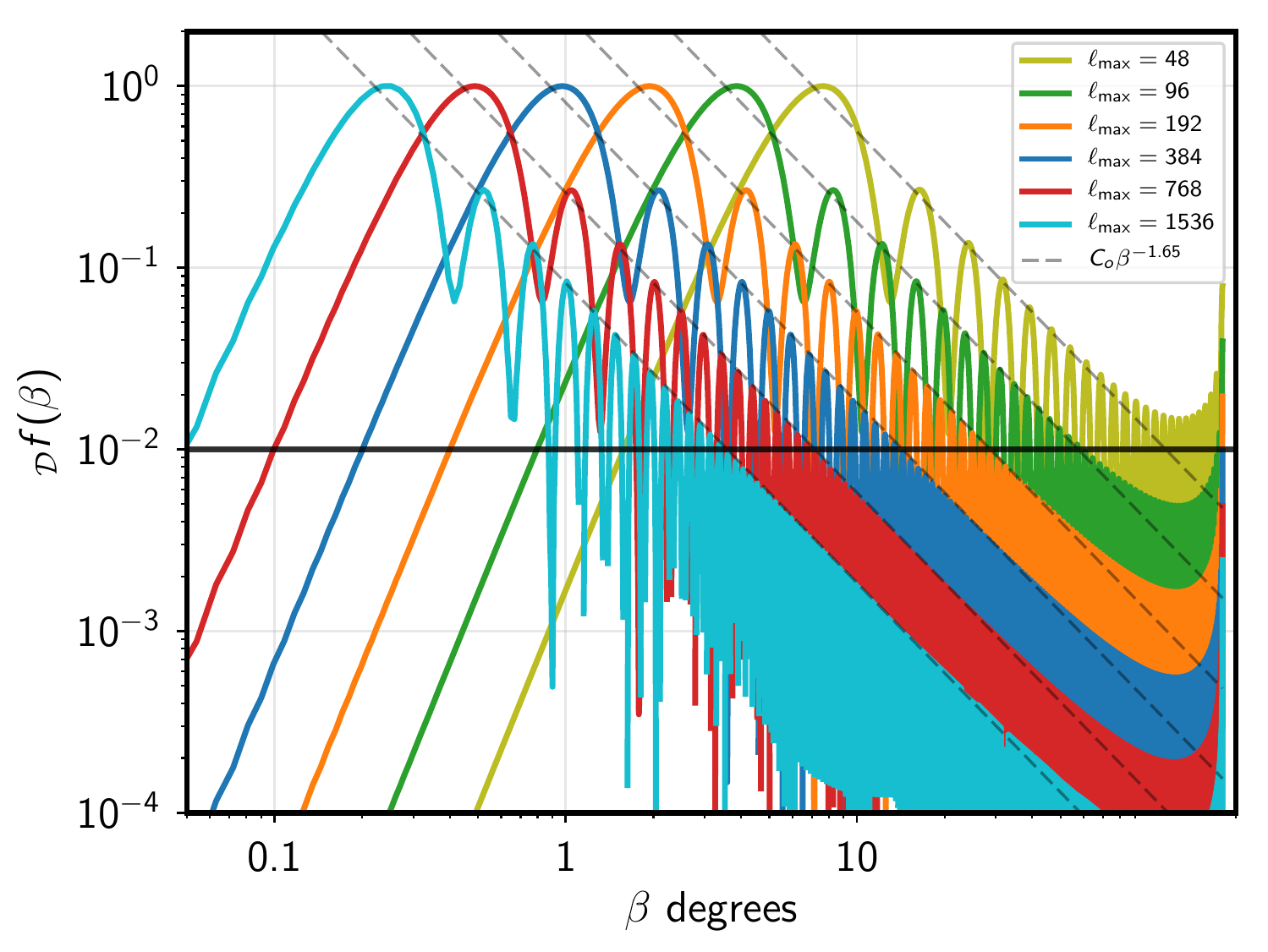}}\hfill
\subfigure[\label{fig:ifbeta}]{\includegraphics[width=0.325\columnwidth]{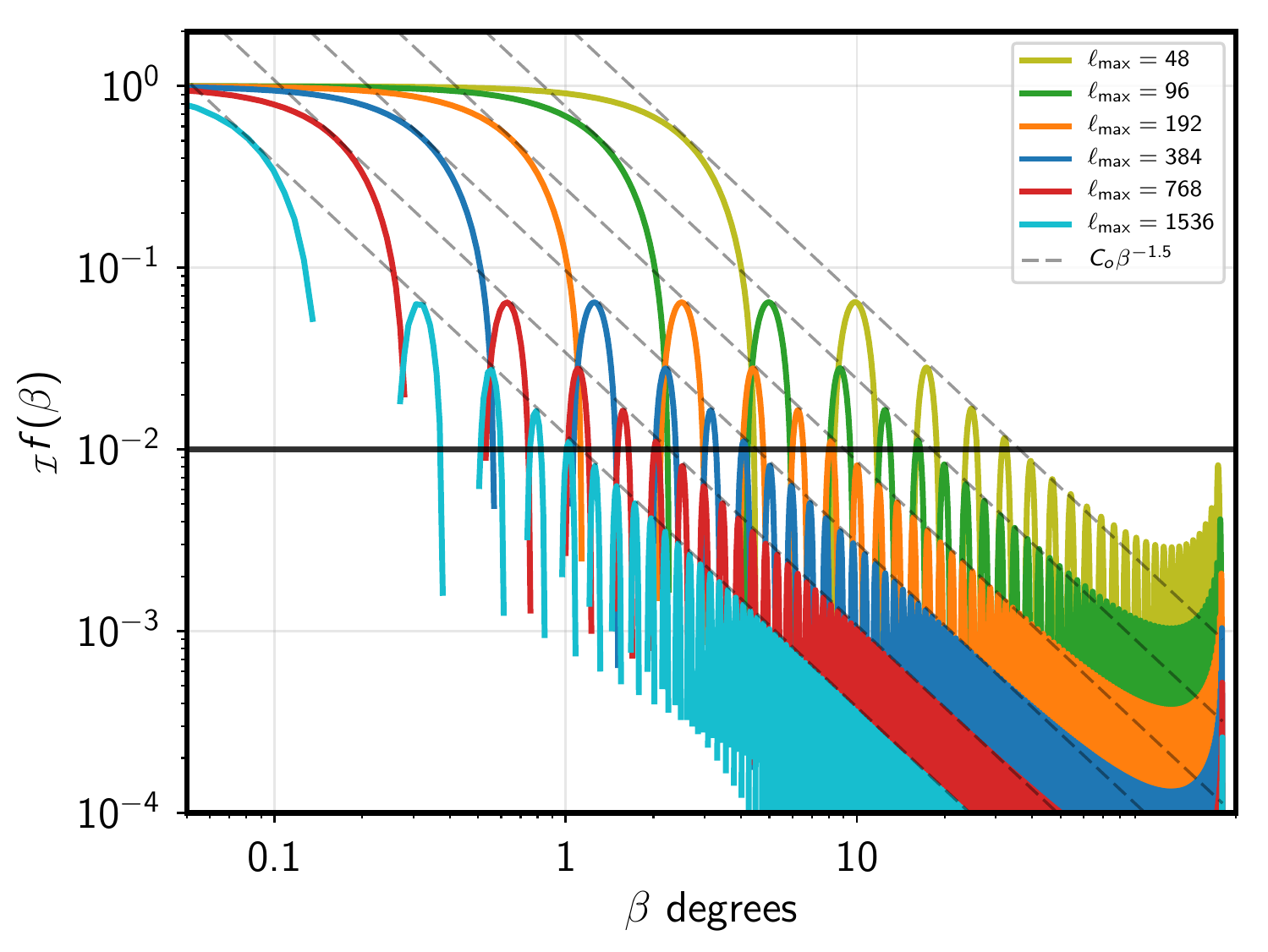}}\hfill
\centering
\subfigure[\label{fig:fbeta_avg}]{\includegraphics[width=0.325\columnwidth]{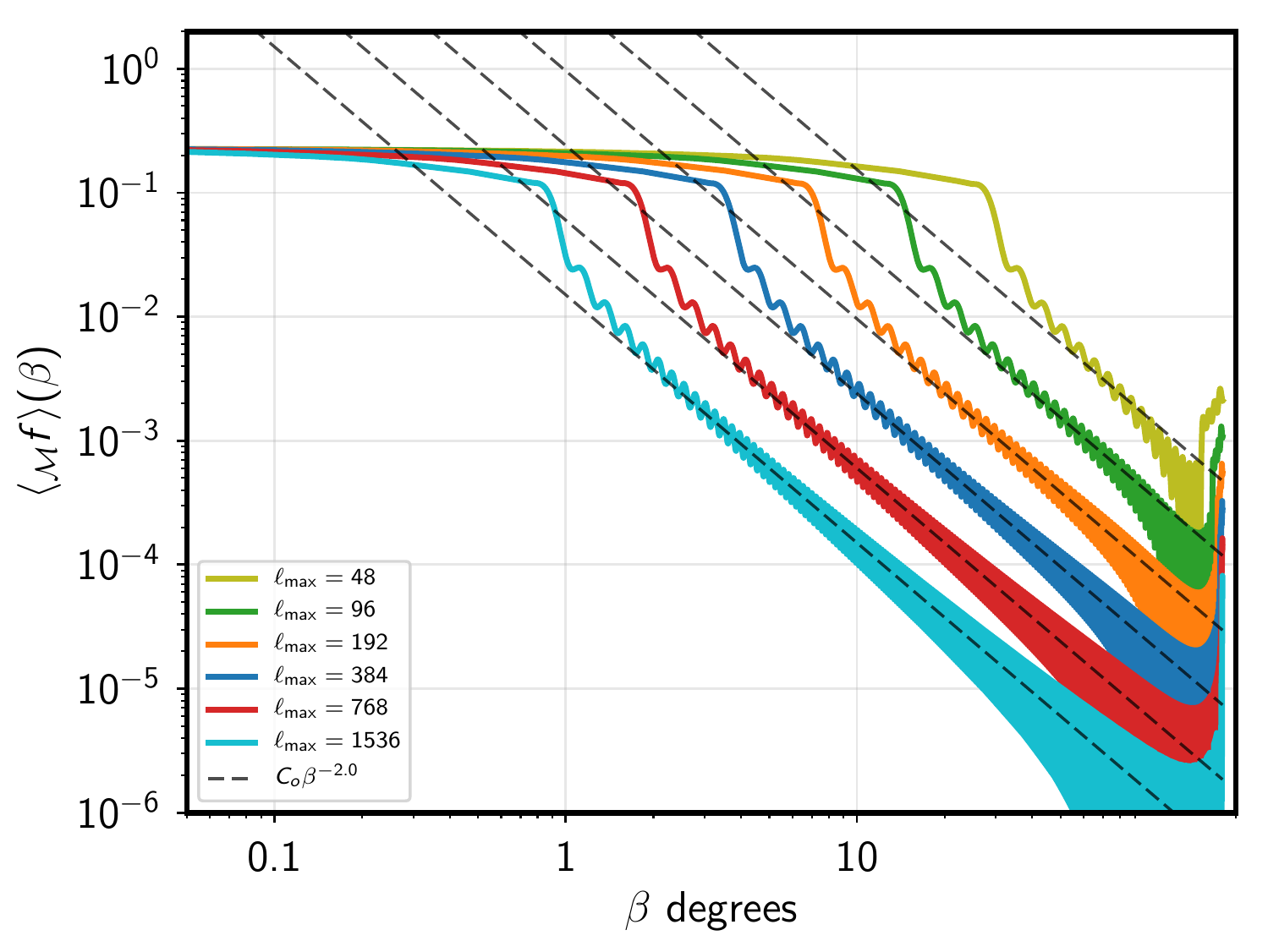}}
\subfigure[\label{fig:dfbeta_avg}]{\includegraphics[width=0.325\columnwidth]{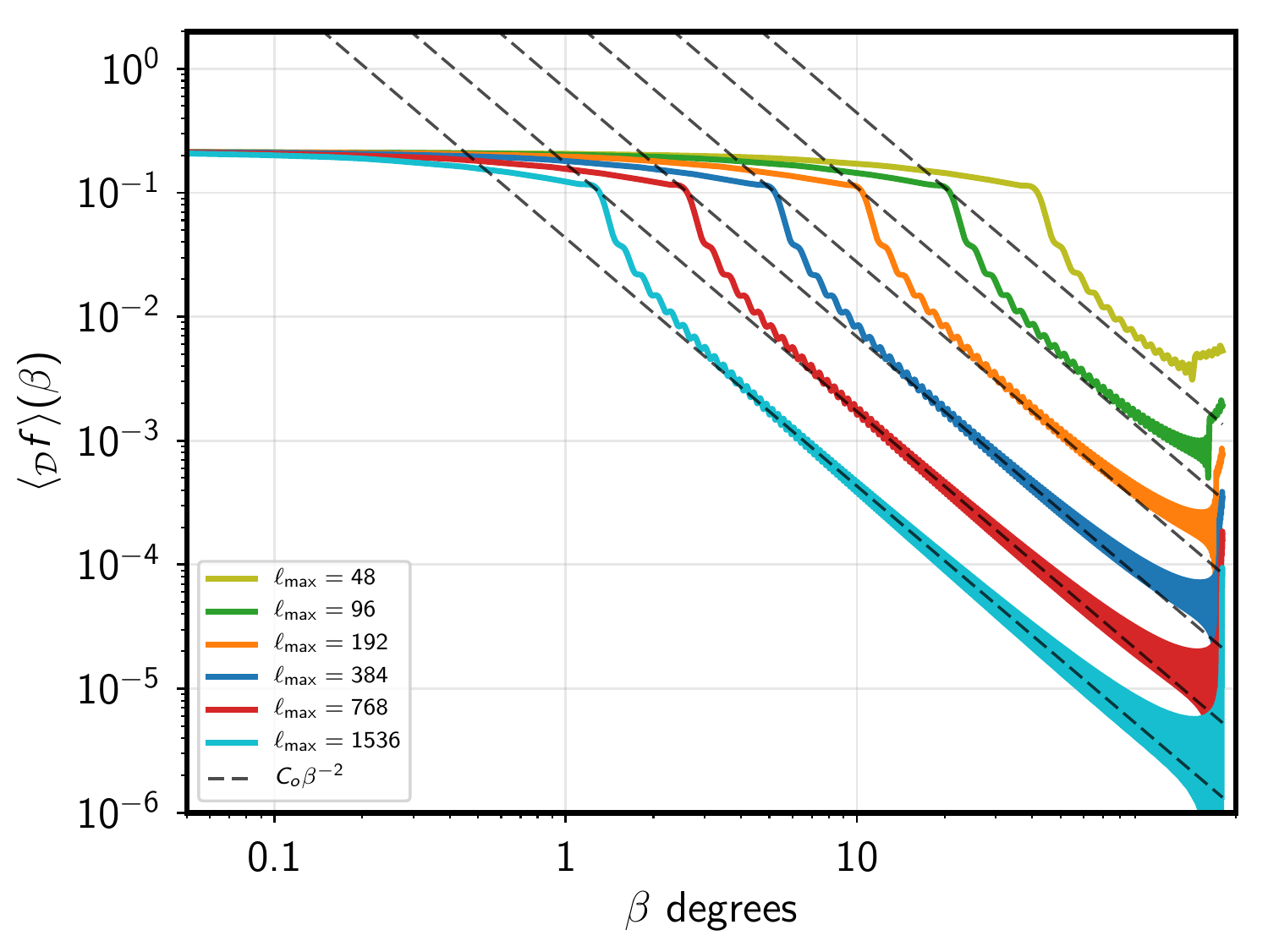}}

\caption{This figures depicts the radial functions for the different kernels and varying band limit, set by fixed $\ell_{\rm min}=2$ and varied $\ell_{\rm max}$ as indicated. \fig{fig:fbeta} depicts ${_{\mm}f}(\beta,\ell_{\rm min},\ell_{\rm max})$ the kernel for $Q/U$ to $E/B$ and vice versa. \fig{fig:dfbeta} depicts ${}_{\md}f(\beta,\ell_{\rm min},\ell_{\rm max})$ and  \fig{fig:ifbeta} depicts ${}_{\mi}f(\beta,\ell_{\rm min},\ell_{\rm max})$ respectively, the Stokes purification kernels. All the curves are normalized such that their maxima is set to unity. The horizontal solid black line marks the location where the amplitude of the respective kernels fall below 1\% of its maximum. The thin slanted dashed gray lines indicate a power law fit (by eye) to the envelope of the radial functions. The thick black short vertical dashed lines indicate the transition points as predicted by the empirically derived relation for the non-locality parameter $\beta_0={\rm min}(180^\circ,180^\circ \times 22/\ell_{\rm max})$.  We show moving averages over oscillations in \fig{fig:fbeta_avg} and  \fig{fig:dfbeta_avg}.  They depict the radial functions $\langle{_{\mm}f}(\beta,\ell_{\rm min},\ell_{\rm max})\rangle$ and $\langle{_{\md}f}(\beta,\ell_{\rm min},\ell_{\rm max})\rangle$ respectively where $\langle \cdots \rangle$ denotes a moving average over a step function centered at $\beta$ with a width $1.4\beta_0$. These smoothed functions are seen to be well fit by a power law $\propto \beta^{-2}$  and depict the average radial response.}
\label{fig:rad_ker_decay}
\end{figure}

The function ${_{\mm}f}$ is the radial part of the kernel that translates the Stokes parameters $Q$/$U$ to scalars $E$/$B$ and vice versa.  It has an oscillatory nature at intermediate angular separations and vanishes as $\beta \rightarrow 0$ and $\beta \rightarrow \pi$ (this is critical to ensure that the derived fields have the necessary spin properties), since the Euler angles $\alpha$, $\gamma$ are not uniquely defined at those separations.

The radial part of the kernel that decomposes the  Stokes parameters into parts that correspond to $E$ and $B$ modes  are also  necessarily non-local.  The function ${_{\mi}f}$ is the radial part of the band limited delta function $\mi$.  It expectedly has its maxima at $\beta=0$ and decays with increasing angular separation.  ${_{\md}f}$ has a vanishing value in the region where $\beta \rightarrow 0$ however it does not vanish at $\beta \rightarrow \pi$ as seen in \fig{fig:beta_kernel}.

\paragraph{Band limit dependence.} 
To quantify the non-locality of the real space operators, we study the radial extent of the respective radial kernels and their dependence on the maximum multipole accessible. We evaluate the radial functions for different values of $\ell_{\rm max}$, while keeping the lowest multipole fixed at $\ell_{\rm min}=2$. 

The resultant set of radial function are depicted in \fig{fig:rad_ker_decay}. The amplitude of these radial function scales up as $\propto \ell_{\rm max}^2$.  For clarity in the plot, their global maxima are normalized to unity.  This normalization highlights the key feature, that on increasing $\ell_{\rm max}$ the radial kernels shift left, attaining their global maxima at progressively small angular distances $\beta$.  At intermediate values of $\beta$, the envelope of the radial functions is fit well by a power law $ \propto \beta^{-n}$.  As the band limit changes, the small angle portion of the radial functions have a similar shape.

These finding are neatly summarized in the observation that the radial functions with different maximum multipoles are approximately self-similar over many oscillations, and are related by this telescoping and scaling property:
\begin{equation}{}_rf(\beta,2,\ell_{\rm max}) \approx \Big[\frac{\ell_{\rm max}}{\ell'_{\rm max}}\Big]^2{}_rf(\beta'=\frac{\ell'_{\rm max}}{\ell_{\rm max}} \beta ,2,\ell'_{\rm max}),\end{equation}
where ${}_rf , r \in [\mm, \md,\mi]$ denotes all the different radial functions. This telescoping property encapsulates both the amplitude scaling and leftward shift of the radial kernels on increasing the maximum multipole. {Specifically, when $\ell_{\rm max} > \ell'_{\rm max}$, the global maxima of the kernel with maximum multipole $\ell_{\rm max}$ is amplified by a factor $(\ell_{\rm max}/\ell'_{\rm max})^2$ with respect to the kernel defined by the maximum multipole $\ell'_{\rm max}$, while the period of the oscillation is compressed by a factor $(\ell'_{\rm max}/\ell_{\rm max})$. The detailed accuracy of this approximation varies with the angular separation $\beta$ and the difference in the maximum multipoles ($\ell_{\rm max}$, $\ell'_{\rm max}$)  used in evaluating the two functions.}

To quantify the non-locality of the scalar modes $E/B$, we can define a characteristic angular radius of the region from which the kernels get most of their contribution.   We define a non-locality parameter $\beta_{0}$ as the angular distance beyond which the function ${_{\mm}f}(\beta,\ell_{\rm min}=2,\ell_{\rm max})$ transitions to being consistently below 1\% of its maximum.
The empirical relation:
\beq
\beta_0= {\rm min}\left(180^\circ,180^\circ \frac{\ell_{0}}{\ell_{\rm max}} \right) \,,
\eeq
with $\ell_{0}=22$ provides a reasonable estimate of this transition point for ${}_{\mm}f$ as depicted by the short dashed vertical black lines in \fig{fig:fbeta}. Setting $\ell_{0}=10$ and $\ell_{0}=32$ predicts the transition points for the functions ${}_{\mi}f$ and  ${}_{\md}f$ respectively.

The radial function ${}_\mm f(\beta)$ is oscillatory, regardless of the maximum $\ell_{\rm max}$. The envelope of the oscillation decays with angular separation at intermediate $\beta$ but increases as $\beta \rightarrow \pi$.  We find that the positive envelope of ${}_{\mm}f$ scales as $\beta^{-1.6}$ at intermediate $\beta$.  However, when we average over several oscillations with a moving window, we find that the average behavior at intermediate $\beta$ scales as $\beta^{-2}$, visible in \fig{fig:fbeta_avg}. We set the width of the smoothing window to $0.3\beta_0$, and this results in the narrowing of the smoothing width as the band limit increases, so as to keep the number of oscillations averaged over roughly the same.  This agrees with the hypothesis for the {flat sky and continuum case}, for which e.g. Zaldarriaga (2001) \cite{Zaldarriaga2001a} argued that form of the radial function has to be ${}_\mm f(\beta>0)\simeq\beta^{-2}$ to ensure that the Fourier modes of Stokes $Q/U$ relate to those of $E/B$ merely by rotations (with no scale dependence). In addition, we find that ${}_{\md}f$ under a moving average shows a similar scaling of $\beta^{-2}$, as seen in \fig{fig:dfbeta_avg}.
Applying a moving average on radial functions with increasing band limit, we see that a larger fraction of angular domain matches the scaling of $\beta^{-2}$.   Extrapolating this trend to the case of a very high band limit one expects this behavior to approach $\beta^{-2}$ across the domain $\beta\in[\epsilon,\pi-\epsilon]$ for $\epsilon \ll 1$. In essence the moving average can be understood as the averaging resulting from discrete sampling of a continuum field (one with an infinite band limit) where the width of the averaging window corresponds to the size of the pixel.

Near the edges of the domain $\beta \rightarrow 0,\pi$ the window function partly falls outside the domain and this is in part responsible for the scaling deviating from the power law, but it is also important to note that ${}_{\mm}f$ and ${}_{\md}f$ need to vanish at $\beta=0$ and therefore the sliding average in the vicinity of these points cannot have the same power law behavior.
\section{Generalized polarization operators and recovering standard power spectra}\label{sec:generalized_operators}  
With our better understanding of the radial part of the kernel for CMB polarization, we can write down generalized $E/B$-like fields that depend on a different radial function, even one that we specify to have compact support.
The spin symmetry constrains the the azimuthal part of the real space kernels to be of the form $\sim e^{\pm i2 \alpha}$.  The radial parts of the standard operators are determined by the sum over spherical harmonics and varies as a function of the band limit. It is here that we may potentially choose alternate forms for the radial functions to suit certain kind of analysis.

We can systematically generalize the real space operator by introducing the following harmonic space filter function:
\beq
\tilde{\mathcal{G}} = {\begin{bmatrix} g_{\ell}^E & 0  \\  0 & g_{\ell}^B \end{bmatrix}} \,,
\eeq
where the functions $g_{\ell}^E$ and $g_{\ell}^B$ represent the harmonic representation of the modified radial functions and can in the most general case be chosen to be different for $E$ and $B$ modes. To simplify discussions, we proceed by setting $g_{\ell}^E = g_{\ell}^B= g_{\ell}$. Given this harmonic function $g_{\ell}$, we can define the real space operator $\bar{O}'$ which translates Stokes $Q/U$ to $E/B$-like scalars (and the inverse operator $\bar{O}'^{-1}$) in the following manner,
\begin{subequations} \label{eq:gen_qu2eb}
\beqry
{\bar O}' &=& {{}_0\mathcal{Y}} \, \tilde T^{-1} \tilde{\mathcal{G}} {{}_2\mathcal{Y}^{\ddag}} \, \bar T \,,\\
{\bar O}'^{-1}&=& \bar{T}^{-1} {{}_2\mathcal{Y}}\, \tilde{\mathcal{G}}^{-1} \tilde T {{}_0\mathcal{Y}^{\ddag}}.
\eeqry
\end{subequations}
The primed notation distinguishes these generalized operators from the default operators defined in \sec{sec:qu2eb} and \sec{sec:eb2qu}. We require both the forward and inverse operators to be well defined.   This constrains the choice of $\tilde{\mathcal{G}}$ to have a valid  inverse, which is important when recovering the standard CMB power spectra. The radial parts of this generalized operator and it's inverse are given by the following expressions,
\begin{subequations}\label{eq:generalized_radial_kernel}
\beqry 
G_{QU \rightarrow EB}(\beta) &=& G(\beta) = \sum _{\ell=2} ^{\ell_{\rm max}} g_{\ell}\frac{2 \ell+1}{4 \pi} \sqrt{\frac{(\ell-2)!}{(\ell + 2)!}} P_{\ell}^2(\cos{\beta}) \, \label{eq:mod_rad_forward} \,, \\
G_{EB \rightarrow QU}(\beta) &=& G^{-1}(\beta) = \sum _{\ell=2} ^{\ell_{\rm max}} g_{\ell}^{-1}\frac{2 \ell+1}{4 \pi} \sqrt{\frac{(\ell-2)!}{(\ell + 2)!}} P_{\ell}^2(\cos{\beta}) \,. \label{eq:mod_rad_inverse}
\eeqry
\end{subequations}
The default radial function is just a special case resulting from the choice $\tilde{\mathcal{G}}=\mathbb{1}$ ($g_{\ell}=1$), in which case $\tilde{\mathcal{G}}^{-1}=\tilde{\mathcal{G}}$ and therefore $G^{-1}(\beta) = G(\beta)={{}_{\mm}f}$.
\begin{figure}[!t] 
\centering
\subfigure[\label{fig:gl_gbeta}]{\includegraphics[width=0.48\columnwidth]{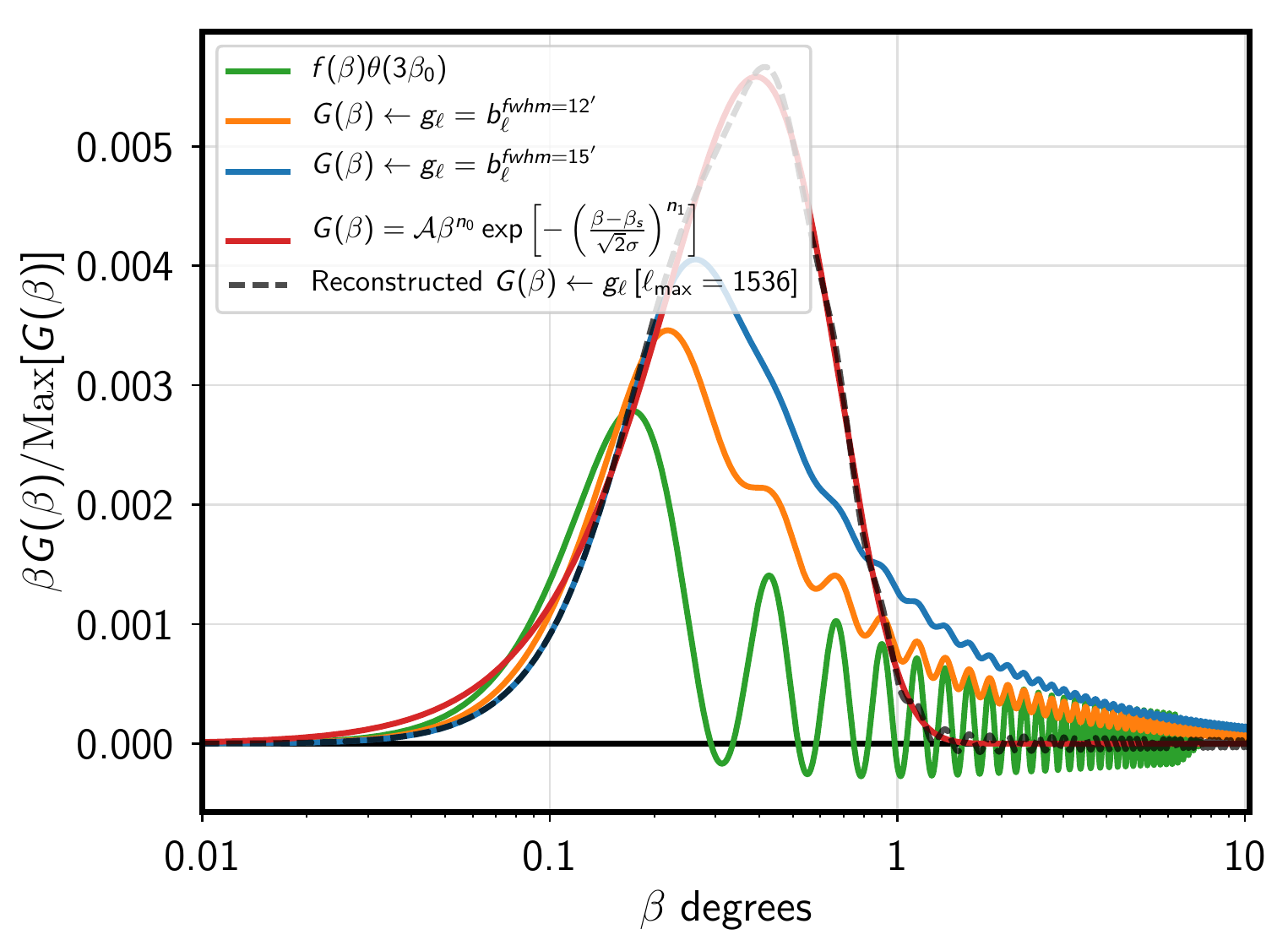}}
\subfigure[\label{fig:glbl}]{\includegraphics[width=0.48\columnwidth]{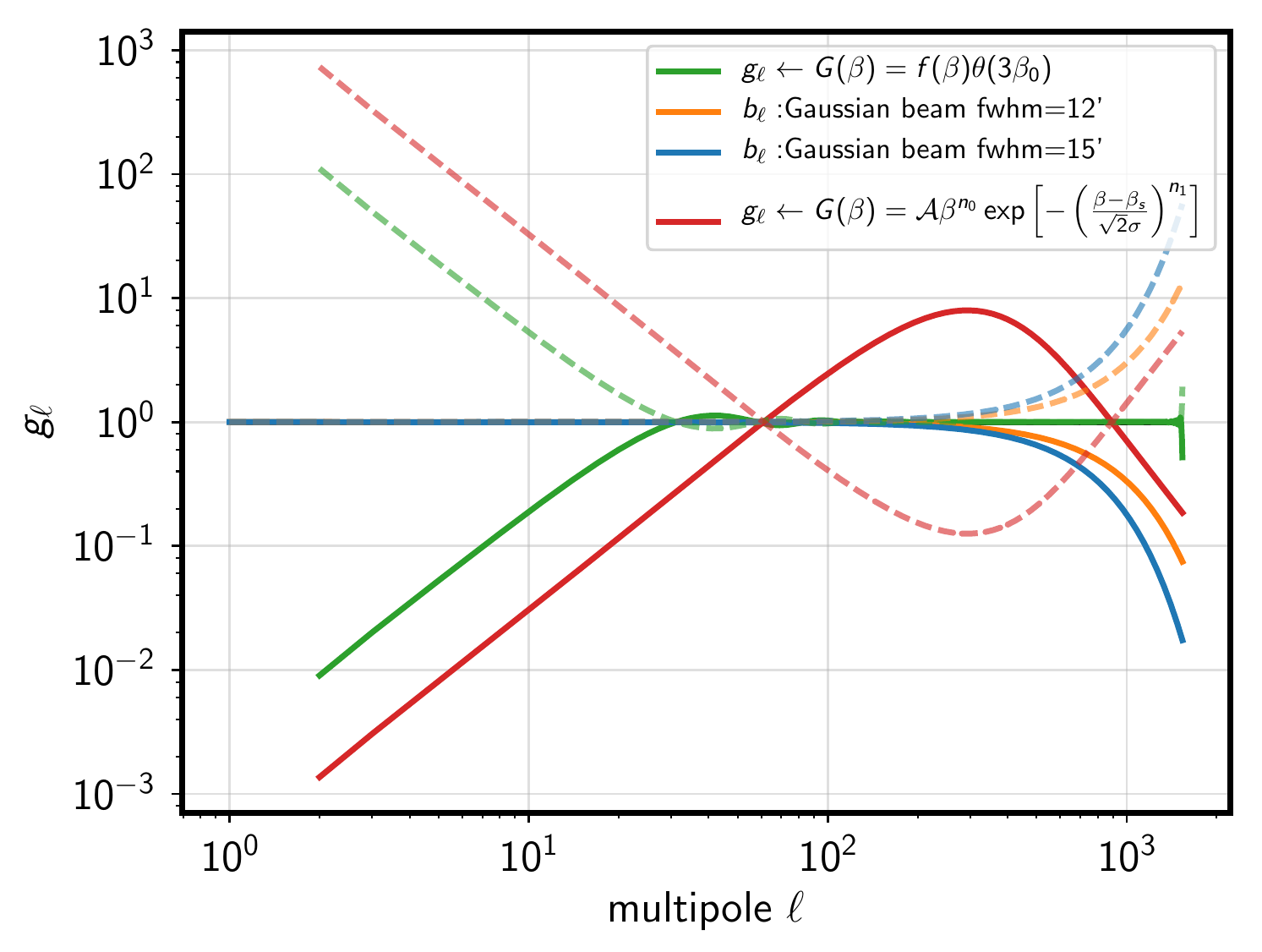}}
\subfigure[\label{fig:gl_bbeta}]{\includegraphics[width=0.48\columnwidth]{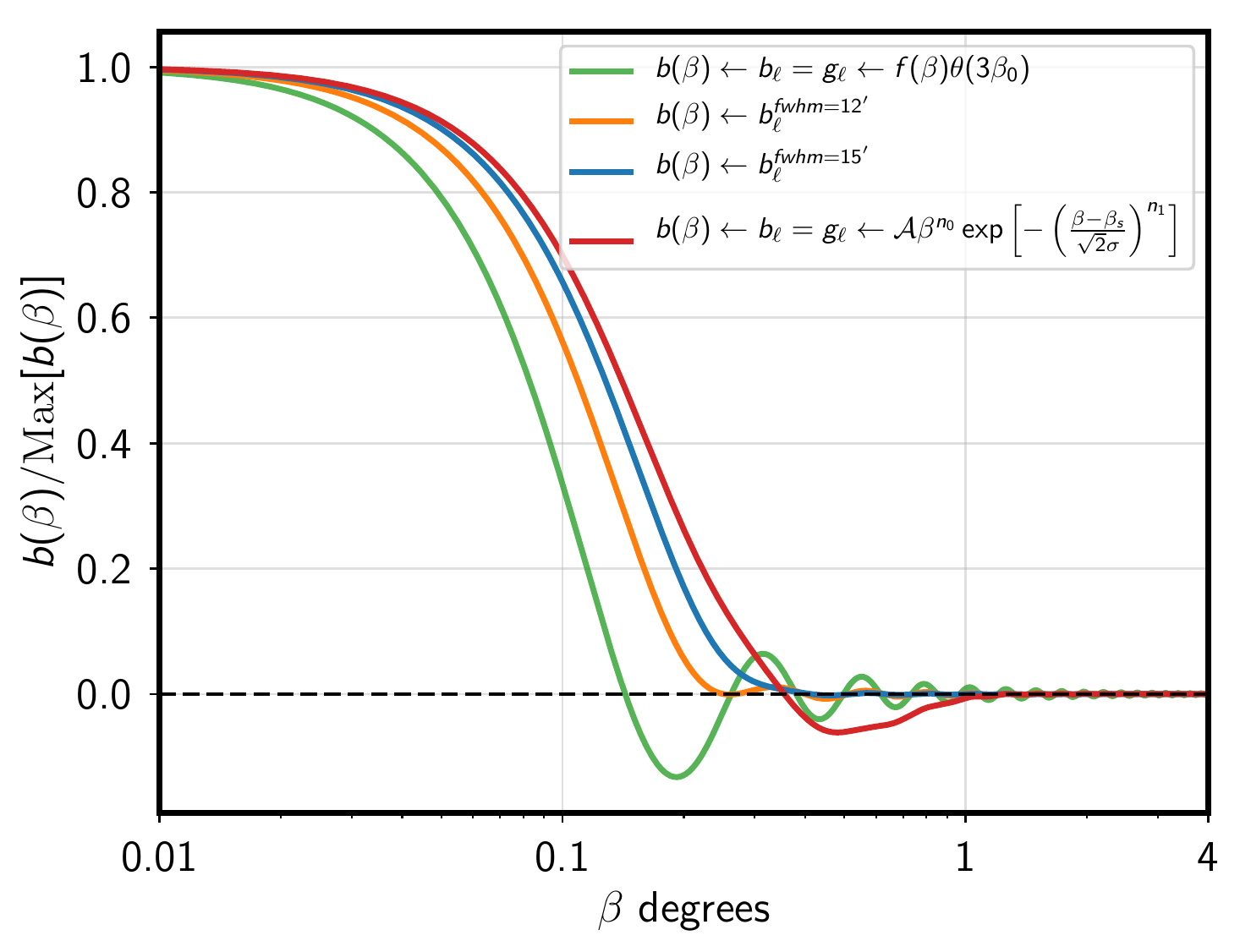}}
\caption{\textit{Top left:} The green line depicts the default radial kernel $f(\beta)$ defined in \eq{eq:qu2eb_gen_kernel}, multiplied by an apodized step function $\theta(3 \beta_0)$. The blue and orange lines depict the modified radial function resulting the beam harmonics $b_{\ell}$ corresponding to Gaussian beams with fwhm=15 \& 12 arc-minutes respectively. The red curve depicts an example modified radial function: $G(\beta)=\mathcal{A} \beta^{n_o} \exp{\left[ -\left( {(\beta-\beta_s)}{/\sqrt{2} \sigma} \right)^{n_1} \right]}$ with parameters set to the following values $[n_0=1;\, \beta_s=0 ;\, \sigma = 0.004 ;\, n_1=1.5]$. The black dashed curve depicts the band limited reconstruction of the modified radial function. \textit{Top right:} The harmonic representation of the respective radial functions as indicated by the legend. The dashed curves of the corresponding color depict the inverse of the harmonic functions. \textit{Bottom:}  The normalized beam function $b(\beta)$ evaluated from interpreting the respective harmonic functions as those corresponding to an effective beam applied to $E/B$. }
\label{fig:example_gbeta}
\end{figure}

While defining these generalized operators, it is more natural to choose the real space function $G(\beta)$ rather than the harmonic space $g_l$, bearing in mind the constraint that $G(\beta=0)=G(\beta=\pi)=0$. Employing the orthogonality property of associated Legendre polynomials it can be shown that the harmonic function $g_{\ell}$ is given by the expression,
\beq
g_{\ell} = 2 \pi \sqrt{\frac{(\ell-2)!}{(\ell+2)!}} \int _{0}^{\pi} G(\beta) P_{\ell}^{2}(\cos{\beta}) d\cos{\beta} \,. \label{eq:gb2bl}
\eeq
An arbirtrary $G(\beta)$ for which $g_{\ell} \neq 1$ can be equivalently thought in terms of the standard $E/B$ fields being convolved with some effective circularly symmetric beam whose radial profile is given by the expression,
\beq
b(\beta) = \sum_{\ell=0}^{\ell_{\rm max}} \frac{2 \ell+1}{4 \pi} g_{\ell} P_{\ell}^{0} (\cos{\beta})\,,
\eeq
where $g_{\ell}$ is the same harmonic function as that appearing in \eq{eq:generalized_radial_kernel}.
In contrast to the radial function $G(\beta)$, a beam function when appropriately normalized has the property $b(\beta) \rightarrow 1$ as $\beta \rightarrow 0$. Though the real space functions $G(\beta)$ (operating on Stokes parameters) and $b(\beta)$ (operating on scalar $E/B$) are different, in harmonic space they play identical roles.

In \fig{fig:example_gbeta}, we examine in more detail the relationship between the modified radial kernels and these beam harmonic coefficients.  
 \fig{fig:gl_gbeta} depicts the radial profile of the  effective beams corresponding to different radial kernels: the standard kernel modified by a radial cutoff, kernels corresponding to Gaussian smoothings of the $E/B$ fields, and a radial function without oscillations and an exponential cutoff.  Note that the smoothing tend to increase the non-locality, indicated by the shifting right of the maxima of the respective kernels, as one may have expected.  The exponential cutoff (red curve) by construction has a very small non-locality (parameterized by $\beta_0$).

\fig{fig:glbl} depicts the harmonic description $g_{\ell}$ for these respective radial kernels and beams.
Finally, the beam that the modified radial kernels effectively apply to the $E/B$ fields are shown in \fig{fig:gl_bbeta}.  Note that the beam function corresponding to the default radial kernel ($g_{\ell}=1$) is merely a band limited representation of a delta-function beam.

The generalized convolution kernels defined in the previous section, when operated on the Stokes vector returns some scalar $E'$ and $B'$ mode maps,
\beq
\bar{S}' = \bar{O}' \bar{P}
\eeq
which are merely filtered versions of the standard $E/B$ modes maps. The filter function is simply $g_{\ell}$, which is easily obtained from the modified radial function $G(\beta)$.  It can be simply interpreted as the set of harmonic coefficients for some azimuthally symmetric beam. The power spectra of the modified scalar fields $E'$ and $B'$ are thus related to the spectra of the standard $E$ and $B$ fields via the following relation, 
 \begin{subequations}
 \beqry
C_{\ell}^{EE,BB,EB} &= &C_{\ell}^{E'E',B'B',E'B'} /   g_{\ell}^2\,,\\
C_{\ell}^{TE,TB}  &=&  C_{\ell}^{TE',TB'} / g_{\ell}\,,
 \eeqry
 \end{subequations}
 where $C_{\ell}$ denotes the angular power spectra and $T$ refers to the temperature anisotropy map. Therefore the standard CMB spectra can always be recovered as long as the $1/g_{\ell}$ and $1/g_{\ell}^2$ are well behaved functions, which can be ensured by making a suitable choice for the modified radial function $G(\beta)$. 

\paragraph{Relation to the spin raising $\eth^2$ and lowering $\bar{\eth}^2$ operators.}
Recall that on operating twice with the spin lowering operator on the Stokes charge ${}_{+2}X$ results in filtered version of $E/B$ maps as in \eq{eq:ebdef}. Now note that it is possible to construct a modified real space operator by choosing the harmonic space function to be $g_{\ell} = [{(\ell+2)!/(\ell-2)!}]^{1/2}$, resulting in similarly filtered $E/B$ maps as follows:
\beq 
[\mathcal{E} + i \mathcal{B}](\hat{n}_e)=- \Delta \Omega\sum_{q=1}^{N_{\rm pix}} \Bigg\lbrace  \left[  \sum_{\ell=\ell_{\rm min}}^{\ell_{\rm max}} \frac{2 \ell+1}{4 \pi} P_{\ell}^2(\beta_{qe}) \right] e^{-i2\alpha_{eq}} {}_{+2}X(\hat{n}_{q}) \Bigg\rbrace \,. \label{eq:bl_ebdef_lower} 
\eeq
Comparing to \eq{eq:ebdef_lower} makes apparent the following mapping:
\beq
\bar{\eth}^2_{\textrm BL} \equiv \Delta \Omega \sum_{q=1}^{N_{\rm pix}} \left[ \sum_{\ell=\ell_{\rm min}}^{\ell_{\rm max}} \frac{2 \ell+1}{4 \pi} P_{\ell}^2(\beta_{qe}) \right] e^{-i2\alpha_{eq}}\,,
\eeq
where we use the notation $\bar{\eth}^2_{\textrm BL}$ to represent the band limited version of the spin lowering operator, that is relevant in the case of a discretely sampled spin-2 field on the sphere. In the limit of $\ell_{\rm max} \rightarrow \infty$, the $\bar{\eth}^2_{\textrm BL} \rightarrow \bar{\eth}^2$.
The band limited version of the spin raising operator $\eth^2$ is derived by simply taking the conjugate of the above equation.
\section{Understanding polarization signatures of magnetized filaments}
\label{sec:pol_filaments}

The real space kernels give us a better intuitive understanding of the $E/B$ modes associated with physical objects.  For example, a simple model for a magnetized filament has the magnetic field threaded along a linear gas over-density.  Precession of the dust grains around the magnetic field leads to a net polarization perpendicular to the magnetic field (and perpendicular to the filament overall).  For a filament aligned North--South, the polarization will be horizontal or $Q<0$, $U=0$ (left pane of \fig{fig:polfilaments}).  The Green's function kernels for horizontal polarization are rotated by 90 degrees relative to the components of $\mm_G$ in \fig{fig:vis_kernel}.

The kernel can be thought of as the orientable nib of a calligraphy pen or paintbrush that we can trace along the filament.  The positive components for the $E$ part of the Green's function align and reinforce along the filament, consequently the filament is highlighted as a segment with $E>0$.  Since the over-density will also have emission in total intensity, this naturally predicts a positive $TE$ correlation for magnetized filaments.  The $E$ pattern is somewhat negative along the outside of the filament, also a consequence of the kernel shape.

The $B$ part of the Green's function, traced along the filament, cancels itself except at the filament ends.  This results in a non-zero $B$ pattern for the filament.  
 For a North--South filament, the $B$-mode pattern is positive on the northeast and southwest, and negative in the northwest and southeast. (\fig{fig:polfilaments} shows East to the left in sky map convention.)  The size of this $B$-mode pattern is set by the dimensions of the filament, chiefly the width.  In contrast to the kernel radial fuctions, which are more compact at higher band limit, the filament $B$-mode  pattern does not depend much on the band limit, provided it is sufficient to resolve the filament structure.  A filament with a more gradual edge has a $B$-pattern that is more spread out, but follows the same general structure.  The work in \cite{Zaldarriaga2001a} correctly argued that linear filaments that are infinite and without ends can only produce $E$-power, but we find that for realistic filaments, the ends produce $B$-mode power with a clear signature. 

The non-zero $B$ result is somewhat surprising given that the polarization pattern is symmetric to both horizontal and vertical reflections through the filament center.  However, unlike a circular ring, this filament is not a configuration with a definite parity.  Since the scalar description of polarization is coordinate independent, the $E/B$ patterns do not depend on the orientation of the filament.  For example, a filament inclined at $45^\circ$ will have a similarly inclined $E/B$ pattern, but different reflection symmetries.  

 Changing the polarization direction within the filament changes the $E/B$ patterns.  A $90^\circ$ rotation of the polarization with respect to the filament changes the sign of both $E$ and $B$.  In a polarization pattern aligned at $45^\circ$ to the filament, the $E$ pattern will swap with the $B$ pattern.  Careful study of the $E/B$ mode power in filaments can provide insights into the orientation of the magnetic field with respect to the axis of the filament, this information could potentially shed light on the internal dynamics of filaments.
\begin{figure}[t]
\includegraphics[width=0.5\columnwidth]{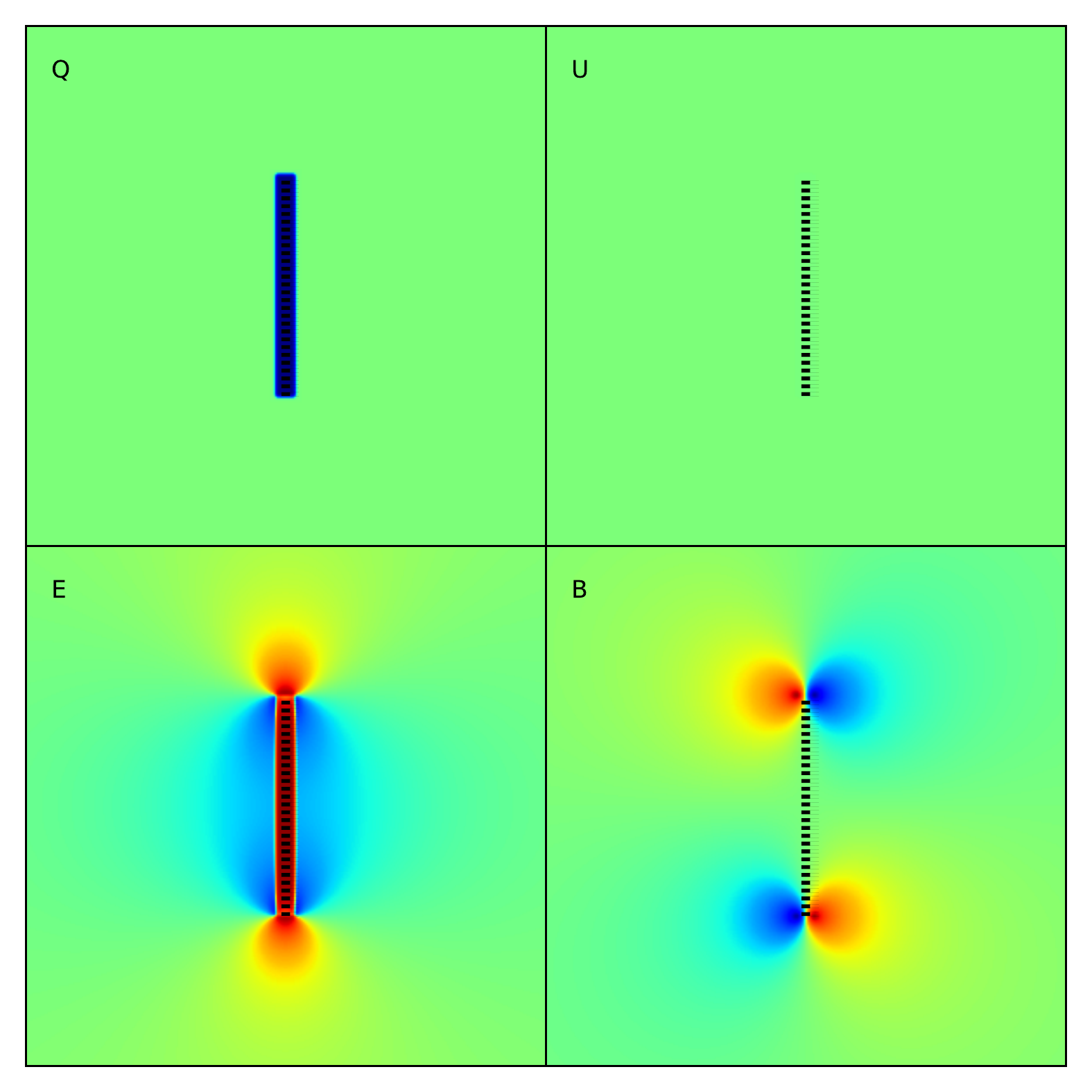}
\includegraphics[width=0.5\columnwidth]{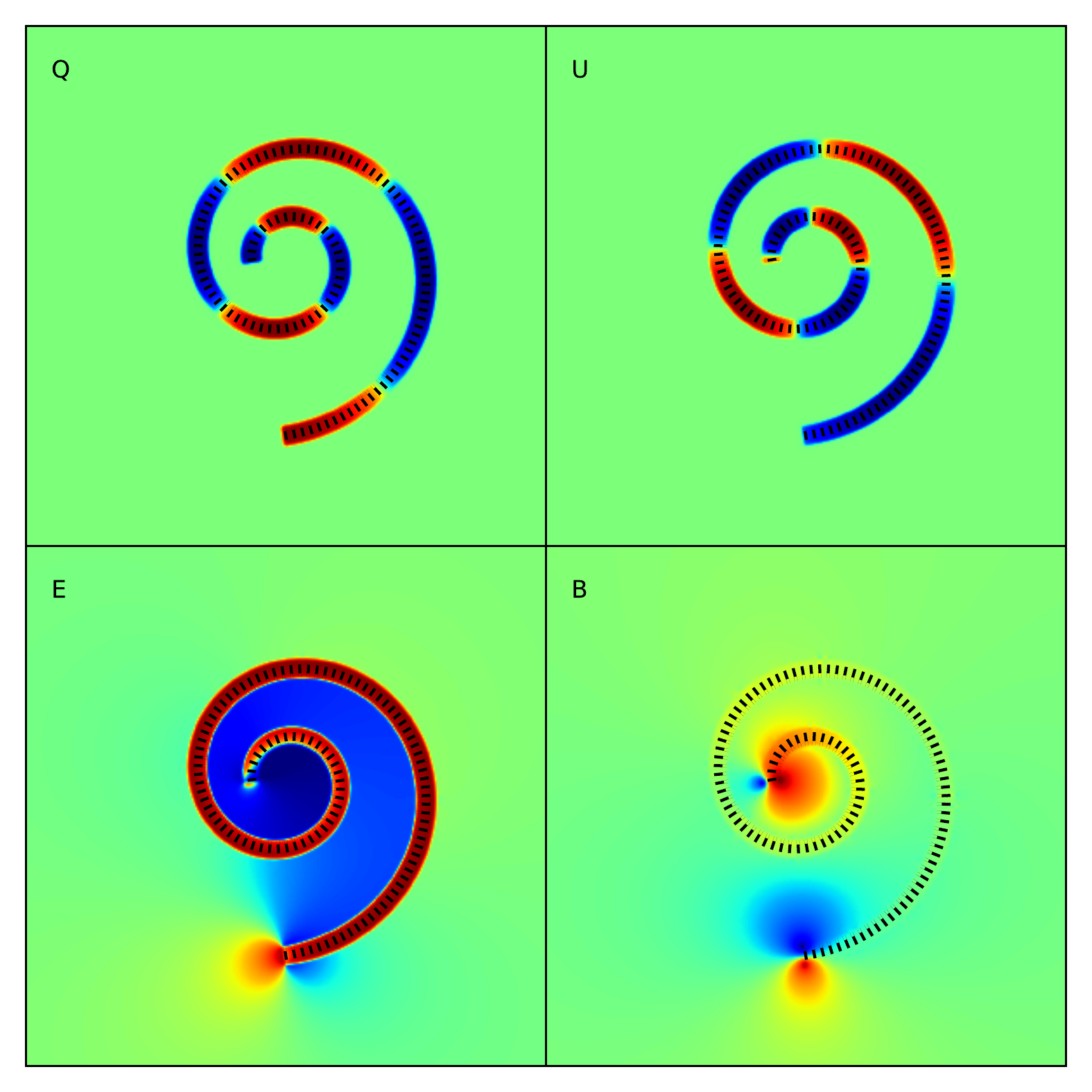}
\includegraphics[width=1.0\columnwidth]{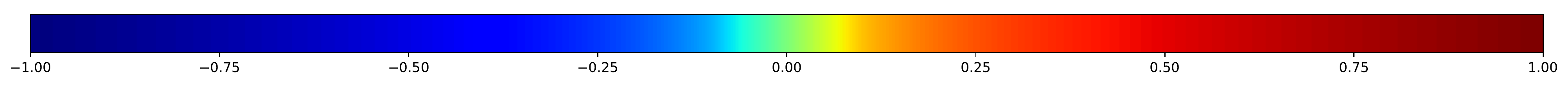}
\caption{ The polarization signals of toy filament structures. In a filament organized perfectly along a magnetic field line, the polarization will be perpendicular to the filament direction.  The $E/B$ modes of filaments are in some ways easier to think about than the Stokes parameters. Left panels: in a straight filament, the $E$-mode is positive along the filament and at the ends, but negative along the sides.  The $B$-modes are only non-zero at the ends.  Right panels: in a curved filament, the $E$-mode is again positive along the filament.  Outside the filament, the $E$-mode is more negative on the interior of the curve than the exterior.  The $B$-modes are again non-zero at the ends, akin to the straight filament case, but also along the filament due to the changing radius of curvature. In all images, the longitude angle increases to the left (which is East in sky convention).  All plot are on a common, arbitrary color scale.}
\label{fig:polfilaments}
\end{figure}

The intuition from the real-space kernels holds also when we distort the shape of the filament.  If the filament were bent around into a circle, the positive and negative parts of the $B$ pattern will cancel, and we are left with a hoop of pure $E$ pattern. Here it is important to note that this cancellation will not happen for an ellipse or any loop of non-constant radius of curvature.
The same general description holds for a spiral-shaped filament, which can can be viewed as the distortion of the straight filament (right panel of  \fig{fig:polfilaments}).  The filament is highlighted by positive $E>0$.  The $E$-pattern is more negative on the interior of a curve than on the exterior, and the concentric rings of filamentary structure make an increasingly negative $E$ value inside. While the $B$-pattern is again concentrated at the ends of the filament in an oriented pair of positive/negative fluctuations, note that it is non-vanishing in the intermediate regions along the filament due to the varying radius of curvature of the spiral.  A spiral turning with the other handedness will have a $B$-pattern with the opposite sign.

A stacking analysis of Planck data \citep{2016A&A...586A.141P} sees $E>0$ along filaments (selected from intensity data), but claims no $B$-mode signal above the noise. We predict that a $B$-mode signal from the filaments should be present, since they have a finite length of a few degrees.  Detecting $B$-modes from filaments will be easier with more signal-to-noise and may require a more careful filament analysis, rescaling and aligning the filament ends.  While the detectability of the $B$-mode signature from stacked filaments in Planck data calls for a more careful assessment, we should see it in higher-fidelity data. 
\section{Discussion}\label{sec:discussion}
In this work, we presented a first derivation of the real space operators on the sphere that transform the Stokes polarization vector to into a vector of scalars and vice versa.  We also presented real space operators that directly decompose the full Stokes vector \vp{} into vectors \vp{E} and \vp{B} that correspond to the respective scalar modes.  To facilitate these derivations we introduced a vector-matrix notation which allows for concise book keeping of all the standard operations involved in the analysis of  polarization maps (or any spin-2 fields) on the sphere. This real-space analysis method trivially generalizes to maps of arbitrary spin.

These real space operators offer a spatially intuitive way of understanding the different decompositions of the Stokes vector on the sphere. We explicitly  demonstrated that all the real space operators are separable into azimuthal and radial parts. While the azimuthal part of the operators is primarily responsible for handling the spin decomposition, the radial weights determine the non-local dependence of the resulting fields on the original fields.  Only the radial part of each kernel depends on the band limit. 
The radial parts of the operator kernels are roughly self-similar in the sense that the radial kernels evaluated with some band limit are related to other radial kernels (evaluated with a different band limit) by an approximate rescaling of the function. We use this property to define a non-locality parameter $\beta_0$ as the angular distance at which the amplitude of the radial kernel falls below one percent of its maxima and it  is a function of the maximum multipole $\ell_{\rm max}$ available for the analysis. We empirically show that the non-locality parameter is approximately given by: $\beta_0 = \mathrm{min}(180^{\circ}, 180^{\circ} \frac{\ell_{0}}{\ell_{\rm max}})$ with $\ell_{0}=22$ for the operator that converts Stokes $Q/U$ to scalars $E/B$ and vice versa (see \fig{fig:rad_ker_decay}). An analysis in  \cite{Zaldarriaga2001a} treated real space $E/B$ operators in the flat sky.  It did not explicitly derive the radial part of the kernel, but argued on geometric grounds that it should fall with angular separation as $\beta^{-2}$.  We find that this agrees with the average behavior of radial functions after averaging oscillations and note that some such averaging always takes place in practice due to pixelization of the signal.  However, for precision reconstruction of $E/B$ on the sphere, the oscillations must be taken into account.

Our careful study of the real space operators show that they can be expressed either as Green's functions or as convolving beam functions.  The convolution interpretation is not a totally new concept.  It guides the discussion in \cite{Zaldarriaga2001a,Chiueh2002} and closely relates to the popular spokes and pinwheel descriptions of the $E/B$ modes. However, the radiation/Green's function interpretation of the operators is a new one and is discussed here for the first time. These two different interpretations of the operators emerge from the expression of the kernels in terms of the forward or inverse rotation Euler angles.  The mathematical forms of the Green's function and convolution kernels swap roles (and are conjugated) when transforming back to $Q/U$ from $E/B$.

The Green's function interpretation provides some useful insights into these operations. In particular it allows us to think of ${}_{+2}X=Q+iU$ as some spin-2 charge which radiates out a complex spin-0 scalar field $E+iB$. The resulting complex scalar maps can be then understood as arising from superposition of the radiated spin-0 scalar fields emanating from all the spin charges on the sphere.  The $E/B$ mode maps are merely the real and imaginary parts of this field. Results that demonstrate the equivalence of these real space methods to the conventional harmonic space methods will be presented in the next paper in this series.

Deeper understanding of the non-locality of the real space operators has allowed us to generalize the real space operators that transform between the spin-2 and spin-0 representations of the CMB polarization. We presented a systematic procedure to modify and generalize the construction of the scalar polarization fields and to control the radial kernels, specifying them with few restrictions.  We argue that these modifications to the radial kernel have the same effect as a smoothing operation on the $E/B$ fields by a circularly symmetric beam.  Therefore it is trivial to recover the standard CMB angular power spectra from the modified scalar polarization maps resulting from the modified kernels.  We noted that the standard spin-raising ($\eth^2$) and spin-lowering ($\bar{\eth}^2$) operators are special cases of these generalized operators which allowed us to present a band limited representation of these operators. 

Modified real-space operators with compact kernels could open several alternative analysis routes in the future.  No spin-harmonic transforms are necessary as the real space operators only rely on computing the Euler angles which can be done on the fly.  The radial functions (depending on $P_{\ell}^{2}$) need be tabulated only once at some determined resolution.  Especially given the Green's function interpretation of these operators, their implementation is trivially parallelizable over a compact domain, since the $E/B$ contribution from the Stokes charges in each pixel can be evaluated independently.  Alternatively, the spatially-varying convolution kernel could be applied as a polarized effective beam, as implemented in \cite{2011ApJS..193....5M} in a parallel scheme.

The real space kernels could in principle be incorporated into the pointing matrices for map making, allowing maps of $E/B$ to be made directly from instrument data, without the need for Stokes parameter maps as intermediate products. The pointing matrix---projecting the maps into the time-ordered data at a point---requires the convolution version of the kernel, while the transpose pointing matrix---projecting the time-ordered data to the map---requires the Green's function kernel. The method would be similar to pixel-based strategies to deconvolve an instrument beam during map making \citep[e.g.][]{2010ApJS..187..212C}. Such a strategy for map making would result in the pointing matrix in being much less sparse, hence making a practical implementation significantly challenging. Using the compact radial kernels might help with restoring some of the sparseness of the pointing matrix, however a real world implementation of this method requires a more careful feasibility study.
 
Since the real space operators let us tune the locality of $E/B$ maps, this can be potentially exploited to eliminate foreground contamination from distant parts of the sky. Such applications can be difficult to implement using conventional harmonic space methods.  For instance, the real space operators can be defined such that the locality of their radial kernels is varied on different portions of the sky, dictated by say the foreground morphology, resulting in some modified scalar $E'/B'$ maps.  While this idea seems interesting, the usefulness of this idea will depend on whether the standard $E/B$ mode spectra are easily recoverable in this fashion.  We will explore some of these possible analysis directions in the future papers of this series.

Finally, the toolbox of real-space operators gives us more intuition about the $E$- and $B$-mode structure of polarized gas and dust filaments in the Milky Way, an important foreground for inflationary science. We demonstrate that filaments with finite length or changing radius of curvature result in $B$-mode patterns in addition to the $E$-mode pattern already expected.  We therefore predict a characteristic $B$-mode pattern from filament sources that should be observable in future polarization measurements.
\section*{Acknowledgments}

This work was supported by NASA Astrophysics Theory Program under grant NNX17AF87G.  We thank David~C.~Collins for useful discussions.  We also thank Anthony Challinor and Mathieu Remazeilles for closely reading a draft of this text and providing useful comments.
\endgroup

\bibliographystyle{JHEP}
\bibliography{ref}
  
\end{document}